\documentclass[aps,prb,reprint,superscriptaddress,nofootinbib,longbibliography]{revtex4-2}

\usepackage{amsmath,amssymb,mathtools,bm}
\usepackage{booktabs}
\usepackage{microtype}
\usepackage{xcolor}
\definecolor{cbBlue}{HTML}{0072B2}\definecolor{cbOrange}{HTML}{E69F00}
\definecolor{cbGreen}{HTML}{009E73}\definecolor{cbVerm}{HTML}{D55E00}
\definecolor{cbPurple}{HTML}{CC79A7}\definecolor{cbSky}{HTML}{56B4E9}
\usepackage{graphicx}
\usepackage{tikz}
\usepackage{pgfplots}
\usetikzlibrary{decorations.pathmorphing,arrows.meta,calc,patterns}
\pgfplotsset{compat=1.18}
\usepackage[colorlinks=true,allcolors=blue!55!black]{hyperref}
\hypersetup{pdftitle={Resonance statistics, Fock-space branching, and long-range pair networks in slowly varying interacting chains},
pdfauthor={Yogeshwar Prasad},
pdfsubject={Many-body localization; slowly varying potentials; long-range interactions},
pdfkeywords={many-body localization, slowly varying potential, long-range interactions, resonances, Fock space}}

\newcommand{\dd}{\mathrm{d}}
\newcommand{\ii}{\mathrm{i}}

\newcommand{\e}{\mathrm{e}}
\newcommand{\order}{\mathcal{O}}

\begin{document}

\title{Resonance statistics, Fock-space branching, and long-range pair networks in slowly varying interacting chains}

\author{Yogeshwar Prasad}
\email[Contact author: ]{yogeshwar@snu.ac.kr}
\affiliation{Department of Physics, Hanyang University, Seoul 04763, Korea}
\affiliation{Research Institute of Basic Sciences, Seoul National University, Seoul 08826, Korea}
\date{\today}

\begin{abstract}
In a slowly varying aperiodic potential $h_i=h\cos(2\pi\beta i^n+\phi)$ with random
power-law density interactions $V_{ij}/|i-j|^\alpha$, the resonant-object ensemble changes
across $\alpha=2n$, wherever Hartree fragmentation is operative, from isolated two-level bonds
to a mixture of fragmented multi-site clusters and isolated wing bonds that survive at finite
density, while the pair-resonance scaling
$x_{\rm pair}^{4-2n-\alpha}\ln x_{\rm pair}\propto h^2/(Vt_0)$ is unchanged~\cite{letter}.
Here we develop the microscopic resonance theory underlying these results, together with its
domain of validity.  We derive the exact phase-averaged resonance statistics --- the
supply $N_0\propto L^{2-n}/h$, the correlated common-phase comb, and the closed-form Hartree
variance --- proving that the interaction leaves the leading supply law unchanged.  Exact
construction of the resonant Fock-space graph at $L\le18$ shows that the order-one
forward-branching scale $h_{\rm FB}\propto L^{2-n}$ carries no giant component: one-step
branching and connectivity are inequivalent.  We separate the fixed-pair matching law from the
shell-averaged $q\ln(1/q)$ law of the bond ensemble and develop the comb into a mesoscopic shell
theory; we map the fragmentation domain, with its support threshold $V_*(\alpha)$ and the
boundary-healing recursion; and we treat the marginal case $\alpha=2$, where shell and matching
marginalities compound into a double logarithm.  Three long-range thresholds emerge with distinct
meanings --- $\alpha=1/2$ (the exact variance threshold of the random Fock-space energy and the
square-summability boundary of the leading LIOM-dressing estimate), $\alpha=2n$ (change of the
local resonant objects), and $\alpha=2$ (marginality of the long-range shell sum) --- none of
which is, by itself, a localization transition.
\end{abstract}

\maketitle

\section{Introduction}

\label{sec:intro}

Many-body localization (MBL) admits complementary formulations in terms of Fock-space localization,
resonant processes, and quasi-local integrals of motion (LIOMs)~\cite{Basko,Huse_2007,Serbyn2013,HuseNO2014,Ros2015,Imbrie2016,Chandran2015,Mierzejewski2018}; see Refs.~\cite{Abanin_rev,Sierant_rev} for reviews.
Long-range interactions complicate the picture because an algebraically small matrix element can be
compensated by a growing number of distant resonant partners; resonant-pair arguments predict an
instability for sufficiently slowly decaying couplings, with a characteristic $\alpha<2d$ threshold
in $d$ dimensions~\cite{Burin2006,Burin2015,Yao2014,Mirlin}.  Long-range interacting MBL has been studied in a variety of settings~\cite{Hauke2015,NandkishoreSondhi2017,DeTomasi2019,Thomson2020,Maksymov2020,Botzung2021,Sierant2019,Vu2022,Defenu2023}, and long-range resonances specifically in quasiperiodic chains~\cite{Padhan2026}.

A separate source of structure arises in deterministic slowly varying (SV) potentials,
\begin{equation}
 h_i=h\cos(2\pi\beta i^n+\phi),\qquad 0<n<1,
 \label{eq:potential}
\end{equation}
whose single-particle spectrum supports mobility edges, with extended states near the band center
for weak amplitude and localized states otherwise~\cite{Fishman,Sarma1990}.  In the strong-potential
regime relevant here the spectrum is localized throughout, but the local detuning scale that controls
nearest-neighbor hopping decreases systematically with position, and it is this inhomogeneity rather
than the localization length that organizes the interacting problem.  Many-body localization in
slowly varying potentials with conventional short-range interactions has recently been studied
numerically in Ref.~\cite{LiTuDasSarma2025}; the present work instead derives the local resonance
supply analytically and couples it to explicitly algebraic density interactions.  Quasiperiodic potentials support MBL with phenomenology distinct from random disorder~\cite{AA,Huse,Mace2019,Khemani2017,Varma2019,Falcao2024,Singh2021,Thomson2023,Prasad2024}, and have been realized experimentally~\cite{Schreiber2015,Smith2016}.

Reference~\cite{letter} establishes the physical outcome.  The slowly varying
potential fixes a parametrically sparse resonance supply: phase averaging gives a bare
resonant-bond count $N_0\propto L^{2-n}/h$, a purely single-particle quantity carrying no
reference to $\alpha$.  The resonant-object ensemble reconstructs across $\alpha=2n$: for
$\alpha\le2n$ resonances are asymptotically isolated two-level bonds (marginally so at the
equality), while for $\alpha>2n$ the bare resonant runs grow without bound and, wherever fragmentation is
operative, are cut by the interaction-induced Hartree field into multi-site clusters ---
but not into a pure island gas, since the run wings retain a finite fraction $f_1=\order(1)$ of
isolated bonds.  The islands' intrinsic matching law is numerically linear over the resolved low-frequency
window, $P^{\rm match}_{\rm isl}\simeq C_{\rm isl}q$, while the surviving bonds keep the
ensemble logarithm alive at strongly reduced weight, $b_{\rm all}=w_{11}b_1\simeq f_1^2b_1$; the long-range network scaling
$x_{\rm pair}^{4-2n-\alpha}\ln x_{\rm pair}\propto h^2/(Vt_0)$ is therefore unchanged across the
reconstruction; and the coupling driving the network survives at $VR^{-\alpha}$ only because the
interaction coefficients fluctuate independently --- spatially uniform couplings cancel to
$VR^{-\alpha-2}$.

This paper develops the microscopic resonance theory underlying those statements, together with
its domain of validity.  The resonance statistics of the slowly
varying potential are obtained exactly: the phase-averaged supply with its discrete corrections
and its nonuniform $n\to1^-$ limit, the correlated common-phase (comb) structure that separates
this problem from a gas of independently disordered two-level systems, and the closed-form
Hartree variance in every $(V,\alpha,L)$ regime, which shows where the interaction leaves the
supply law intact and where the same field instead fragments the resonant runs.  The comb is then
developed into a continuum shell theory, which is what distinguishes matching logarithms that are
properties of a physical pair from those of the ensemble.  The reconstruction of the object
ensemble at $\alpha=2n$ is derived, with its support threshold, its operative-fragmentation
condition, and the surviving wing-bond sector that keeps the ensemble logarithm alive.  And three
long-range thresholds --- $\alpha=1/2$, $\alpha=2n$, $\alpha=2$ --- are kept strictly separate,
since conflating them is how the generic $\alpha<2d$ counting is misapplied to this model.

Treating the resonant many-body
configurations as a graph, the natural one-step quantity is a forward branching number $B$; this
Fock-space viewpoint on MBL resonances has a long
history~\cite{Altshuler1997,Mossi2017,Subroto_corr}, and the resonant network has been
formulated explicitly as a percolation problem~\cite{RoyChalkerLogan2019,Prelovsek2021}, with
the correlated structure of Fock-space energies playing a central role~\cite{RoyLogan2020}.
Here the construction is supplied by an analytically determined resonance count: $B$ rises
smoothly and slightly superlinearly with $N_{\rm res}$, placing the criterion $B\sim1$ at
$N_{\rm res}\simeq2$, and because $N_0$ is $\alpha$-blind, so is the leading exponent of the resulting
forward-branching scale $h_{\rm FB}\propto L^{2-n}$ (its amplitude retains a weak $\alpha$
dependence, Sec.~\ref{sec:crossover}).  Exact construction of the resonant Fock graph at $L\le18$
(Sec.~\ref{sec:crossover}) then shows that one-step branching and connectivity are inequivalent:
at $B=1$ the graph carries no giant component, and the data are incompatible with ordinary
percolation at fixed $\order(1)$ resonance count.  Forward branching and Fock-space percolation
are different phenomena in this model, and conflating them misreads what finite-size numerics can
see.

Section~\ref{sec:liom} connects the resulting scale structure to the finite-size spectral
phenomenology of Ref.~\cite{yp_nee}, where localization-like behavior was reported even for
$\alpha<1$ after a broad nonergodic regime, and where uniform long-range interactions in closely
related models behave very differently~\cite{Nag2019}.

Our conclusions bear on a broader and unsettled question: whether finite-size numerics can be
extrapolated to the thermodynamic limit at all, given the avalanche instability and the
size-dependence of apparent MBL transitions~\cite{DeRoeck2017,avalanche2018,Suntajs2020,Tu2023}.

\section{Microscopic model and scope}
\label{sec:model}

We consider spinless fermions on an open chain,
\begin{align}
H={}&-t_0\sum_{i=1}^{L-1}\left(c_i^\dagger c_{i+1}+c_{i+1}^\dagger c_i\right)
+\sum_{i=1}^{L}h_i n_i \nonumber\\
&+\sum_{j>i}\frac{V_{ij}}{|i-j|^\alpha}n_i n_j,
\label{eq:H}
\end{align}
where $n_i=c_i^\dagger c_i$, $h_i$ is given by Eq.~\eqref{eq:potential}, and the interaction coefficients are independent random variables uniformly distributed on $[-V,V]$.  The earlier numerical study used $n=1/2$ and $\beta=(\sqrt5-1)/2$~\cite{yp_nee}; we keep $0<n<1$ general until specializing when useful.

Two structural properties are useful at the outset.  First, $[H_{\rm int},n_i]=0$, so the density
continuity equation contains only the nearest-neighbor hopping current,
\begin{equation}
 \dot n_i=j_{i-1}-j_i,
\qquad
j_i=-\ii t_0(c_i^\dagger c_{i+1}-c_{i+1}^\dagger c_i).
\label{eq:continuity}
\end{equation}
Thus the long-range interaction does not directly transport charge.  Second, this fact does \emph{not} protect LIOMs: after local resonant bonds are diagonalized, the same density interaction contains transverse pseudospin terms that can exchange excitation energy over long distances.

Our goal is to determine how the density of those local pseudospins and their mutual coupling depend on position, system size, $h$, $V$, $n$, and $\alpha$.

\section{Exact local resonance statistics}
\label{sec:resbonds}

\subsection{Resonant bond density}

Write $\theta_i=2\pi\beta i^n+\phi$ and $\Delta\theta_i=2\pi\beta[(i+1)^n-i^n]$.  The bare
nearest-neighbor mismatch then factorizes exactly,
\begin{equation}
 \delta_i^{(0)}\equiv h_{i+1}-h_i
 =-A_i^{\rm d}\,\sin\!\left(\theta_i+\tfrac{1}{2}\Delta\theta_i\right),
\label{eq:bare-delta-exact}
\end{equation}
with envelope $A_i^{\rm d}=2h|\sin(\Delta\theta_i/2)|$.  We call a bond a microscopic hopping
\emph{resonant bond} when $|\delta_i^{(0)}|<t_0$.  Because the only random variable is the global phase,
averaging over $\phi$ gives the marginal bond-resonance probability in closed form,
\begin{equation}
 p_i^{(0)}=\frac{2}{\pi}\arcsin\!\left[\min\!\left(1,\frac{t_0}{A_i^{\rm d}}\right)\right].
\label{eq:pi-discrete}
\end{equation}

\begin{figure}[t]\centering
\includegraphics{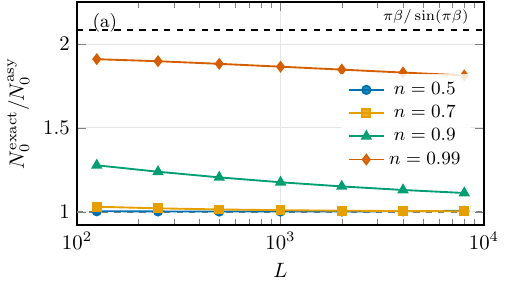}\\[3pt]
\includegraphics{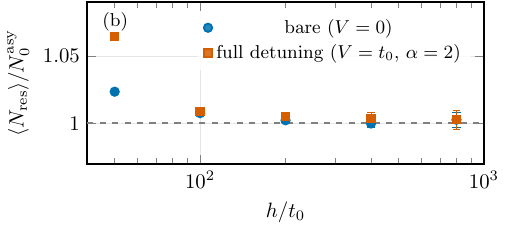}
\caption{Exactness and robustness of the resonance supply.  (a) Ratio of the exact
phase-averaged resonant-bond count, the sum of Eq.~\eqref{eq:pi-discrete}, to the asymptotic law
Eq.~\eqref{eq:Nres}, at $h/t_0=200$.  For $n=1/2$ the two agree to better than $0.6\%$ over the
entire range.  For every fixed $n<1$ the ratio ultimately approaches unity, since
$\Delta\theta_i\to0$; the approach is however strongly nonuniform as $n\to1^-$ --- at $n=0.99$
the ratio is still $\simeq1.8$ at $L=8000$ --- and taking $n\to1^-$ \emph{before} the slowly
varying limit gives the noncommuting value $\pi\beta/\sin(\pi\beta)=2.08$ (dashed).  (b) Measured phase-averaged resonant-bond count at
$n=1/2$, $L=2000$ ($400$ realizations per point), divided by the asymptotic law: the bare count
(circles) and the count with the full detuning including the Hartree field (squares, $V=t_0$,
$\alpha=2$) both collapse onto the supply law; the small positive deviation at $h/t_0=50$ is the
predicted $\order((t_0^2+\sigma_{\rm int}^2)/A^2)$ correction of Eq.~\eqref{eq:p-int-robust}.
The supply of local resonances is fixed by the potential alone.}
\label{fig:supply}
\end{figure}

At large $i$ the envelope becomes $A_i^{\rm d}\to2\pi\beta n h\,i^{n-1}$, which defines the
resonance-starvation scale $i_*=(2\pi\beta nh/t_0)^{1/(1-n)}$.  For $i\ll i_*$ the resonant bonds are dilute,
$p_i^{(0)}\simeq(2/\pi)(i/i_*)^{1-n}$, and summing along an open chain gives the quantity that
organizes everything below,
\begin{equation}
 N_0(L,h)\simeq\frac{1}{\pi^2\beta n(2-n)}\,\frac{t_0}{h}\,L^{2-n},
\label{eq:Nres}
\end{equation}
or $N_0\simeq0.219\,(t_0/h)L^{3/2}$ at $n=1/2$.  Equation~\eqref{eq:Nres} holds for every fixed
$0<n<1$ in the asymptotic slowly varying regime $\Delta\theta_i\ll1$, but the limit is not uniform
as $n\to1^-$: at $L=2000$, $h/t_0=200$ its ratio to the exact sum is $0.998$ at $n=1/2$ but only
$0.541$ at $n=0.99$; taking $n\to1^-$ before the slowly varying limit gives the noncommuting
discrepancy $\pi\beta/\sin(\pi\beta)=2.08$, while at any fixed $n<1$ the ratio ultimately
returns to unity --- Fig.~\ref{fig:supply}(a) shows this ratio directly.  Equation~\eqref{eq:Nres} contains no
reference to the interaction: the potential alone fixes how many local resonances a chain of a given
length possesses.  At the sizes reached by exact diagonalization the discrete sum of
Eq.~\eqref{eq:pi-discrete}, not its asymptotic form, should be used.

\subsection{Conditional detuning}

For uniform phase $\delta=A_i^{\rm d}\sin\vartheta$ follows the arcsine law, and conditioning on
the resonance criterion gives a truncated arcsine density (Supplemental Material~\cite{SM}).  In the
resonance-starved limit it flattens to $\rho_i\to1/2t_0$ up to relative corrections
$\order(t_0^2/(A_i^{\rm d})^2)$, so the detuning of a resonant bond is uniform on $(-t_0,t_0)$ to leading order.

\subsection{The common-phase comb}
\label{subsec:comb}

The probabilities above are marginals of a one-parameter ensemble: every bond is a deterministic
function of the same $\phi$.  We distinguish the bare and interacting resonance indicators,
\begin{equation}
 \chi_i^{(0)}(\phi)=\mathbf 1\big(|\delta_i^{(0)}(\phi)|<t_0\big),\qquad
 \chi_i=\mathbf 1\big(|\Delta_i|<t_0\big);
\label{eq:chidef}
\end{equation}
their marginals agree to leading order by Eq.~\eqref{eq:p-int-robust}.  With $\chi_i^{(0)}$ the exact
two-bond probability is
\begin{equation}
 P_{ij}^{\rm res}=\frac{1}{2\pi}\int_0^{2\pi}\dd\phi\,\chi_i^{(0)}(\phi)\chi_j^{(0)}(\phi),
\label{eq:jointres}
\end{equation}
which is not $p_i^{(0)}p_j^{(0)}$.  Resonant bond windows recur whenever the local phase advances by $\pi$, so
they are spaced by
\begin{equation}
 \ell_{\rm comb}(x)\simeq\frac{\pi}{\partial_x(2\pi\beta x^n)}=\frac{x^{1-n}}{2\beta n},
\label{eq:comb}
\end{equation}
and a fractional separation $\kappa=R/x$ is resolved only to
$\Delta\kappa\sim x^{-n}/(2\beta n)$.  A smooth continuum weight in $\kappa$ is therefore meaningful
only after coarse graining over this scale.  This correlation, absent in a gas of independently
disordered two-level systems, is what distinguishes the slowly varying problem throughout.

\subsection{Robustness to interactions}
\label{sec:resrobust}

Equation~\eqref{eq:pi-discrete} describes the bare potential.  The physical hop mismatch is
\begin{equation}
 \Delta_i=\delta_i^{(0)}+\eta_i=-A_i^{\rm d}\sin\psi_i+\eta_i,
 \qquad \psi_i=\theta_i+\tfrac12\Delta\theta_i,
\label{eq:full-detuning}
\end{equation}
where $\eta_i=\sum_{j\neq i,i+1}(U_{i+1,j}-U_{i,j})n_j$ is the Hartree shift, with
$U_{ij}=V_{ij}/|i-j|^\alpha$.  It is a sum of independent contributions
and its variance is not uniformly $\order(V^2)$: for $V_{ij}$ uniform on $[-V,V]$ at half filling,
\begin{align}
 \sigma_{\rm int}^2&=\frac{V^2}{3}\!\!\sum_{j\neq i,i+1}\!\!n_j
 \Big(|i{+}1{-}j|^{-2\alpha}+|i{-}j|^{-2\alpha}\Big)\nonumber\\
 &\sim\begin{cases}
  V^2, & \alpha>\tfrac12,\\[2pt]
  V^2\ln L, & \alpha=\tfrac12,\\[2pt]
  V^2L^{1-2\alpha}, & \alpha<\tfrac12
 \end{cases}
 \quad\text{(half filling)},
\label{eq:sigmaint}
\end{align}
a distinction that matters because the model of interest includes $\alpha\le1/2$.  Nevertheless,
the leading resonant-bond scaling is unchanged.  Along the $\order(1)$-resonant-bond trajectory
$h\propto L^{2-n}$ the bare envelope
grows as $A(L)\sim2\pi\beta nh\,L^{n-1}\propto L$, so
$\sigma_{\rm int}/A\sim L^{-1}\to0$ for every $\alpha>1/2$; at $\alpha=1/2$ exactly
$\sigma_{\rm int}/A\sim\sqrt{\ln L}/L\to0$; and only for $0<\alpha<1/2$ does the ratio take the
slower form $\sigma_{\rm int}/A\sim L^{-(1+2\alpha)/2}$, which still vanishes down to $\alpha=0$.  The ratio therefore
vanishes on the relevant trajectory for the whole range of interest.  Although $\eta_i$ can be comparable to the resonance window $t_0$, its effect on the marginal
resonance probability is suppressed by $A_i^{-2}$.  For fixed $\eta$, and provided both shifted
windows lie strictly inside the arcsine support, $|t_0\pm\eta|<A$ (outside which the corresponding
$\arcsin$ saturates at $\pi/2$), the phase average is
$p(A,\eta)=\pi^{-1}[\arcsin\frac{t_0-\eta}{A}+\arcsin\frac{t_0+\eta}{A}]$; expanding for
$A\gg t_0,|\eta|$ and averaging over a zero-mean shift of variance $\sigma_{\rm int}^2$,
\begin{equation}
 p_i=\frac{2t_0}{\pi A_i}
 \left[1+\frac{t_0^2}{6A_i^2}+\frac{\sigma_{\rm int}^2}{2A_i^2}+\order(A_i^{-4})\right].
\label{eq:p-int-robust}
\end{equation}
The correction is positive and of relative order $(t_0^2+\sigma_{\rm int}^2)/A_i^2$, so
$p_i\propto A_i^{-1}\propto i^{1-n}/h$ and Eq.~\eqref{eq:Nres} carry over to the interacting
problem.

The distinction follows from the different scales governing one- and two-bond statistics, and
recurs in Sec.~\ref{sec:broadening}.  One-point statistics are controlled by $\sigma_{\rm int}/A_i$, which is
small; two-point statistics, conditioned on both bonds lying inside a window of width $t_0$, are
controlled by $\sigma_{\rm int}/t_0$, which need not be.  As $t_0\ll A_i$ throughout the resonance-starved
regime, the interaction can leave the number of resonant bonds essentially unchanged while altering their
joint statistics substantially.

\section{Sparse resonances in Fock space: forward branching without percolation}

\label{sec:crossover}

We distinguish three quantities.  $N_0(L,h)=\sum_ip_i^{(0)}$ is the
\emph{bare} supply of Sec.~\ref{sec:resbonds}, a property of the potential alone.
$N_{\rm res}(L,h,V,\alpha)=\langle\#\{i:|\Delta_i|<t_0\}\rangle$ is the \emph{interacting}
resonant-bond count, defined by the full detuning Eq.~\eqref{eq:full-detuning} and carrying no
flippability constraint; it is the abscissa of Fig.~\ref{fig:collapse} and the variable in which the
forward-branching scale is quoted.  Finally $K(S)=\sum_iF_i(S)\,\Theta(t_0-|\Delta_i|)$ is the
resonant Fock degree of a configuration, which does carry the flippability mask.  By
Eq.~\eqref{eq:NresV}, $N_{\rm res}=N_0[1+o(1)]$ along the trajectory of interest, but the two differ
by up to $10\%$ at the smallest sizes used here, which is larger than several effects discussed
below --- hence the separate symbols.

The resonant-bond count determines the available local resonances, and the accessible system
sizes are those on which only a few of them exist.  A chain with only one or two resonant bonds is
far too sparse to support a macroscopic real-space pair network, yet Fock space has exponentially
many vertices, and it is natural to ask when the resonant graph acquires an order-one forward
degree per vertex.  We note at the outset what the answer does and does not mean: the graph
construction of Sec.~\ref{sec:fockgraph} shows that the graph remains far from any
giant-component regime at that point, so $B=1$ defines a one-step forward-branching scale, not a
percolation or connectivity threshold.

\subsection{The resonant Fock graph}

For a configuration $S$ define an active edge on bond $i$ by
$e_i(S)=F_i(S)\,\Theta(t_0-|\Delta_i(S)|)$,
where $F_i(S)=1$ if bond $i$ is flippable, i.e.\ $n_i\neq n_{i+1}$, and $\Delta_i$ is the full
interacting detuning of Eq.~\eqref{eq:full-detuning}.  The number of active edges leaving $S$ is
$K(S)=\sum_i e_i(S)$.

The flippability combinatorics are elementary.  For $M$ particles on $L$ sites,
\begin{equation}
 q_L\equiv P(F_i=1)=\frac{2M(L-M)}{L(L-1)}\;\xrightarrow[\ M=L/2\ ]{}\;\frac{L}{2(L-1)},
\label{eq:qL}
\end{equation}
and, conditioned on a bond $A$ being flippable, at half filling
$P(F_{i+1}{=}1\mid F_i{=}1)=\tfrac12$ exactly, while
$P(F_B{=}1\mid F_A{=}1)=(L-2)/[2(L-3)]$ for disjoint bonds.  The first identity is independent of $L$ and of
filling fraction.  These corrections are individually of order $5$--$10\%$ at accessible sizes, but
they carry opposite signs and there are only two adjacent bonds against $L-4$ disjoint ones, so they
cancel to better than $0.4\%$: the geometric baseline
\begin{equation}
 B_{\rm geom}\simeq q_L(L-2)\,\bar p
\label{eq:Bgeom}
\end{equation}
holds throughout the range of interest.  A derivation is given in Appendix~\ref{app:flip}.

\subsection{Forward branching}

The natural one-step quantity is the non-backtracking forward degree: the expected number of
\emph{other} active edges available after one active edge has been traversed,
\begin{equation}
 B=\frac{\displaystyle\sum_A\Big\langle e_A(S)\!\!\sum_{B\neq A}\!e_B(S_A)\Big\rangle}
        {\displaystyle\sum_A\big\langle e_A(S)\big\rangle},
\label{eq:Bdef}
\end{equation}
where $S_A$ denotes the configuration reached by hopping across $A$.  This automatically incorporates
the spatially varying resonant-bond density, open boundaries, the position dependence of the comb, the
flippability constraints, and --- through $\Delta_B(S_A)$ --- the conditional detuning correlations.
It requires no reference to the exponentially small many-body level spacing.

One further ingredient enters $\Delta_B(S_A)$.  For non-overlapping bonds the hop shifts the detuning
of every other bond by
\begin{equation}
 \ \delta\Delta_B=\pm4J^{zz}_{AB}\ \qquad (R\ge2),
\label{eq:hopshift}
\end{equation}
with $J^{zz}_{AB}$ the mixed second difference of Eq.~\eqref{eq:Jzz}; the restriction $R\ge2$ is
necessary because adjacent bonds share a site and $J^{zz}$ is then undefined.  For $R=1$ the three-site problem must be treated separately, the hop shift then being a
\emph{first} difference $\mp(U_{a,a+2}-U_{a,a+1})$ rather than Eq.~\eqref{eq:hopshift}.  As a separate diagnostic --- a property of the static shifts that does not enter
$\Delta_B(S_A)$ --- the Hartree shifts on two adjacent bonds are intrinsically anticorrelated, with
an exact $\alpha$-dependent coefficient given in the Supplemental Material~\cite{SM}.

\subsection{The forward-branching scale and finite-size scaling}

\begin{figure}[t]\centering
\includegraphics{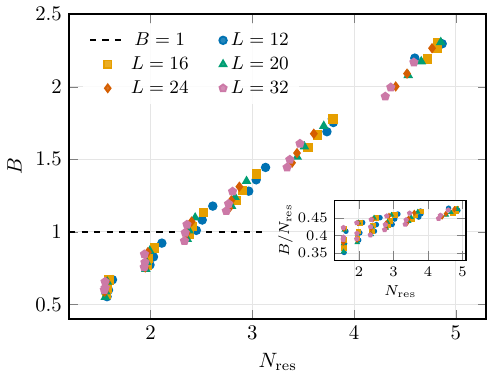}
\caption{Forward branching of the resonant Fock graph against resonant-bond number.  Each point is one
$(L,\alpha,h)$ evaluation of Eq.~\eqref{eq:Bdef}, spanning $12\le L\le32$ and $0.5\le\alpha\le2$.
The dashed line marks the branching criterion $B=1$; the scale $N_c$ is read off by
interpolating each $(L,\alpha)$ series through it, with no fit.  Inset: the same data as
$B/N_{\rm res}$, which would be constant if $B$ were proportional to $N_{\rm res}$.  It is not --- it
rises monotonically with $N_{\rm res}$ in every series --- so no single constant of proportionality is used anywhere.
The Monte Carlo error on each point is $0.12\%$--$0.44\%$, well inside the markers, so the curvature
is systematic.  The finite-$L$ independent-edge baseline is $B_{\rm ind}/N_{\rm res}=q_L(L-2)/(L-1)$,
Eq.~\eqref{eq:Nc0}, which tends to $1/2$ only asymptotically and assumes uniform marginals.
Each $L$ marker overlays the $\alpha=0.5$, $1$, and $2$ series, whose near-coincidence is the
stated range-blindness; the $\alpha$-resolved curves are separated in the Supplemental
Material~\cite{SM}.  Because the curves for different $\alpha$ lie close together, $B\sim1$
reduces to a condition on $N_{\rm res}$ alone; Sec.~\ref{sec:fockgraph} shows that this condition marks order-one one-step
branching, not percolation of the graph.  Protocol and error analysis: Appendix~\ref{app:numerics}.}
\label{fig:collapse}\end{figure}

Evaluating Eq.~\eqref{eq:Bdef} for the full model across $0.5\le\alpha\le2$ and $12\le L\le32$
gives Fig.~\ref{fig:collapse}.  $B$ is a smooth, monotone, and slightly \emph{superlinear} function
of $N_{\rm res}$.  We do not fit it: the branching criterion $B=1$ is applied by interpolating each
$(L,\alpha)$ curve directly through it.  $B$ is \emph{not}
proportional to $N_{\rm res}$ --- the ratio $B/N_{\rm res}$ rises monotonically with
$N_{\rm res}$ in every one of the fifteen series (equivalently, falls as $h$ increases), from
$0.352$ at $N_{\rm res}\simeq1.6$ to $0.478$ at $N_{\rm res}\simeq4.6$, a
$12$--$34\%$ variation against a Monte Carlo uncertainty of $0.12\%$--$0.44\%$
(Appendix~\ref{app:numerics}), so a single constant of proportionality would misrepresent the data by up to
$22\%$ over the plotted range, and by about $10\%$ in the threshold window itself.

The subsequent scaling uses only $N_c=\order(1)$: if $B=1$ at $N_{\rm res}=N_c$, then
$N_{\rm res}\propto L^{2-n}/h$ gives $h_{\rm FB}\propto L^{2-n}$, whatever the shape of
$B(N_{\rm res})$.

The branching calculation uses no spectral-data fitting.  The measured scale contains a combinatorial
baseline and two competing correlation corrections.

The \emph{independent-edge baseline} follows from Eq.~\eqref{eq:Bgeom} alone.  A bond carrying
$|\Delta|<t_0$ contributes to $B$ only if it is also flippable, which at half filling happens with the
exact probability $q_L\to1/2$ of Eq.~\eqref{eq:qL}; excluding the traversed bond,
\begin{equation}
 \ \frac{B_{\rm ind}}{N_{\rm res}}=q_L\frac{L-2}{L-1}\longrightarrow\frac12,
 \qquad N_c^{\rm ind}\longrightarrow2\ ,
\label{eq:Nc0}
\end{equation}
equal to $0.496$ at $L=12$ and $0.4995$ at $L=32$.  This is the \emph{uniform-marginal} baseline:
pure combinatorics, treating the bond-resonance probabilities as identical and the active edges as
statistically independent.  Neither holds here.  Because $p_i\propto i^{1-n}$ is strongly position
dependent, the exclusion correction is really controlled by $\sum_ip_i^2/N_{\rm res}^2$ rather than
exactly $1/(L-1)$; at these sizes that is a sub-percent shift, and we quote
Eq.~\eqref{eq:Nc0} as the asymptotically exact baseline for the threshold scale rather than as an
exact finite-$L$ value.  The departure from independence is what we quantify next.

The baseline receives two competing corrections.  \emph{(i) The common-phase comb suppresses
branching.}  Setting $V=0$, where the only correlation left is the shared phase of
Eq.~\eqref{eq:jointres}, we measure $B/N_{\rm res}=0.3794(7)$ at $L=16$ --- a $24\%$ reduction
below Eq.~\eqref{eq:Nc0}, with no interaction present at all.  The common-phase correlations are strongly nonuniform in separation --- at some comb offsets joint
resonances are enhanced, at others forbidden --- and their net effect is to reduce the
position-averaged forward degree below the independent-edge baseline.  \emph{(ii) Interactions progressively relax that suppression.}  As $V$ is turned on,
$B/N_{\rm res}$ rises monotonically back toward the independent-edge value:
\begin{center}
\footnotesize
\setlength{\tabcolsep}{3.5pt}
\begin{tabular}{l|cccccc}
\hline\hline
$V/t_0$ & $0$ & $0.25$ & $0.5$ & $1$ & $1.5$ & $2$\\
$B/N_{\rm res}$ & $0.3794$ & $0.3838$ & $0.3958$ & $0.4290$ & $0.4538$ & $0.4655$\\
$N_c^{\rm res}$ ($B{=}1$) & $2.57$ & $2.55$ & $2.49$ & $2.35$ & $2.25$ & $2.17$\\
\hline\hline
\end{tabular}
\end{center}
($L=16$, $\alpha=1$.  The $B/N_{\rm res}$ row is evaluated at the single field $h_0(L)$, with
batch-mean statistical errors $\le0.0015$ on the ratio; the $N_c^{\rm res}$ row comes from
independent $h$ scans at each $V$, interpolated to $B=1$, and is quoted to two decimals.)  The ratio rises monotonically from $0.379$ toward the baseline $0.498$.  The change is weak through $V/t_0=0.25$ and becomes pronounced by $V/t_0=0.5$, consistent with the
two-point scale $\sigma_{\rm int}/t_0$ of Sec.~\ref{sec:resrobust}: the interaction leaves the resonant bond
\emph{number} nearly unchanged while altering the resonant bond \emph{joint statistics}.

All quantities here are referred to the \emph{measured} interacting count
$N_{\rm res}=\langle\#\{a:|\Delta_a|<t_0\}\rangle$, the same abscissa as Fig.~\ref{fig:collapse} and
the same count for which the independent-edge baseline Eq.~\eqref{eq:Nc0} is derived; mixing it with
the asymptotic form Eq.~\eqref{eq:Nres} would compare a threshold in one convention against a
baseline in another.  In that single convention, solving $B(L,\alpha,h)=1$ defines
$N_c^{\rm res}(L,\alpha,V)\equiv N_{\rm res}|_{B=1}$ and gives $N_c^{\rm res}\simeq2.25$--$2.46$,
essentially \emph{independent of $L$}: $2.264\to2.255$ for $\alpha=0.5$, $2.356\to2.352$ for
$\alpha=1$ and $2.430\to2.459$ for $\alpha=2$, over $12\le L\le32$.
Over these sizes the threshold is nearly $L$-independent; we do not claim $N_c\to2$
asymptotically: the comb correction survives at $V=0$, so the independent-edge value
Eq.~\eqref{eq:Nc0} is a baseline rather than a limit.  It does mean that $N_c$ retains an
$\order(1)$ dependence on interaction strength, and hence so does the amplitude --- though not the
exponent --- of $h_{\rm FB}$; we return to this in Sec.~\ref{sec:Vdep}.
$B\sim1$ defines a forward-branching scale; Sec.~\ref{sec:fockgraph} shows that it is
\emph{not} a percolation threshold of the resonant graph, and its connection to spectral
statistics remains a conjecture (Sec.~\ref{sec:reinterp}).  The same solution gives $h_{\rm FB}\propto L^{1.41\hbox{--}1.44}$, slightly below the asymptotic
$L^{3/2}$ law, consistent with the finite-$L$ difference between the discrete interacting
resonance count and its asymptotic form.

Fig.~\ref{fig:collapse} and the scale quoted above show why \emph{a few} resonant bonds
suffice to give the resonant Fock graph an order-one one-step forward branching number: the
relevant object is a forward degree on an exponentially large graph, not a density in real space.
They do not suffice to \emph{connect} the graph, as the exact construction below shows.  The construction also produces only weak
range dependence of the threshold.  The
residual $\alpha$ dependence of the threshold is small: in the measured-count convention the ratio
$N_c(\alpha{=}2)/N_c(\alpha{=}0.5)$ is $1.073$ at $L=12$ and $1.090$ at $L=32$ --- an $\simeq8\%$
effect showing no decrease with $L$, and equally consistent with a small persistent
$\alpha$-dependent offset.  Hence $N_c(L,\alpha)=\order(2)$ with weak range dependence,
the zeroth-order value Eq.~\eqref{eq:Nc0} being $\alpha$-independent by construction, with all
$\alpha$ dependence entering through the conditional-detuning correction.  Whether that residual dependence vanishes as $L\to\infty$ or tends to distinct $\order(1)$
constants is not resolved at the available sizes: the marginal resonant-bond law becoming
$\alpha$-blind does not by itself make the \emph{conditional} forward branching $\alpha$-blind, and
the $R=1$ Hartree anticorrelation~\cite{SM} is explicitly $\alpha$-dependent and $\order(1)$.  If the physical spectral crossover develops an order-unity $\alpha$ spread while the active-edge
threshold remains weakly range dependent at the same sizes, the active-edge description is
incomplete.  The full $N_c(L,\alpha)$ is given in the Supplemental
Material~\cite{SM}; being derived from the same evaluation of Eq.~\eqref{eq:Bdef} as
Fig.~\ref{fig:collapse}, it is a presentation of that calculation and not independent evidence for
it.

\subsection{The exact resonant graph carries no giant component at $B=1$}
\label{sec:fockgraph}

The forward branching number is a one-step quantity, and identifying $B=1$ with a connectivity
transition of the graph would require the graph to be locally tree-like and its edges effectively
independent --- neither of which holds here, where short loops from commuting hops, occupation
constraints, and the shared phase are all present.  We therefore construct the resonant Fock graph
\emph{exactly} for $L\le18$: for each realization $(\phi,\{V_{ij}\})$ every one of the
$\binom{L}{L/2}$ half-filled configurations is a vertex, every active edge [flippable and
$|\Delta_a(S)|<t_0$, evaluated from Eq.~\eqref{eq:Delta-numeric}] is inserted, and the connected
components are found exactly; $B$ is evaluated from the identical samples, so the two quantities
are directly comparable.  Protocol and the full data table are given in the Supplemental
Material~\cite{SM}.

\begin{figure}[t]\centering
\includegraphics{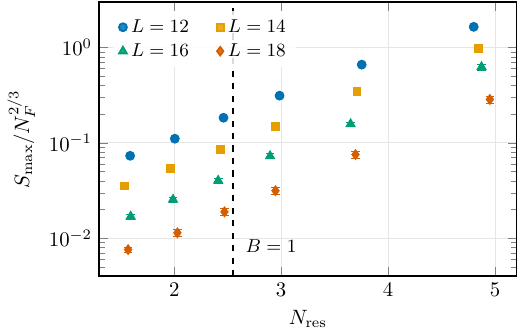}
\caption{The exact resonant Fock graph carries no giant component at the forward-branching scale.  Largest
connected component of the exactly constructed graph, scaled by $N_F^{2/3}$ (the mean-field/
random-graph critical reference scaling), against the measured resonance count,
at $\alpha=1$, $V=t_0$, for $L=12$--$18$ ($200$--$50$ realizations; batch-mean errors).  At every
fixed $N_{\rm res}$ the diagnostic falls with $L$: the graph moves \emph{away} from percolation
with increasing size at fixed resonance count.  The dashed line marks the $B=1$ crossing at this
$\alpha$ ($N_{\rm res}\simeq2.5$--$2.55$); the same behavior holds at $\alpha=0.5$ and $2$
(Supplemental Material~\cite{SM}).}
\label{fig:fock}\end{figure}

The $B=1$ crossings land at $N_{\rm res}=2.41$--$2.65$ across
$12\le L\le18$ and $0.5\le\alpha\le2$, consistent with the production values above.  But at those
crossings the largest connected component holds only $13$--$53$ of up to $48\,620$ vertices, and the
mean finite-cluster size $\chi_{\rm cl}\simeq3$--$5$ is essentially independent of $L$.  More
sharply, Fig.~\ref{fig:fock} shows the diagnostic $S_{\rm max}/N_F^{2/3}$, normalized by the
critical-cluster scaling of \emph{ordinary} (mean-field) percolation --- our graph is correlated
and loop-rich, so this is a reference scale, not a theorem: at every fixed $N_{\rm res}$ in the
range studied --- including values twice the branching scale --- it \emph{decreases} with $L$.
The finite-size graphs are far from a giant-component regime at $B=1$, and the data are
incompatible with ordinary percolation at fixed $\order(1)$ resonance count; extrapolating to
strict thermodynamic subcriticality would exceed what $L\le18$ can establish.  A plausible structural reason (a hypothesis,
not a derived result) is that at $N_{\rm res}=\order(1)$ the resonant graph is approximately a
union of low-dimensional sub-hypercubes generated by the few active bonds, whose interconnection is
gradual rather than percolative.

First, $B=1$ is an order-one forward-branching scale and nothing more; we
write the associated field scale as
\begin{equation}
 h_{\rm FB}(L)\propto L^{2-n},
\label{eq:hFB}
\end{equation}
and we avoid the notation $h_\times$ precisely because $\times$ suggests a physical crossover.
Second, any phenomenon tied to a fixed number of local resonant objects inherits the supply
exponent $L^{2-n}$ --- that statement is robust --- but the graph calculation does not establish
that the finite-size spectral crossover of Ref.~\cite{yp_nee} is such a phenomenon
(Sec.~\ref{sec:reinterp}).

\section{Reconstruction of the resonant-object ensemble}
\label{sec:islands}

The supply law of Sec.~\ref{sec:resbonds} counts resonant bonds; it does not by itself determine
whether the elementary resonant \emph{object} is a two-site bond.  This section derives when it is.
The answer divides the interaction range into three regimes at $\alpha=2n$, and in the long-range
regime $\alpha>2n$ it is the Hartree field --- not the bare potential --- that determines the local
object.

\subsection{The two-site pseudospin as minimal illustration}
\label{sec:tls}

In the one-particle subspace of a bond, $\{|10\rangle,|01\rangle\}$, the local Hamiltonian,
its eigenbasis rotation, and the induced transverse coupling of a density-density term
$J_{AB}^{zz}\tau_A^z\tau_B^z$ are
\begin{align}
 H_i&=\frac{\Delta_i}{2}\tau_i^z-t_0\tau_i^x,
 &\Omega_i&=\sqrt{\Delta_i^2+4t_0^2},\nonumber\\
 \tau_i^z&=\cos\vartheta_i\,\sigma_i^z+\sin\vartheta_i\,\sigma_i^x,
 &\sin\vartheta_i&=\frac{2t_0}{\Omega_i},\nonumber\\
 J_{AB}^{\perp}&=J_{AB}^{zz}\sin\vartheta_A\sin\vartheta_B,&&
\label{eq:localpseudo}
\end{align}
the last exchanging the two pseudospin excitations when $|\Omega_A-\Omega_B|$ is sufficiently
small.  With the Pauli-matrix convention used here, the matrix element between $|e_Ag_B\rangle$ and $|g_Ae_B\rangle$ generated by this term is $J_{AB}^{\perp}$; rewriting the pseudospins as spin-$1/2$ operators changes the numerical convention but not the resonance condition.  This is the microscopic route by which a longitudinal density interaction drives the long-range resonant-pair mechanism~\cite{Burin2006,Yao2014} even though Eq.~\eqref{eq:continuity} contains no long-range charge current.  The two-site pseudospin is the \emph{minimal illustration} of that route; whether it is also the
physical local object at the scales of interest is decided next.

\subsection{Bare turning-point runs and the three regimes}
\label{sec:runs}

Near a zero of $\sin\psi$ the bare detuning crosses zero linearly, advancing per bond by
$s(x)=A^{\rm d}(x)\,\psi'(x)\simeq h\,\psi'(x)^2$ with $\psi'(x)=2\pi\beta n\,x^{n-1}$, so the bare
resonance condition $|\delta^{(0)}_i|<t_0$ is satisfied on a \emph{run} of
\begin{equation}
 w_0(x)=\frac{2t_0}{s(x)}
 =\frac{2t_0}{(2\pi\beta n)^2}\,\frac{x^{2-2n}}{h}
\label{eq:w0}
\end{equation}
consecutive bonds around each turning point, where resonant bonds pair up once the envelope
drops below $t_0$ [Fig.~1(a) of Ref.~\cite{letter} shows a chain snapshot].  At fixed $h$ the runs lengthen with
position; the physically relevant question is how $w_0$ behaves on the pair-network trajectory of
Sec.~\ref{sec:cascade}, $h^2\sim Vt_0\,x_{\rm pair}^{4-2n-\alpha}$ (times the matching
logarithm), along which
\begin{equation}
 w_0(x_{\rm pair})\sim\sqrt{\frac{t_0}{V}}\;
 \frac{x_{\rm pair}^{(\alpha-2n)/2}}{(\ln x_{\rm pair})^{1/2}},\qquad 0<\alpha<2,
\label{eq:w0traj}
\end{equation}
Because the matching logarithm of Sec.~\ref{sec:matching} is present throughout $0<\alpha<2$
[Eq.~\eqref{eq:Pmix}], Eq.~\eqref{eq:w0traj} holds without case distinction.  The exponent
changes sign at $\alpha=2n$, and the local-object structure divides accordingly:
\begin{equation}
\begin{aligned}
 &\alpha<2n:\ \ w_0\to0\quad\text{(isolated bonds)},\\
 &\alpha=2n:\ \ w_0\sim(\ln x_{\rm pair})^{-1/2}\to0\quad\text{(marginal)},\\
 &\alpha>2n:\ \ w_0\to\infty\quad\text{(growing runs if $p_{\rm br}=0$;}\\
 &\hphantom{\alpha>2n:\ \ }\text{fragmented mixture if $p_{\rm br}>0$, $w_0\gg\xi_{\rm frag}$)}.
\end{aligned}
\label{eq:regimes}
\end{equation}
For $\alpha\le2n$ the isolated-bond asymptotics of Sec.~\ref{sec:tls} is self-consistent ---
including at the equality, where the argument closes on itself: the matching logarithm enters the
boundary, and precisely that logarithm drives
$w_0\sim\sqrt{t_0/V}\,(\ln x_{\rm pair})^{-1/2}\to0$, confirming the assumption.  The
approach to isolation at $\alpha=2n$ is only logarithmic, so at accessible parameters the
marginal case retains amplitude-dependent multi-bond structure even though its asymptotics is
bond-like.  For $n=1/2$ this places $\alpha=1$ asymptotically on the bond side.  For $\alpha>2n$
a purely two-level description of the turning-point region fails asymptotically, and the
composition of the resonant-object ensemble is decided by the Hartree field, as follows.  It
will \emph{not} become a pure island gas: Sec.~\ref{sec:islandmc} shows that a finite fraction of
isolated bonds survives in the run wings at every $w_0$.

\subsection{Hartree fragmentation: $p_{\rm br}$, $V_*$, and $\xi_{\rm frag}$}
\label{sec:frag}

The physical resonance condition uses the full detuning $\Delta_i=\delta^{(0)}_i+\eta_i$ of
Eq.~\eqref{eq:full-detuning}.  Within a bare run the deterministic part is nearly constant, while
the Hartree sequence $\eta_i$ is statistically stationary, of rms $\sigma_{\rm int}$
[Eq.~\eqref{eq:sigmaint}], and \emph{summably correlated} along the chain --- adjacent Hartree
detunings are in fact anticorrelated, with the exact $\alpha$-dependent coefficient given in the
Supplemental Material~\cite{SM}, and the cross-bond covariance decays algebraically but summably
for $\alpha>1/2$.  A run is therefore cut wherever the local Hartree shift exceeds the remaining
window margin.  For a bond near the run center ($\delta^{(0)}\simeq0$) the breaking probability is
\begin{equation}
 p_{\rm br}(V,\alpha)=\Pr\big(|\eta_i|>t_0\big),
 \qquad
 \xi_{\rm frag}\equiv p_{\rm br}^{-1},
\label{eq:pbr}
\end{equation}
and since the Hartree sequence is summably correlated along the chain --- with
$\xi_{\rm frag}=p_{\rm br}^{-1}$ quantitatively reproducing the measured fragmentation scale
(Sec.~\ref{sec:islandmc}) --- $\xi_{\rm frag}$ is the
characteristic fragmentation length: it depends on $(V/t_0,\alpha)$ only, \emph{not} on position or
scale.

Whether $p_{\rm br}$ can be nonzero at all is controlled by the support of the Hartree field.  For
$\alpha>1$ the spatial sums converge and $\eta_i$ has strictly bounded support,
\begin{equation}
 \operatorname{ess\,sup}|\eta_i|=S(\alpha)\,V,\qquad
 S(\alpha)=2\big[2\zeta(\alpha)-1\big],
\label{eq:etasupport}
\end{equation}
so that $p_{\rm br}=0$ identically below the support threshold
\begin{equation}
 V_*(\alpha)=\frac{t_0}{S(\alpha)},
\label{eq:Vstar-frag}
\end{equation}
equal to $0.118\,t_0$ at $\alpha=3/2$ and $0.218\,t_0$ at $\alpha=2$.  Below $V_*$ the central
portion of a bare run can never be cut, no matter how large $w_0$ becomes.  Two remarks prevent
misuse of Eq.~\eqref{eq:Vstar-frag}.  First, it applies to $\alpha>1$; for $\alpha\le1$ the column
sums diverge with system size (the essential supremum grows as $V\ln L$ at $\alpha=1$), so
$p_{\rm br}>0$ at any fixed $V>0$ on large enough scales.  Note that at $n=1/2$ the fragmented
regime $\alpha>2n=1$ is precisely where the support is finite, so the amplitude threshold is a
structural feature of that regime.  More generally, the island classification as stated requires
the fragmented branch to lie at $\alpha>1/2$, i.e.\ $n\ge1/4$, so that $\sigma_{\rm int}$ is
$L$-independent [Eq.~\eqref{eq:sigmaint}] and a scale-independent $\xi_{\rm frag}(V,\alpha)$
exists; this covers the principal case $n=1/2$.  In the corner $n<1/4$, $2n<\alpha\le1/2$ the
Hartree width itself grows with system size, the strong Hartree field can move resonances rather
than merely cut pre-existing bare runs, and a separate scale-dependent fragmentation analysis
would be required; that corner is outside the present island theory.  Second --- and this is the condition the island theory actually
requires --- $V>V_*$ separates only $p_{\rm br}=0$ from $p_{\rm br}>0$; it does not by itself make
fragmentation operative.  Just above threshold $p_{\rm br}$ is large-deviation suppressed: at
$\alpha=1.25$, $V=0.25\,t_0$ one has $V/V_*\simeq4.1$ yet $p_{\rm br}<5\times10^{-7}$, i.e.\
$\xi_{\rm frag}\gtrsim10^6$ --- mathematically fragmented, operationally uncut over enormous
scales.  The statement that holds is
\begin{equation}
 p_{\rm br}>0
 \qquad\text{and}\qquad
 w_0(x_{\rm pair})\gg\xi_{\rm frag}(V,\alpha),
\label{eq:fragcond}
\end{equation}
which for $\alpha>2n$ is satisfied asymptotically along the trajectory whenever $p_{\rm br}>0$,
since $w_0\to\infty$ there.  Measured values at $V=t_0$ are
$p_{\rm br}=0.146$ ($\alpha=1.5$) and $0.113$ ($\alpha=2$), i.e.\ $\xi_{\rm frag}\simeq7$--$9$
bonds; the full $p_{\rm br}(V,\alpha)$ table is given in the Supplemental Material~\cite{SM}.

\subsection{Fragmented islands: cluster statistics}
\label{sec:islandmc}

We test this picture by direct classical counting: chains of $4\times10^4$ bonds around
$x=10^6$, with the exact potential, random couplings, and half-filled configurations, scanning the
bare width $w_0$ at fixed $(V,\alpha)$ --- which, by Eq.~\eqref{eq:w0traj}, is equivalent to moving
along the pair-network trajectory (protocol in the Supplemental Material~\cite{SM}).  Maximal runs
of consecutive bonds with $|\Delta_i|<t_0$ define the physical islands.  Two size measures must be
distinguished: the count-weighted mean island size $\langle\ell\rangle_{\rm isl}$, and the
bond-weighted mean
$\langle\ell\rangle_B=\langle\ell^2\rangle_{\rm isl}/\langle\ell\rangle_{\rm isl}$ --- the island
size seen by a typical resonant bond, which is the measure relevant for matrix elements and
matching.

\begin{figure}[t]\centering
\includegraphics{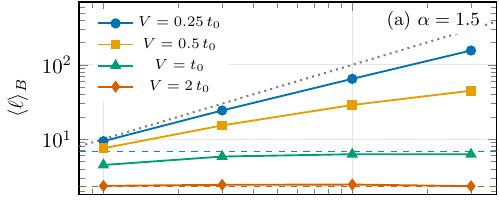}\\[2pt]
\includegraphics{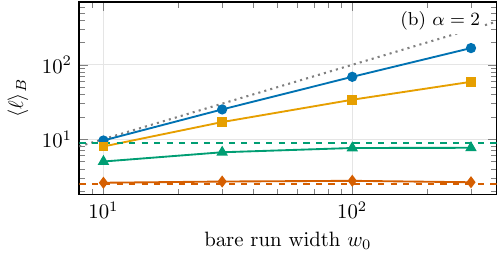}
\caption{Hartree fragmentation of the turning-point runs over the tested fragmentation domain,
for the \emph{snapshot (unhealed) partition}: bond-weighted mean island size
$\langle\ell\rangle_B$ against the bare run width $w_0$ at (a)
$\alpha=1.5$ and (b) $\alpha=2$ (increasing $w_0$ at fixed $V$ is equivalent to moving outward
along the pair-network trajectory).  For $V\ge t_0$ the island size saturates at a
$w_0$-independent value consistent with $\xi_{\rm frag}=p_{\rm br}^{-1}$ (dashed lines at
$6.8$ and $2.3$ for $\alpha=1.5$, and $8.8$ and $2.5$ for $\alpha=2$, for $V/t_0=1$ and $2$,
respectively).  After recursive boundary healing (Sec.~\ref{sec:islandmc}, Table~\ref{tab:heal})
the bond-weighted mean at $V=t_0$ instead drifts upward with $w_0$ while the count-weighted mean
and $A_{\rm isl}$ stay fixed; the saturation shown here is a property of the snapshot partition.
For $V=0.25\,t_0$
--- above both support thresholds $V_*(1.5)=0.118\,t_0$ and $V_*(2)=0.218\,t_0$, though only
slightly so at $\alpha=2$ --- fragmentation is large-deviation suppressed
($\xi_{\rm frag}\gtrsim10^6$) and the islands track the bare width (dotted), and $V=0.5\,t_0$
($\xi_{\rm frag}\sim10^2$--$10^3$) is preasymptotic over the plotted range.  Chains of
$4\times10^4$ bonds at $x=10^6$; numerical values in the Supplemental Material~\cite{SM}.}
\label{fig:islands}\end{figure}

Figure~\ref{fig:islands} shows $\langle\ell\rangle_B$ against $w_0$ at $\alpha=1.5$ and $2$.  For
$V\ge t_0$ the island size \emph{saturates}: it becomes independent of $w_0$ (hence of position
along the trajectory) at a value quantitatively consistent with $\xi_{\rm frag}$ ---
$\langle\ell\rangle_B\to7.7$ against $\xi_{\rm frag}=8.8$ at $V=t_0$, and $2.7$ against $2.5$ at
$V=2t_0$ (at $\alpha=1.5$: $6.3$ against $6.8$).  For $V=0.25\,t_0$ --- above both support thresholds, and only slightly so at $\alpha=2$
($V/V_*\simeq1.15$), with $p_{\rm br}$ unresolvably small and $\xi_{\rm frag}\gtrsim10^6$ in
both cases --- the islands simply track the bare width,
$\langle\ell\rangle_B\simeq0.85\,w_0$, over the entire accessible range: fragmentation is
mathematically allowed but operationally absent, exactly as anticipated below
Eq.~\eqref{eq:fragcond}.  $V=0.5\,t_0$, with $\xi_{\rm frag}\simeq5\times10^2$, is
preasymptotic throughout the plotted range.  Three further facts carry over to the rest
of the paper.  First, the resonant-bond \emph{density} is unchanged by fragmentation to within
$0.3\%$ in every case studied --- an independent confirmation of the supply robustness of
Sec.~\ref{sec:resrobust}: the interaction reorganizes the joint statistics of the resonances
without changing their number.  Second, roughly half of the islands are \emph{frozen} (their
sampled particle number is $0$ or maximal, leaving no internal doublet); they do not participate
in pair matching and enter only as an $\order(1)$ renormalization of the active-object density.
Third, at the physical $V=t_0$ the multi-site component of the fragmented ensemble is small:
a typical island contains only a few bonds,
$\langle\ell\rangle_{\rm isl}\simeq2.8$ at $\alpha=2$, while bond weighting emphasizes the
broad tail of the size distribution, $\langle\ell\rangle_B\simeq7.6$ --- a typical resonant
bond sits in an island of $8$--$9$ sites.

Fragmentation does not produce a pure island gas, and this matters for the matching theory.  A
turning-point region has \emph{wings}, and the way to see that they
carry an extensive population is to rescale.  At offset $m$ from the turning point the bare
mismatch is $\delta^{(0)}(m)\simeq-(2t_0/w_0)\,m$, so in the scaled coordinate
\begin{equation}
 y=\frac{2m}{w_0},\qquad \delta^{(0)}/t_0\simeq-y,
\label{eq:wingscale}
\end{equation}
the resonance condition $|\delta^{(0)}+\eta|<t_0$ becomes $|\eta/t_0-y|<1$.  The Hartree sequence
is stationary with a $w_0$-independent distribution, so
\begin{equation}
 p_{\rm res}(y)=\Pr\big(|\eta/t_0-y|<1\big)
\end{equation}
is an $\order(1)$ function of $y$ alone, positive for $|y|-1<S(\alpha)V/t_0$ by
Eq.~\eqref{eq:etasupport}.  Since an interval $\dd y$ contains $(w_0/2)\dd y$ bonds, the wing
population is
$N_{\rm wing}=(w_0/2)\int\dd y\,p_{\rm res}(y)\propto w_0$ --- the same linear growth as the
fragmented interior, which contributes $\propto w_0/\langle\ell\rangle_{\rm isl}$ objects.  The
same rescaling controls the \emph{singletons}: with
$p_{\rm sing}(y)=\Pr(\chi_i=1,\chi_{i\pm1}=0\mid y)$, neighbouring bonds share the same $y$ to
$\order(1/w_0)$ and differ only through $\eta_i$ versus $\eta_{i\pm1}$, whose joint distribution
is non-degenerate --- indeed adjacent Hartree shifts are \emph{anti}correlated, with the exact
coefficient computed in the Supplemental Material~\cite{SM}, which favours the neighbours falling
outside the window.  Hence $p_{\rm sing}(y)>0$ on a finite interval and
$N^{\rm wing}_{\ell=1}=(w_0/2)\int\dd y\,p_{\rm sing}(y)\propto w_0$.  Both populations grow
linearly in $w_0$, so their ratio does not vanish: the fraction of \emph{active} objects that are
single bonds tends to a constant,
\begin{equation}
 f_1(w_0)\ \longrightarrow\ f_1^{\infty}=\order(1).
\label{eq:f1}
\end{equation}

The logarithm requires one step beyond $f_1^\infty>0$: the surviving singletons must retain
spectral weight at the band edge.  Since $\Omega-2t_0\simeq\Delta^2/4t_0$, the edge singularity
$f_{\rm sing}(\Omega)\propto[\rho_{\rm sing}(0)]\,(\Omega-2t_0)^{-1/2}$ survives if and only if
the singleton-conditioned detuning density is nonzero at zero detuning,
\begin{equation}
 \rho_{\rm sing}(0)=\int\!\dd y\;
 p\big(\eta_i=t_0y,\;\chi_{i-1}=0,\;\chi_{i+1}=0\big)>0 .
\label{eq:b1inf}
\end{equation}
For $\alpha>1$ this holds throughout the operative fragmented regime, which includes the
entire fragmented branch for the principal case $n=1/2$.  Write the integrand of
Eq.~\eqref{eq:b1inf} as $F(y)$, the joint density for the central bond to sit at $\Delta=0$ while
both neighbours are non-resonant; at scaled position $y$ the three consecutive detunings are
$\Delta_{i,i\pm1}\simeq t_0(-y)+\eta_{i,i\pm1}$ to $\order(1/w_0)$, so
$F(y)=p\big(\eta_i=t_0y,\ |\eta_{i\pm1}-t_0y|>t_0\big)$.

\emph{(i) $F(0)>0$.}  The condition $\eta_i=0$ is a single constraint on the couplings and
carries no measure by itself; what is needed is a positive \emph{conditional density} at
$\eta_i=0$.  Single out one dimensionless coefficient $u\equiv V_{i+1,j_*}/V\in[-1,1]$ with
$n_{j_*}=1$, $j_*\ne i,i+1$, so that $U_{i+1,j_*}=cu$ with $c=V|i+1-j_*|^{-\alpha}\neq0$; $u$
enters $\eta_i$ linearly, and we write
$\eta_i=cu+G(\mathbf z)$ with $\mathbf z$ the remaining variables.  Then
$\delta(\eta_i)=|c|^{-1}\delta(u-u_*)$, $u_*(\mathbf z)=-G(\mathbf z)/c$, and
\begin{equation}
 F(0)=\frac{1}{|c|}\int\!\dd\mathbf z\,p(\mathbf z)\,p_u\big(u_*(\mathbf z)\big)\,
 \mathbf 1\big[|\eta_{i\pm1}(\mathbf z,u_*)|>t_0\big],
\label{eq:F0}
\end{equation}
with $p_u$ the uniform density on $[-1,1]$.  It suffices to exhibit an open set of $\mathbf z$ on
which $u_*\in(-1,1)$ strictly and both neighbours lie outside the window.  Because
$\eta_{i-1}$, $\eta_i$, $\eta_{i+1}$ draw on partly disjoint coupling sets --- only $\eta_i$ and
$\eta_{i-1}$ contain $U_{i,\cdot}$, only $\eta_i$ and $\eta_{i+1}$ contain $U_{i+1,\cdot}$, and
$U_{i-1,\cdot}$, $U_{i+2,\cdot}$ appear in $\eta_{i-1}$, $\eta_{i+1}$ alone --- the four
partial sums $\Sigma_a\equiv\sum_jU_{a,j}n_j$, each ranging over
$[-\tfrac12S(\alpha)V,\tfrac12S(\alpha)V]$ by Eq.~\eqref{eq:etasupport}, can be steered
separately: assign to each $\Sigma_a$ a disjoint finite set of occupied spectator sites whose
couplings $V_{a,j}$ are used for the steering.  Because the series defining
$S(\alpha)$ converges for $\alpha>1$ and $S(\alpha)V>t_0$ strictly, a sufficiently large finite
subset reaches every inequality below with a finite margin, and all remaining couplings may then
vary in a neighbourhood without closing it.  Take $\Sigma_{i+1}$ and $\Sigma_i$ in the interior near $-\tfrac12S(\alpha)V$ with
$\Sigma_i=\Sigma_{i+1}$ at $u=u_*$; since the contribution of $u$ to $\Sigma_{i+1}$ is
$cu$ with $|c|<\tfrac12S(\alpha)V$, this places $u_*$ strictly inside $(-1,1)$.  Take
$\Sigma_{i+2}$ and $\Sigma_{i-1}$ near $+\tfrac12S(\alpha)V$, which is free because neither
enters $\eta_i$; then $|\eta_{i\pm1}|\simeq S(\alpha)V$, and $p_{\rm br}>0$ means precisely
$S(\alpha)V>t_0$, so both exceed $t_0$ \emph{strictly}.  All inequalities are strict, so
$\mathbf z$ may be moved by a finite amount: the integrand of Eq.~\eqref{eq:F0} is positive on an
open set of positive measure, and $F(0)>0$.  The construction uses the finite support
Eq.~\eqref{eq:etasupport}, i.e.\ $\alpha>1$.

\emph{(ii) From a point to the integral.}  A single $y$ carries no measure, so $F(0)>0$ alone
does not bound Eq.~\eqref{eq:b1inf}.  But $F$ is continuous at $y=0$ --- the deterministic ramp
enters only through the continuous shift $t_0y$ of a continuous joint density --- so there are
$\epsilon,c_0>0$ with $F(y)>c_0$ for $|y|<\epsilon$, whence
\begin{equation}
 \rho_{\rm sing}(0)=\int\!\dd y\,F(y)\ \ge\ 2\epsilon c_0\ >\ 0 .
\label{eq:rhosing}
\end{equation}
Hence $b_1^\infty>0$ throughout the operative domain $p_{\rm br}>0$ for $\alpha>1$ --- the
same condition that makes fragmentation operative, with no stronger support requirement --- which
at $n=1/2$ is the entire fragmented branch.  For $1/2<\alpha\le1$, reached only when $n<1/2$,
the support of $\eta$ is unbounded and the argument above does not apply; the statement is not
proved there.  Direct measurement confirms it and shows no drift:
the singleton-conditioned density at $|\Delta|<0.1\,t_0$ is $0.66,0.61,0.61,0.59$ at
$(\alpha,V)=(2,t_0)$ for $w_0=10,30,100,300$ and $1.06,0.90,0.97,0.90$ at $(1.5,2t_0)$, in units
where the full window has unit weight --- depleted relative to $|\Delta|\to t_0$, which is why
$b_1<2$, but bounded away from zero.
Measured over $w_0=10$--$300$: $f_1=0.35,0.37,0.38,0.40$ at $(\alpha,V)=(2,t_0)$ and
$0.58,0.55,0.57,0.60$ at $(1.5,2t_0)$, with no downward trend, and $f_1=0.32$--$0.50$ /
$0.54$--$0.59$ respectively after the partition is iterated to a fixed point against internal
rearrangements (below).  Crossing $\alpha=2n$ therefore \emph{reconstructs} the resonant-object
ensemble --- from asymptotically bond-like to a mixture of Hartree-fragmented multi-site islands
and surviving isolated wing bonds --- rather than replacing one elementary object by another.

The partition must finally be tested against the islands' own dynamics: internal hopping changes
occupations, occupations feed the Hartree detunings of the boundary bonds, and one must ask
whether cut bonds stay cut.  Enumerating the internal Fock configurations of both islands
adjacent to each cut, and iterating the resulting merges to a fixed point under a deliberately
permissive rule --- a cut reopens if \emph{any} internal configuration returns it to the window,
with no energetic accessibility imposed --- the reopening probability is substantial
($\simeq0.6$ at $V=t_0$) but $w_0$-independent, and the recursion converges in a few passes
(Table~\ref{tab:heal}).

The outcome must be read in the right observable.  The \emph{bond-weighted} mean
$\langle\ell\rangle_B=\langle\ell^2\rangle/\langle\ell\rangle$ drifts upward with $w_0$ wherever
fragmentation is weak, growing roughly as $w_0^{1/2}$ at $V=t_0$ while saturating at $V=2t_0$.
That drift is a \emph{tail} effect: $\langle\ell\rangle_B$ is dominated by the rare largest
clusters, whereas the quantity entering $\rho_{\rm act}$ is the \emph{count}-weighted mean, and
that saturates --- as does the active-object density $A_{\rm isl}=\rho_{\rm act}/p$ itself, with
cluster medians of one to two bonds throughout.  The recursively healed ensemble is therefore a
$w_0$-independent density of $\order(1)$ objects plus a rare, slowly broadening tail.

The tail's influence on the network coefficients is bounded in two distinct senses.  A sum rule controls the \emph{total} transition weight: for any cluster
eigenstate $\sum_a\sum_{s\neq r}|\langle r|n_a|s\rangle|^2=\sum_a\langle n_a\rangle_r
(1-\langle n_a\rangle_r)\le(\ell+1)/4$, so the weight per cluster grows at most linearly in size
while the number of clusters per unit length falls as $1/\langle\ell\rangle_{\rm count}$; the
total weight per unit length is therefore $\order(1)$ uniformly in cluster size, and cannot
diverge.  That is a statement about the integrated weight, not about its distribution: a large
cluster could in principle satisfy it while concentrating weight at ever smaller frequencies, and
it is the low-frequency density that sets $\widetilde C_{\rm rec}$.  The numerical evidence
addresses exactly that and shows no such concentration --- the doublet transition norm
\emph{decreases} with cluster size, $D(w)$ is flat over three decades including the separately
measured $\ell=11$--$20$ sector, and the infinite-temperature all-transition ensemble is flat as
well --- but these calculations reach only the cluster sizes accessible numerically.  We
therefore take uniform boundedness of the low-frequency matching density along the rare
asymptotic tail as an assumption, supported by but not proved by the data.  The
matching-law result of Sec.~\ref{sec:matching}, which needs only that the object be multi-site
with island-specific avoided crossings, is if anything reinforced by healing.  Protocols,
parameter tables and realization counts are in the Supplemental Material~\cite{SM}.

\begin{table}[t]
\caption{Recursive boundary healing at the fixed point of the permissive merge rule.  The
bond-weighted mean $\langle\ell\rangle_B$ drifts with $w_0$ where fragmentation is weak, but the
count-weighted mean and the active-object density $A_{\rm isl}=\rho_{\rm act}/p$ --- the quantity
that enters the pair density --- are $w_0$-independent.}
\label{tab:heal}
\begin{ruledtabular}
\begin{tabular}{lccccc}
$(\alpha,V/t_0)$ & quantity & $w_0{=}10$ & $30$ & $100$ & $300$\\
\colrule
$(2,1)$   & $\langle\ell\rangle_B$ & $8.8$ & $18.3$ & $33.1$ & $47.5$\\
$(2,1)$   & $\langle\ell\rangle_{\rm count}$ & $4.6$ & $4.9$ & $4.8$ & $4.6$\\
$(2,1)$   & $A_{\rm isl}$ & $0.169$ & $0.145$ & $0.142$ & $0.150$\\
\colrule
$(1.5,1)$ & $\langle\ell\rangle_B$ & $8.6$ & $17.1$ & $28.0$ & $36.0$\\
$(1.5,1)$ & $\langle\ell\rangle_{\rm count}$ & $4.3$ & $4.5$ & $4.3$ & $4.2$\\
$(1.5,1)$ & $A_{\rm isl}$ & $0.184$ & $0.161$ & $0.162$ & $0.165$\\
\colrule
$(2,2)$   & $\langle\ell\rangle_{\rm count}$ & $2.5$ & $2.4$ & $2.4$ & $2.2$\\
$(2,2)$   & $A_{\rm isl}$ & $0.284$ & $0.279$ & $0.274$ & $0.289$\\
$(1.5,2)$ & $\langle\ell\rangle_{\rm count}$ & $2.1$ & $2.2$ & $2.1$ & $2.0$\\
$(1.5,2)$ & $A_{\rm isl}$ & $0.301$ & $0.294$ & $0.307$ & $0.317$\\
\end{tabular}
\end{ruledtabular}
\end{table}  The matching theory of Sec.~\ref{sec:matching} must be
built accordingly.

\section{Long-range coupling of local resonant objects}
\label{sec:pseudospin}

\label{sec:randunif}

The interaction between two distant local resonant objects is controlled by an exact second
difference of the couplings.  We derive it for bonds, where it is sharpest, and note at the end how
finite islands modify it: they renormalize matrix-element \emph{amplitudes} by $\order(1)$
island-dependent factors while leaving the random-coupling \emph{power} $R^{-\alpha}$ untouched.

\subsection{Exact second-difference coupling}

For bonds $A=(a,a+1)$ and $B=(b,b+1)$,
\begin{equation}
 J_{AB}^{zz}=\frac14\left(
 U_{ab}-U_{a,b+1}-U_{a+1,b}+U_{a+1,b+1}
 \right),
\label{eq:Jzz}
\end{equation}
where $U_{ij}=V_{ij}/|i-j|^\alpha$.  Equivalently, Eq.~\eqref{eq:Jzz} is obtained by comparing the four local bond configurations: all terms independent of the opposite bond state cancel, leaving precisely the mixed discrete second difference.  If $R=b-a\ge2$ and the four random coefficients are independent,
\begin{align}
 {\rm Var}\,J_{AB}^{zz}&=\frac{V^2}{48}
 \left[2R^{-2\alpha}+(R-1)^{-2\alpha}+(R+1)^{-2\alpha}\right],\nonumber\\
 J_{AB}^{zz}&\simeq\frac{V}{4R^\alpha}\varsigma,\qquad
 \varsigma=X_1-X_2-X_3+X_4,
\label{eq:varJ}
\end{align}
with independent $X_a\sim U[-1,1]$, so $J^{zz}_{\rm rms}\simeq VR^{-\alpha}/(2\sqrt3)$; two exact
moments are $\mathbb E|\varsigma|=14/15$ and
$\mathbb E[|\varsigma|\ln|\varsigma|]=\ln4-539/450$, so
$\mathbb E|J_{AB}^{zz}|=(7V/30)R^{-\alpha}$.

The random-versus-uniform distinction developed next is a coherent multipole cancellation effect
already familiar from long-range localization theory~\cite{Yao2014}; here it supplies the
microscopic control channel.

For two finite islands the same structure appears summed over the island sites with
transition-density weights: projecting $\sum_{a\in A,b\in B}U_{ab}n_an_b$ onto the two island
doublets gives $J^{\perp}_{AB}=\sum_{a\in A,b\in B}U_{ab}\,m^A_a m^B_b$ with
$m_a=\langle g|n_a|e\rangle$.  For random couplings the terms are independent, so the sum performs
no coherent cancellation: conditional on the islands, $J^{\perp}_{AB}$ has rms
$\propto VR^{-\alpha}\,\lVert m^A\rVert\lVert m^B\rVert$.  Finite island size therefore
renormalizes the coupling \emph{amplitude} by $\order(1)$ island-dependent form factors while the
random-coupling \emph{power} $R^{-\alpha}$ is untouched.

The uniform-coupling cancellation also survives fragmentation, and for a structural reason.  Both island
states lie in the same particle-number sector, so both transition densities have zero monopole,
$\sum_a m^A_a=\langle g_A|N_A|e_A\rangle=0$ and likewise for $B$.  Expanding the uniform
coupling, $V/|R+b-a|^{\alpha}=VR^{-\alpha}\big[1-\alpha(b-a)/R
+\tfrac12\alpha(\alpha+1)(b-a)^2/R^2+\cdots\big]$, the $R^{-\alpha}$ and $R^{-\alpha-1}$ terms
vanish identically by these monopole sums, and the leading survivor is the dipole--dipole term,
\[
 J^{\perp}_{AB,\,\rm unif}\simeq-V\alpha(\alpha+1)\,R^{-\alpha-2}\,d_Ad_B,
 \qquad d_X=\sum_a a\,m^X_a
\]
--- the same two-power suppression as for bonds, Eq.~\eqref{eq:uniformJ2}, now with dipole
weights.

\subsection{Uniform interactions: the cancellation and its consequences}
\label{sec:uniform}

For uniform couplings the mixed second difference is parametrically suppressed.  For
$U_{ij}=V/|i-j|^\alpha$,
Eq.~\eqref{eq:Jzz} is a smooth discrete second difference and
\begin{equation}
 \ J^{zz}_{AB,\,\rm unif}\simeq-\frac{V}{4}\alpha(\alpha+1)R^{-\alpha-2}\ ,
\label{eq:uniformJ2}
\end{equation}
i.e.\ two powers faster than the random case at the same bare interaction strength; Fig.~3 of
Fig.~3 of Ref.~\cite{letter} places the two side by side over the measured range
$2\le R\le200$, where the gap reaches four orders of magnitude.  Randomness in $V_{ij}$ removes
the neighboring-site cancellation.
A thermodynamic caveat applies to this control: for all-positive couplings $V/r^\alpha$ with
$\alpha\le1$ the unnormalized interaction energy is superextensive, and a comparison at identical
bare $V$ forgoes Kac normalization.  The second-difference cancellation itself is unaffected --- it
is a statement about local energy differences --- but the random-versus-uniform contrast at fixed
bare $V$ should be read as a statement about matrix elements, not about thermodynamically
normalized models.

Anticipating the shell geometry of Secs.~\ref{sec:cascade} and~\ref{sec:alpha2}, inserting
$\alpha\to\alpha+2$ gives
$G^{\rm unif}_{n,\alpha}(\kappa)=(1-\kappa)^{1-n}\kappa^{-\alpha}$, for which
$\dd\ln G^{\rm unif}/\dd\kappa=-(1-n)/(1-\kappa)-\alpha/\kappa<0$
for all $\kappa\in(0,1)$ and $\alpha>0$,
so $G^{\rm unif}$ is monotonically decreasing and possesses no interior maximum: the long-distance
saddle that organizes the random case does not exist.  Moreover, the boundary obtained by
formally proceeding anyway would satisfy $h\propto L^{1-n-\alpha/2}$, whence
$(L/i_*)^{2(1-n)}\propto L^{\alpha}\to\infty$ for every $\alpha>0$: the putative uniform boundary
leaves the resonance-starved regime and thereby violates its own domain of validity, in contrast to the
random case, where $L^{\alpha-2}\to0$ for $\alpha<2$ keeps it inside ($\alpha=2$ is deferred to
Sec.~\ref{sec:alpha2}).

  With uniform couplings the Hartree
shifts become deterministic functions of the configuration rather than random variables, so the
comparison is not a clean one-parameter change.  Evaluating the forward branching of
Sec.~\ref{sec:crossover} for both cases at matched $h$ and $L=16$, we find that the $\alpha$ trend
\emph{reverses}: $B(\alpha{=}0.5)/B(\alpha{=}2)=1.12$ at $h=0.7\,h_0(L)$ for random couplings (the ratio runs from $1.05$ to $1.17$ across the field range), meaning longer range
yields more branching, against $0.93$ for uniform couplings.  The magnitudes differ by only about
$20\%$, and the comparison is made at a single matched $(L,h)$; we therefore present it as an
illustrative control rather than a quantitative result.  The calculation therefore shows a qualitative reversal of the range dependence, together
with the strong parametric suppression of Eq.~\eqref{eq:uniformJ2}, rather than the absence of any
effect.
\section{Matching laws: bonds versus islands}

\label{sec:matching}

Two local resonant objects exchange excitation energy when their splittings match to within the
transverse coupling.  This section derives the matching probability in both local-object regimes of
Eq.~\eqref{eq:regimes}: for isolated bonds ($\alpha\le2n$) the two-level band edge produces
$P^{\rm match}\sim q\ln(1/q)$, while for the fragmented islands of the physical
$\alpha>2n$ regime no logarithm is resolved and $P^{\rm match}\simeq C_{\rm isl}\,q$ over the accessible window.  We
establish the first analytically with numerical verification, and the second by exact
diagonalization of the physical island ensemble.

\subsection{Isolated bonds: the two-level edge}

For one resonant bond the conditional detuning of Sec.~\ref{sec:resbonds} is uniform to leading order, which implies the normalized splitting density
\begin{equation}
 f(\Omega)=\frac{\Omega}{t_0\sqrt{\Omega^2-4t_0^2}},
 \qquad 2t_0<\Omega<\sqrt5\,t_0,
\label{eq:fOmega}
\end{equation}
in the resonance-starved limit.  The square-root lower-edge singularity is the Jacobian of the map $\delta\mapsto\Omega$ and therefore survives smooth deformations of the single-bond detuning distribution.

\subsection{Independent-bond benchmark}

If two resonant bonds are drawn \emph{independently} from Eq.~\eqref{eq:fOmega}, the physical mixing-angle resonance condition
\begin{equation}
 |\Omega_1-\Omega_2|<|J^{zz}|\,
 \frac{2t_0}{\Omega_1}\frac{2t_0}{\Omega_2}
\label{eq:mixcond}
\end{equation}
gives, for $q=|J^{zz}|/t_0\ll1$,
\begin{align}
 P_{\rm mix}^{\rm(ind)}(q)&=2q\left[\ln\frac1q+c_{\rm mix}\right]+o(q),\nonumber\\
 c_{\rm mix}&=1+\ln\!\left[\frac{16(5\sqrt5-8)}{28+13\sqrt5}\right]=0.8853\ldots
\label{eq:Pmix-ind}
\end{align}
If the mixing factors are removed, the corresponding exact constant is
$c_\Omega=1.1214\ldots$ (Appendix~\ref{app:matching}).  The convention matters: widening the
resonance window to $|\Omega_1-\Omega_2|<\eta|J^{zz}|s_1s_2$, with $s_i=2t_0/\Omega_i$ the mixing
factors, gives
$P^{\rm(ind)}_{\rm mix}=2\eta q[\ln(1/q)+c_{\rm mix}-\ln\eta]+o(q)$, so \emph{both} the leading
coefficient and the constant change.  We use $\eta=1$ throughout, so the
slope $2$ tested in Fig.~\ref{fig:pmatch} is a prediction of the equidistribution hypothesis
\emph{together with} this window convention, not a convention-independent law.  Throughout,
the coefficient $2$ and the constants $c_{\rm mix}$, $c_\star$ are exact for this decorrelated
benchmark and window convention; they are benchmark constants of the isolated-bond regime, not
universal physical numbers.  The same caveat
applies to the choice of resonance window $|\delta|<t_0$ itself: $N_{\rm res}$ and $N_c$ both scale
with it, so the numerical value of $N_c$, and amplitudes generally, are convention dependent; the scaling
exponents are the robust quantities.

\subsection{Fixed-pair linearity and the coarse-grained logarithm}

\label{sec:broadening}

Two physical bonds in the SV chain are not independent draws from Eq.~\eqref{eq:fOmega}: at $V=0$
their splittings are deterministic functions of the same $\phi$.  The mixing factors are kept \emph{explicit} in the resonance condition, as in
Eq.~\eqref{eq:mixcond}.  The matching probability is defined \emph{conditionally on both bonds being
resonant bonds}, since the bond-resonance probabilities are supplied separately in Eq.~\eqref{eq:nR}:
\begin{align}
 P_{ij}^{\rm match}(q\,|\,{\rm res})=\frac{\displaystyle\int_0^{2\pi}\!\!\dd\phi\,
 \chi_i^{(0)}\chi_j^{(0)}\,\Theta\!\big(qt_0s_is_j-|\Omega_i-\Omega_j|\big)}
 {\displaystyle\int_0^{2\pi}\!\!\dd\phi\,\chi_i^{(0)}\chi_j^{(0)}},
\label{eq:exact-pairmatch}
\end{align}
the denominator being $2\pi P^{\rm res}_{ij}$ of Eq.~\eqref{eq:jointres}.  This is a
one-dimensional phase integral, not the two-dimensional product measure that produces
Eq.~\eqref{eq:Pmix-ind}.  At a simple root $\phi_a$ of
$g_{ij}(\phi)=\Omega_i(\phi)-\Omega_j(\phi)$ the coarea expansion gives
\begin{align}
 P_{ij}^{\rm match}(q\,|\,{\rm res})=\frac{q\,t_0}{\pi P^{\rm res}_{ij}}
 \sum_{\phi_a}\frac{\chi_i^{(0)}\chi_j^{(0)}\,s_is_j}{|g_{ij}'(\phi_a)|}\bigg|_{\phi_a}+o(q).
\label{eq:coarea}
\end{align}
\emph{For a fixed pair the bare matching law is therefore linear in $q$}, with no logarithm.  We
verify this directly in Appendix~\ref{app:pmatch}: at $V=0$ and fixed $(A,B)$, $P/q$ is flat over
four decades in $q$.

This per-pair statement does not, however, survive coarse graining --- and the pair-scale construction of Sec.~\ref{sec:cascade} uses the matching law only in coarse-grained form.  Two distinct routes
produce a two-dimensional joint measure for $(\Delta_A,\Delta_B)$, and either one restores the
logarithm.  The first is interaction broadening, via the scale separation of
Sec.~\ref{sec:resrobust}.  The two Hartree shifts share only the cross-bond coefficients, whose contribution decays with
separation while their individual variances are set by the remaining bath coefficients: for
$R\gg1$,
$\mathrm{Cov}(\eta_A,\eta_B)=\order(V^2R^{-2\alpha})$ while
$\mathrm{Var}\,\eta_A$ is given by Eq.~\eqref{eq:sigmaint} --- $\order(V^2)$ for $\alpha>1/2$, with
the logarithm at $\alpha=1/2$ --- so in every case the correlation vanishes with separation and
$(\Delta_A,\Delta_B)$ acquires a smooth two-dimensional joint density near the origin.  Near the
pseudospin edge, $\Omega-2t_0\simeq\Delta^2/4t_0$, so the matching condition becomes
$|\Delta_A^2-\Delta_B^2|\lesssim4t_0J$.  The area of this hyperbolic strip in a smooth
two-dimensional measure is $\order(J\ln(t_0/J))$, the logarithm arising from the pinch at the origin.
Hence $P^{\rm match}(q)\sim q\ln(1/q)$ at $q=|J^{zz}|/t_0$.
The second route requires no interaction \emph{at fixed} $q$; in the physical limit
$q_R\sim R^{-\alpha}\to0$ its control is more delicate, and for $\alpha>1$ it must be supplemented
by the interaction broadening discussed in Sec.~\ref{sec:cascade}.  Write the
two resonance conditions in terms of the local phases of Eq.~\eqref{eq:full-detuning},
$\Delta_A=-A^{\rm d}_A\sin\psi_A$ and $\Delta_B=-A^{\rm d}_B\sin\psi_B$, with
\begin{equation}
 \psi_B-\psi_A=\Delta\gamma_{AB}
 =2\pi\beta\big[x^n-(x{-}R)^n\big]+\tfrac12(\Delta\theta_{x}-\Delta\theta_{x-R}),
\label{eq:offset}
\end{equation}
a constant independent of $\phi$.  For a \emph{fixed} pair, therefore, $(\psi_A,\psi_B)$ traverses
the line $\psi_B=\psi_A+\Delta\gamma_{AB}$ on the torus as $\phi$ varies: a one-dimensional measure,
transverse crossings, and hence $P^{\rm match}\propto q$ with no logarithm.

Coarse graining changes the measure rather than the dynamics.  Averaging over a mesoscopic shell,
$\ell_{\rm comb}(x)\ll\Delta R\ll\min\{R,x-R,x\}$, varies $\Delta\gamma_{AB}$, and under the same equidistribution
hypothesis that fixes the correlation factor in the Supplemental Material~\cite{SM}, $\Delta\gamma_{AB}$ is
uniform mod $\pi$ over the shell.  The change of variables
$(\phi,\Delta\gamma_{AB})\mapsto(\psi_A,\psi_B)$ is then measure preserving with both images
uniform, so shell averaging is asymptotically equivalent
to drawing the two resonant-bond phases \emph{independently}.  The step is controlled because the
shell varies the phase offset and the envelopes at very different rates: over
$\Delta R\sim\ell_{\rm comb}(x)$ at $R\sim\kappa x$, the offset $\Delta\gamma_{AB}$ sweeps a full
period by construction, whereas the local amplitudes $A^{\rm d}_B$, the marginal $p_B$ and the
coupling $q(R)$ change only by $\order(\Delta R/x)=\order(x^{-n})$.  Averaging over the shell
therefore randomizes the phase offset while holding the envelopes fixed to leading order.  In the resonance-starved limit that is precisely the
independent two-bond ensemble of Appendix~\ref{app:matching}, and therefore
\begin{equation}
 \ \big\langle P^{\rm match}\big\rangle_{\rm shell}\longrightarrow
 P^{\rm(ind)}_{\rm mix}(q)=2q\Big[\ln\frac1q+c_{\rm mix}\Big]\ .
\label{eq:shell-to-ind}
\end{equation}
Under the independent-phase limit this fixes not only the $q\ln(1/q)$ form but the coefficient $2$
and the constant $c_{\rm mix}=0.8853\ldots$ of Eq.~\eqref{eq:Pmix-ind}, with no fitted parameter.  More
generally, any smooth joint detuning density gives
$P^{\rm match}(q)=A_{\rm mix}\,q\ln(1/q)+\order(q)$: the power and the logarithm are robust, while
$A_{\rm mix}=2$ and the finite constant hold under the decorrelated limit and are measured as
$1.95(4)$ and within $6\%$ at $V=t_0$.

The $24\%$ \emph{suppression} of forward branching at $V=0$ in Sec.~\ref{sec:crossover} concerns a
different average over the same comb.  The two observables average over the comb differently.  The finite-chain branching calculation samples a small set of
specific offsets, whereas the pair-scale construction uses a mesoscopic shell average over many offsets.  The comb
therefore remains resolved in the former and becomes an effective phase coordinate in the
latter.

\begin{figure}[t]\centering
\includegraphics{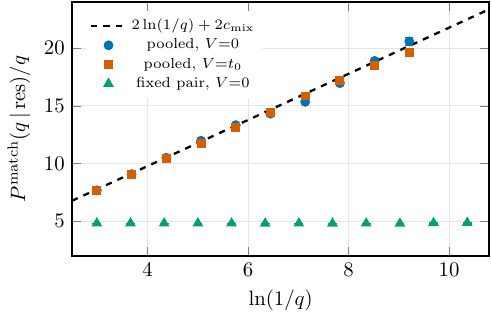}
\caption{One pair is one-dimensional; the pair ensemble is not.  For a \emph{fixed} physical pair at
$V=0$ (triangles, four pairs pooled, $6\times10^7$ phases each) the conditional matching probability
divided by $q$ is flat, fitting $4.795+0.0033\ln(1/q)$ --- the coarea result Eq.~\eqref{eq:coarea}.
Averaging over the pair ensemble (circles and squares; $L=100$, $h/t_0=25$, $\alpha=1$, all
separations $R\ge2$) instead
reproduces the independent-bond law Eq.~\eqref{eq:Pmix-ind}, dashed, with no free parameter: the
fitted slopes are $2.00(5)$ at $V=0$ and $1.95(4)$ at $V=t_0$ against the predicted $2$.  The bare
and interacting ensembles have leading logarithmic slopes that agree within errors, so at these
fixed-$q$ parameters the logarithm is supplied by averaging over the comb offset alone.  The sharper, shell-resolved test of the mesoscopic
hypothesis actually used in Eq.~\eqref{eq:joint-shell} is given in Appendix~\ref{app:pmatch}.  Pooled-ensemble error bars are batch means over $12$ independent
phase/disorder/configuration batches, not Poisson, since events within a chain are correlated by the
shared phase.}
\label{fig:pmatch}
\end{figure}

Figure~\ref{fig:pmatch} compares the fixed-pair and pooled matching laws.  The fixed-pair law is flat; the pooled pair-ensemble
law follows Eq.~\eqref{eq:shell-to-ind} with fitted slope $2.00(5)$ at $V=0$ and $1.95(4)$ at
$V=t_0$ against the predicted $2$, using batch-mean errors over independent
phase/disorder/configuration batches.  Together with the shell-resolved tests of Appendix~\ref{app:pmatch}, these results support the
shell-equidistribution construction; the logarithm itself follows analytically from the smooth
two-dimensional matching measure.  Two
consequences follow.  The logarithm entering Eq.~\eqref{eq:hcapp} is already present at $V=0$ and
does not rest on the system being in any interaction-``resolved'' regime; and the constant
$c_{\rm mix}$, derived for independent draws, is the exact constant of the coarse-grained matching
law \emph{at fixed} $q$ --- the benchmark constant $c_\star$ of Eq.~\eqref{eq:cstar} follows
from it after averaging over the random coupling amplitude.  At fixed $q$ the logarithm does not need the interaction; the role of interaction broadening is to
control the shrinking-window limit $q_R\to0$, where the deterministic route fails for $\alpha>1$
(Sec.~\ref{sec:cascade}).  We therefore do not attempt to locate a threshold in $V/t_0$.

\subsection{Fragmented islands: loss of the logarithm}
\label{sec:islmatch}

The $q\ln(1/q)$ law rests on the square-root edge of Eq.~\eqref{eq:fOmega}, which in turn rests on
every local object sharing the \emph{same} deterministic band edge $\Omega_{\rm min}=2t_0$.  A
fragmented island does not: its lowest splitting is set by an island-specific avoided crossing,
distributed over the ensemble.  Whether any singularity survives is a question about the physical
island ensemble, and we answer it by exact diagonalization of the islands harvested from the
full-detuning counting of Sec.~\ref{sec:islandmc}, up to $\ell=20$ bonds (Hilbert-space dimensions
up to $352\,716$; islands beyond $\ell=20$ carry $5$--$7\%$ of the resonant-bond weight at $V=t_0$
and none at $V=2t_0$; protocol in the Supplemental Material~\cite{SM}).

The island splitting distribution $f_{\rm isl}(\Omega)$ is regular: for $\ell\ge3$ islands the
median splitting is $\simeq t_0$, between $37\%$ and $49\%$ of them have $\Omega<t_0$, and no
accumulation at $2t_0$ remains.  The direct test of the matching law is the pair-density
diagnostic
\begin{equation}
 D(w)=\frac{\Pr\big(|\Omega_A-\Omega_B|<w\big)}{2w}
 \;\xrightarrow[w\to0]{}\;\int\!\dd\Omega\,f_{\rm isl}^2(\Omega),
\label{eq:Dw}
\end{equation}
which saturates as $w\to0$ if and only if $P^{\rm match}\propto q$ without a logarithm.
Table~\ref{tab:Dw} shows $D(w)$ for island pairs with $\ell\ge3$: it is flat over three decades in
every case studied, with the sharpest test --- exact all-pairs counting over $8868$ islands at
$\alpha=1.5$, $V=2t_0$ --- flat to better than $1\%$.  The $\ell=11$--$20$ sector, measured
separately, is also flat, with a \emph{larger} finite value ($D\simeq1.7$--$2.1$): large islands
match more easily, but boundedly.  Hence, for genuine islands,
\begin{equation}
 \ P^{\rm match}(q)\simeq C_{\rm isl}\,q,\qquad 10^{-4}\lesssim q\lesssim10^{-1}\ ,
\label{eq:Pisl}
\end{equation}
with $C_{\rm isl}$ an $\order(1)$, cluster-size-dependent constant and no logarithmic
singularity resolved over the accessible window.

\begin{figure}[t]\centering
\includegraphics{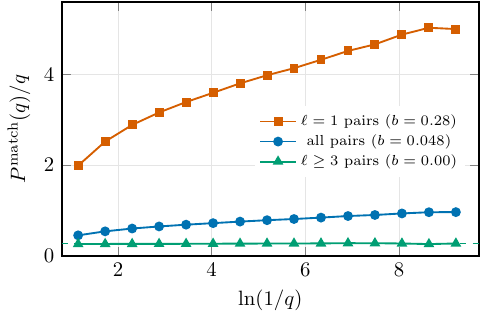}
\caption{$\ell$-resolved decomposition of the matching law in the fragmented regime ($\alpha=2$,
$V=t_0$).  Pairs of surviving single bonds ($\ell=1$, squares) retain a logarithmic slope
$b_{\ell=1}=0.28$ (reduced from the ideal benchmark of Appendix~\ref{app:matching} by the
reshaped conditional detuning); genuine island pairs ($\ell\ge3$, triangles) are flat over the
full range, $b=0.00\pm0.02$ (dashed guide); and the full ensemble (circles) has slope
$b_{\rm all}=0.048$, quantitatively the surviving single-bond contribution
$f_{\ell=1}^2\,b_{\ell=1}=0.41^2\times0.28=0.047$.  Slopes for the other $(\alpha,V)$ sets are
tabulated in the Supplemental Material~\cite{SM}.}
\label{fig:lresolved}
\end{figure}

Equation~\eqref{eq:Pisl} is a statement about the islands, not about the physical ensemble.
Because a finite fraction $f_1$ of active objects remains single bonds at every $w_0$
[Eq.~\eqref{eq:f1}], the ensemble is a mixture, and only bond--bond pairs share the deterministic
edge:
\begin{equation}
\begin{aligned}
 P^{\rm match}&=w_{11}P_{11}+w_{1I}P_{1I}+w_{II}P_{II}\\
 &\simeq b_{\rm all}\,q\ln(1/q)+C_{\rm all}\,q,\quad b_{\rm all}=w_{11}b_1>0.
\end{aligned}
\label{eq:Pmix}
\end{equation}
Two points must be checked before Eq.~\eqref{eq:Pmix} may be used inside the shell theory of
Sec.~\ref{sec:cascade}.  First, the sector weights: the comb correlates objects, so $w_{11}$ need not equal $f_1^2$.
Measured directly on the mesoscopic shells of Sec.~\ref{sec:cascade} (center window at $x_0$,
shell at $\bar R$, active objects only), the ratio $w_{11}/[f_1(x_0)f_1(x_0-\bar R)]$ is
$1.058$ at $\alpha=1.5$ and $1.027$ at $\alpha=2$ ($V=t_0$; $5\times10^4$ and $4\times10^4$
pairs), so $w_{11}\simeq f_1^2$ within a few percent and $b_{\rm all}\simeq f_1^2b_1$ is an
accurate numerical approximation for these shells.  Second, the sector slopes.  Fitting
$P/q=a+b\ln(1/q)$ confirms the decomposition sector by sector: pairs of $\ell=1$ objects
give a finite slope ($b_1=0.28$--$0.75$ depending on $(\alpha,V)$, reduced from the ideal
benchmark because the surviving wing bonds sample a reshaped conditional detuning); $\ell\ge3$
pairs give $b=0.00\pm0.02$ in all cases; and the full ensemble obeys
$b_{\rm all}\simeq f_{\ell=1}^2\,b_{\ell=1}$ ($0.41^2\times0.28=0.047$ against the measured
$0.048$ at $\alpha=2$, $V=t_0$), with $b_{\rm all}=0.048$--$0.24$ over the parameter sets studied;
Fig.~\ref{fig:lresolved} displays the decomposition directly.  Since $q\ln(1/q)\gg q$ as
$q\to0$, this residual channel cannot be discarded asymptotically: with $C_{\rm all}\simeq0.5$--$0.8$
it already dominates for $R\gtrsim10$--$300$ sites, far inside the pair-resonance scale.  Fragmentation
therefore removes the two-level singularity from the multi-site objects themselves and suppresses
the ensemble coefficient by a factor of $8$--$40$ relative to the isolated-bond
benchmark $A_{\rm mix}=2$, but does not remove the logarithm from the physical ensemble; the
pair-resonance scale of Sec.~\ref{sec:cascade} is built on Eq.~\eqref{eq:Pmix}.

Two further checks.  First, the two-bond sector: $\ell=2$ islands make
up $f_{\ell=2}=0.23$--$0.25$ of the population, and their matching slope vanishes,
$b_{\ell=2}=-0.008$ and $-0.005$ at $(\alpha,V)=(2,t_0)$ and $(1.5,2t_0)$ respectively (over
$10^6$ pairs each), so every genuine island class $\ell\ge2$ is slope-free
($b_{\ell\ge2}=-0.001$) and the logarithm is carried exclusively by $\ell=1$.  Second, the
lowest doublet is not special.  The parent problem is a high-energy one, so we repeated the
diagnostic including \emph{every} island transition with nonvanishing local transition density
$\sum_a|\langle r|n_a|s\rangle|^2$, at infinite-temperature weighting (initial eigenstate
uniform, channel weighted by its transition density; $7.0\times10^6$ transitions from $10\,852$
densely diagonalized islands at $\alpha=2$, $V=t_0$): the pair-density diagnostic remains flat
over three decades, $D=0.270$--$0.274$ with small-$w$ slope $+0.0001$ (and $0.283$--$0.292$ at
$\alpha=1.5$, $V=2t_0$).  No singularity appears in the weighted infinite-temperature
all-transition ensemble; the protocol is given in the Supplemental Material~\cite{SM}.
Finally, the sector structure is not an artefact of the snapshot partition: diagonalizing the
recursively healed fixed-point clusters themselves (Sec.~\ref{sec:islandmc}) at $w_0=30$ and
$100$ for $(\alpha,V)=(2,t_0)$ and $(1.5,2t_0)$ gives $b_{\ell\ge2}=0.00\pm0.01$, $D(w)$ with no
systematic low-frequency growth over three decades, and $b_{\rm all}=f_1^2b_{\ell=1}$ to within $3\%$ with the healed
$f_1$ ($0.055$ against $0.055$ at $(2,t_0)$, $w_0=100$; $0.237$ against $0.235$ at
$(1.5,2t_0)$), the residual logarithm being if anything slightly weaker at the fixed point~\cite{SM}.

\begin{table}[t]
\caption{Saturation of the pair-density diagnostic $D(w)$, Eq.~\eqref{eq:Dw}, for genuine islands
($\ell\ge3$), from exact island diagonalization; $D$ is quoted in units of $t_0^{-1}$.  A
logarithmic matching law would make $D$ grow as $w\to0$; instead $D$ is flat over three decades.
The $\ell=2$ sector, measured separately, is likewise slope-free (see text).  The $(\alpha{=}1.5,V{=}2t_0)$ row is the
exact all-pairs evaluation (statistical errors below $1\%$); the others are sampled with
$6\times10^6$ pairs.  Full protocol and the $\ell$-resolved decomposition in the Supplemental
Material~\cite{SM}.}
\label{tab:Dw}
\begin{ruledtabular}
\begin{tabular}{lccc}
$(\alpha,\,V/t_0)$ & $D(10^{-4})$ & $D(10^{-3})$ & $D(10^{-2})$\\
\colrule
$(1.5,\,1)$ & $0.73$ & $0.79$ & $0.78$\\
$(2.0,\,1)$ & $0.74$ & $0.75$ & $0.76$\\
$(1.5,\,2)$ & $0.739(10)$ & $0.737(3)$ & $0.738(1)$\\
$(2.0,\,2)$ & $0.78$ & $0.77$ & $0.75$\\
\end{tabular}
\end{ruledtabular}
\end{table}

\section{The pair-resonance scale and the visibility scale}

\label{sec:cascade}

\subsection{Pair density on a logarithmic shell}

Because of Eq.~\eqref{eq:jointres}, a pair density cannot be written as a product of two marginal
bond-resonance probabilities.  The microscopic object required is the \emph{joint} indicator that
both bonds are resonant \emph{and} matched,
\begin{equation}
\begin{aligned}
 \mathcal I_{AB}&=\chi_A\chi_B\,
 \Theta\!\big(|J^{zz}_{AB}|s_As_B-|\Omega_A-\Omega_B|\big),\\
 n_R(x)&\sim R\,\big\langle\mathcal I_{x-R,\,x}\big\rangle_{\phi,\,R\in{\rm shell},\,V}.
\end{aligned}
\label{eq:nR}
\end{equation}
We do not factorize this into a correlation factor times a separately averaged matching
probability.  Such a factorization would be an independence assumption, and a poor one: both factors
are largest at the same configurations, the comb coincidences where the arcs of
the Supplemental Material~\cite{SM} overlap and where $|g_{ij}'|$ is anomalously small, so the product would
systematically underestimate $n_R$.

Instead, evaluate the joint average directly using the shell mapping of Sec.~\ref{sec:matching}.
For the deterministic $V=0$ shell construction, the step that follows is conditional on the
equidistribution hypothesis of Sec.~\ref{sec:matching}; it is tested independently in two ways in
Appendix~\ref{app:pmatch} and in the Supplemental Material~\cite{SM}.  The physical $V>0$ route
through Hartree broadening, introduced below, does not require it.
Choose a \emph{mesoscopic} shell,
\begin{equation}
 \ell_{\rm comb}(x)\ll\Delta R\ll\min\{R,\,x-R,\,x\},
\label{eq:mesoscopic}
\end{equation}
a window that exists asymptotically for every fixed $n>0$ because
$\ell_{\rm comb}(x)/x\sim x^{-n}\to0$.  Across such a shell the offset $\Delta\gamma_{AB}$ sweeps
through $\order(1)$ phase while the envelopes and the coupling vary only by
$\delta A/A,\ \delta p/p,\ \delta q/q=\order(\Delta R/x)=o(1)$.  Holding $p_A$, $p_B$ and $q$ fixed
and averaging the phase offset, the map of Sec.~\ref{sec:broadening} is measure preserving with
both images uniform, so the \emph{full} joint indicator --- not its pieces separately --- obeys
\begin{equation}
 \ \big\langle\chi_A\chi_B\Theta\big\rangle_{\rm shell}
 =p_Ap_B\,P^{\rm(ind)}_{\rm mix}(q)
 +o\!\Big(p_Ap_B\,q\ln\frac1q\Big).
\label{eq:joint-shell}
\end{equation}
Because the matching window shrinks with scale, $q_R\sim R^{-\alpha}$, ordinary equidistribution of
the offsets is not sufficient for Eq.~\eqref{eq:joint-shell}: resolving the matching scale alone
needs $D_{N_R}\ll q_R\ln(1/q_R)$, and with the optimistic $D_N\sim1/N$ this demands
$\Delta R\gg x^{\alpha}/(\alpha\ln x)$, compatible with $\Delta R\ll x$ only for
$\alpha\le1$.  This limitation is now harmless.  For the $n=1/2$ case emphasized numerically the isolated-bond
regime lies at $\alpha\le1$, where the deterministic shell is resolvable; for general $n>1/2$
the portion $1<\alpha\le2n$ of the bond regime instead relies on the continuous Hartree-broadened
joint measure at fixed nonzero $V$.  In either case the Hartree fluctuations supply a continuous joint detuning density at fixed $R$ ---
the smoothing ratio near the pseudospin edge is
$\sigma_{\rm int}/\sqrt{t_0J(R)}\sim\sqrt{V/t_0}\,R^{\alpha/2}$, growing with separation ---
so both routes to the smooth two-dimensional measure are available.  In the fragmented regime
$\alpha>2n$ the genuine islands contribute the linear law Eq.~\eqref{eq:Pisl}, which requires no
shrinking-target resolution at all --- only the smooth $\order(1)$ density of island splittings
enters --- while the surviving wing bonds contribute the logarithmic channel with weight $f_1^2$.

The shell pair density is therefore
\begin{equation}
\begin{aligned}
 n_R(x)&\sim R\,\rho(x)\,\rho(x-R)\,P^{\rm match}\big(q_R\big),\\
 P^{\rm match}(q)&=b_{\rm all}\,q\ln\frac1q+C_{\rm all}\,q+o(q),
\end{aligned}
\label{eq:nR-final}
\end{equation}
with $b_{\rm all}=w_{11}b_1\simeq f_1^2b_1$ throughout $0<\alpha<2$: below $\alpha=2n$, $f_1\to1$ and
$b_{\rm all}\to A_{\rm mix}$; above it, $f_1\to f_1^\infty=\order(1)$ and $b_{\rm all}$ is
suppressed by a factor of $8$--$40$ but remains positive, while $C_{\rm all}$ collects
the island and mixed sectors.  In the isolated-bond regime, Eq.~\eqref{eq:nR} is used as
written: $p$ is the resonant-bond density, $A_{\rm mix}$ is the finite coefficient of the smooth
joint measure (equal to $2$ in the decorrelated independent-phase limit), and the coupling is
$q_R=|J^{zz}(R)|/t_0=V|\varsigma|/(4t_0R^{\alpha})$, with $\varsigma$ the signed four-term
amplitude of Eq.~\eqref{eq:varJ}.  In the fragmented regime the physical nodes are the
\emph{active islands} of Sec.~\ref{sec:islandmc}, of density
\[
 \rho_{\rm act}(x)=\frac{f_{\rm act}}{\langle\ell\rangle_{\rm isl}}\,p(x)
 \equiv A_{\rm isl}\,p(x),
\]
where $f_{\rm act}$ excludes the frozen clusters; the coupling between two islands is
$J^{\perp}_{AB}=VR^{-\alpha}\Gamma_{AB}$ with $\Gamma_{AB}$ the random form-factor sum of
Sec.~\ref{sec:randunif}, so the matched-pair density reads
\begin{align*}
 n_R^{\rm isl}(x)&\sim R\,\rho_{\rm act}(x)\,\rho_{\rm act}(x-R)\,
 C_{\rm match}\,\frac{V}{t_0R^{\alpha}},\\
 C_{\rm match}&=C_{\rm isl}\big\langle|\Gamma|\big\rangle,
\end{align*}
with $C_{\rm match}$ a nonuniversal $\order(1)$ island constant.  Because
$\rho_{\rm act}\propto p$ with $\order(1)$ coefficients, every scaling statement below is
common to the two regimes; only the amplitude bookkeeping differs.  The shell-averaged correlation factor of
the Supplemental Material~\cite{SM}, $\langle\mathcal C\rangle_{\rm shell}\to1$, is then not an ingredient of
Eq.~\eqref{eq:nR-final} at all: it is an \emph{independent} diagnostic of the same equidistribution
hypothesis, as is the matching law measured in Fig.~\ref{fig:pmatch}.  

Averaging over the random amplitude is closed (Appendix~\ref{app:Jmoments}).  In the
isolated-bond regime,
\begin{equation}
 \ \big\langle P^{\rm match}\big\rangle_{\varsigma}
 =\frac{7}{15}\frac{V}{t_0R^{\alpha}}
 \Big[\alpha\ln R+\ln\frac{4t_0}{V}+c_\star\Big],
\label{eq:Pmatch-random}
\end{equation}
with the benchmark constant $c_\star=c_{\rm mix}-\tfrac{15}{14}(\ln4-\tfrac{539}{450})=0.68334\ldots$
derived there.  In the fragmented regime the four-term amplitude $\varsigma$ does not apply
--- the island coupling carries the form-factor sum $\Gamma_{AB}$ --- and the island--island
\emph{sector} averages to
\begin{equation}
 \big\langle P_{II}\big\rangle
 =C_{\rm match}\,\frac{V}{t_0R^{\alpha}},
 \qquad C_{\rm match}=C_{\rm isl}\big\langle|\Gamma|\big\rangle=\order(1),
\label{eq:Pmatch-isl}
\end{equation}
a nonuniversal island constant: no exact coefficient is claimed for this sector.  Equation
\eqref{eq:Pmatch-isl} is \emph{not} the fragmented-regime matching probability: by
Eq.~\eqref{eq:Pmix} the physical ensemble also contains the $11$ and $1I$ sectors, and only the
$11$ sector carries the shared band edge.  Averaging Eq.~\eqref{eq:Pmix} over the coupling
amplitude at separation $R$ therefore gives, throughout $0<\alpha<2$,
\begin{equation}
 \big\langle P^{\rm match}\big\rangle
 =\frac{V}{t_0R^{\alpha}}\Big[\widetilde B_{\rm rec}\,\alpha\ln R+\widetilde C_{\rm rec}\Big],
 \qquad \widetilde B_{\rm rec}>0 .
\label{eq:Pmatch-rec}
\end{equation}
The distinction from Eq.~\eqref{eq:Pmatch-random} matters.  In the asymptotic isolated-bond
branch, $f_1\to1$ and the objects are the ideal ensemble of Appendix~\ref{app:matching}, so
$\widetilde B_{\rm rec}\to\tfrac{7}{15}$ and $\widetilde C_{\rm rec}$ carries the exact constant
$\ln(4t_0/V)+c_\star$.  In the reconstructed regime neither constant applies: the surviving wing
bonds are a \emph{selected} subpopulation whose conditional detuning is reshaped, with measured
$b_1=0.28$--$0.75$ rather than the ideal benchmark $2$ (Sec.~\ref{sec:islmatch}).  We therefore
keep $\widetilde B_{\rm rec}\propto w_{11}b_1$ and $\widetilde C_{\rm rec}$ as nonuniversal
$\order(1)$ coefficients of the reconstructed ensemble; no exact prefactor is claimed there.  All
that the scaling theory requires is $\widetilde B_{\rm rec}>0$, which is what
Eq.~\eqref{eq:b1inf} establishes.  Within the decorrelated matching measure the bond-regime
average preserves the $R^{-\alpha}\ln R$ structure and shifts only the finite constant,
$c_{\rm mix}\to c_\star$; $c_{\rm mix}$ is the exact constant of the matching law \emph{at
fixed} $q$, for the window convention $\eta=1$ discussed after Eq.~\eqref{eq:Pmix-ind}.  The same cross-bond coefficients enter both
$J^{zz}$ and the Hartree detunings, so $q_R$ and the detunings are not microscopically independent;
those correlations decay with $R$ and can modify finite $\order(1)$ terms, but not the leading
$R^{-\alpha}\ln R$ structure.

\subsection{Why the pair-scale parameter is geometric}

A resonant pair of size $R$ has an internal splitting $\epsilon_R\sim c_1VR^{-\alpha}$, while two
such pair pseudospins a distance $r$ apart interact as $M(r)\sim c_2Vr^{-\alpha}$.  Both inherit
the power $R^{-\alpha}$ from the same second difference, Eq.~\eqref{eq:Jzz}, but they
are different matrix elements of it --- one internal to a pair, one between two pairs --- and there is
no symmetry forcing $c_1=c_2$.  If the density of size-$R$ pair pseudospins is $n_R$, their typical
spacing is $\ell_R\sim n_R^{-1}$, and the second generation becomes resonant when
$M(\ell_R)\gtrsim\epsilon_R$, i.e.
\begin{equation}
 \ n_RR\gtrsim\big(c_1/c_2\big)^{1/\alpha}=\order(1)\ .
\label{eq:geometric-takeoff}
\end{equation}
The explicit overall factor $V$ cancels, and this is the geometric origin of the
iterated-pair criterion; the numerical threshold on the right is $\order(1)$ but is not derived to be
unity.  This argument neglects correlations among overlapping random coefficients and does not establish
all-generation percolation.

Using Eq.~\eqref{eq:nR-final}, define $\Lambda(x,R)\equiv n_R(x)R$.  For a macroscopic shell
$R=\kappa x$ with $\kappa$ bounded away from $0$ and $1$,
\begin{align}
 \Lambda(x,\kappa x)&\sim
 \frac{V}{t_0}
 \frac{x^{4-2n-\alpha}}{i_*^{2(1-n)}}
 (1-\kappa)^{1-n}\kappa^{2-\alpha}\,
 \mathcal L(x,\kappa),\nonumber\\
 \mathcal L(x,\kappa)&=
 \widetilde B_{\rm rec}\big[\alpha\ln x+\alpha\ln\kappa\big]+\widetilde C_{\rm rec},
\label{eq:Lambda-coarse}
\end{align}
$\mathcal L$ being Eq.~\eqref{eq:Pmatch-rec} at $R=\kappa x$, with $\widetilde B_{\rm rec}>0$ on
both sides of the reconstruction and reducing to the exact bond-regime bracket of
Eq.~\eqref{eq:Pmatch-random} as $f_1\to1$.  The exponent $4-2n-\alpha$ follows from the shell average alone; the
reconstruction of the object ensemble at $\alpha=2n$ enters only through the coefficients
$B_{\rm rec}$ and $C_{\rm rec}$, not through the presence of the logarithm.

The coarse-grained shell envelope in Eq.~\eqref{eq:Lambda-coarse} has a formal maximum at
$\kappa_*=(2-\alpha)/(3-n-\alpha)$.  The shell average is what removes the deterministic comb; before it, the joint measure is resolved only on
$\Delta\kappa\sim x^{-n}/(2\beta n)$, as in Sec.~\ref{subsec:comb}.
We use the continuum maximum $\kappa_*$ only as a shell-envelope diagnostic.

\subsection{Visibility scale and apparent finite-size drift}

Within the resonance-network criterion $\Lambda\sim1$, any fixed macroscopic shell with nonvanishing coarse-grained weight gives
$x_{\rm pair}^{\,4-2n-\alpha}\,\mathcal L(x_{\rm pair},\kappa)\propto
(t_0/V)(2\pi\beta nh/t_0)^2$, i.e., at fixed $(n,\alpha)$, for $0<\alpha<2$,
\begin{equation}
 \ x_{\rm pair}^{\,4-2n-\alpha}\,\ln x_{\rm pair}\propto \dfrac{h^2}{Vt_0}\ ,
 \qquad 0<\alpha<2 .
\label{eq:xres-ratio}
\end{equation}
Both the power \emph{and} the logarithm are common to the two sides of the reconstruction; what
the reconstruction changes is the coefficient, through $b_{\rm all}=w_{11}b_1$.  At the boundary
$\alpha=2$ the shell sum is marginal as well and the two logarithms compound,
$x_{\rm pair}^{2(1-n)}(\ln x_{\rm pair})^2\propto h^2/(Vt_0)$, Sec.~\ref{sec:alpha2}.
If the system size is the limiting visibility scale, $x_{\rm pair}\sim L$, then
\begin{equation}
 h_{\rm pair}(L)\propto \sqrt{Vt_0}\;L^{(4-2n-\alpha)/2}\,\sqrt{\ln L},
 \qquad 0<\alpha<2,
\label{eq:hcapp}
\end{equation}
with $(4-2n-\alpha)/2=(3-\alpha)/2$ at $n=1/2$.  The logarithm is the
coarse-grained matching logarithm of Sec.~\ref{sec:broadening}, which Appendix~\ref{app:pmatch}
shows is present already at $V=0$ and is quantitatively the independent-bond law of
Appendix~\ref{app:matching}; it is not a per-pair effect, the fixed-pair law being strictly
linear.  This scale is \emph{not} the forward-branching scale of Sec.~\ref{sec:crossover}: it refers to
a regime containing many resonant bonds, and is parametrically larger.

\subsection{Typicality of the pair count}
\label{sec:typicality}

A mean pair density of order one does not by itself guarantee a network: if the mean were carried
by rare comb coincidences, the pair-network criterion $\Lambda\sim1$ would describe exceptional samples
rather than typical chains.  We therefore measure the full distribution of the shell pair count at
the level of the physical objects: active islands (frozen clusters excluded, each island one node,
with its exact diagonalized doublet), pairs counted between a central window and a mesoscopic
shell, with the coupling scale tuned so that the mean count sweeps through the order-one value
(protocol in the Supplemental Material~\cite{SM}).  Table~\ref{tab:typ} shows the result at
$\alpha=1$, $V=t_0$: the median tracks the mean, the probability of at least one matched pair is
$\simeq85\%$ of the Poisson value $1-\e^{-\langle N\rangle}$, and the variance exceeds the Poisson
benchmark by less than a factor of two.  The excess variance reflects multiplicity clustering ---
when the comb produces one match it tends to produce several --- not rarity of having any match.
At these representative parameters the order-one pair density is therefore a property of the typical
sample; we find no evidence that the criterion is carried by rare comb coincidences.  The test at
$\alpha=1$ sits at the marginal point $\alpha=2n$ of Eq.~\eqref{eq:regimes}; repeating the
identical protocol in the clean isolated-bond regime ($\alpha=0.5$) and in the clean fragmented
regime ($\alpha=1.5$), at matched mean $\langle N\rangle\simeq1.4$, gives closely comparable
distributions --- median $1$, $P(N\ge1)=0.63$--$0.69$,
${\rm Var}/\langle N\rangle^2=1.13$--$1.24$ (Supplemental Material~\cite{SM}) --- and the result
is essentially unchanged when the Gaussian amplitude of the matching window is replaced by the
physical island form-factor coupling $\Gamma_{AB}$~\cite{SM}, and when the $\alpha=0.5$ scan is
repeated in the one--two-bond regime ($w_0(x_0)\simeq1.9$, so the sampled objects are
one--two-bond; median $1$, $P(N\ge1)=0.61$, ${\rm Var}/\langle N\rangle^2=1.44$ at matched
mean~\cite{SM}): the same typical-not-rare behavior
holds on both sides of the reconstruction, not only at the marginal point.  This behavior also distinguishes the present mechanism
from rare-eigenstate long-distance resonances in quasiperiodic chains~\cite{Padhan2026}.

\begin{table}[t]
\caption{Distribution of the shell pair count $N_{\rm pair}$ at the island level ($\alpha=1$,
$V=t_0$, $600$ realizations), as the coupling scale is tuned through the pair-resonance scale.  Poisson columns give
$1-\e^{-\langle N\rangle}$ and $1/\langle N\rangle$ for comparison.}
\label{tab:typ}
\begin{ruledtabular}
\begin{tabular}{cccccc}
$\langle N_{\rm pair}\rangle$ & median & $P(N{\ge}1)$ & Poisson & ${\rm Var}/\langle N\rangle^2$ & Poisson\\
\colrule
$1.41$ & $1$ & $0.64$ & $0.76$ & $1.24$ & $0.71$\\
$2.52$ & $2$ & $0.80$ & $0.92$ & $0.83$ & $0.40$\\
$3.56$ & $3$ & $0.87$ & $0.97$ & $0.66$ & $0.28$\\
\end{tabular}
\end{ruledtabular}
\end{table}

\subsection{The two scales are parametrically separated}
\label{sec:separation}

The resonance-supply scale and the pair-network scale can now be compared directly, with no numerics and no appeal to
the accessible system sizes.  Setting $N_0(x)\sim1$ in Eq.~\eqref{eq:Nres} defines the
length at which local resonances first appear, $x_{\rm bond}\sim(h/t_0)^{1/(2-n)}$,
while Eq.~\eqref{eq:xres-ratio}, with the logarithm dropped since it cannot affect a comparison of
powers, gives $x_{\rm pair}\sim(h^2/Vt_0)^{1/(4-2n-\alpha)}$.  The difference of the $h$
exponents is
\begin{equation}
\begin{aligned}
 &\frac{2}{4-2n-\alpha}-\frac{1}{2-n}
 =\frac{\alpha}{(2-n)(4-2n-\alpha)}>0,\\
 &\Longrightarrow\
 \frac{x_{\rm pair}}{x_{\rm bond}}\sim
 \Big(\frac{h}{t_0}\Big)^{\!\alpha/[(2-n)(4-2n-\alpha)]}
 \Big(\frac{V}{t_0}\Big)^{\!-1/(4-2n-\alpha)}
\end{aligned}
\label{eq:separation}
\end{equation}
diverging in the strong-potential limit at fixed $V/t_0$.  The formal power comparison remains positive for every
$0<\alpha<4-2n$; the resonance-starved construction itself is controlled for $\alpha<2$ --- in the
isolated-bond regime directly, and in the fragmented regime under the operative-fragmentation
condition Eq.~\eqref{eq:fragcond} --- with the marginal case $\alpha=2$ treated separately in
Sec.~\ref{sec:alpha2}, so the controlled statement covers $0<\alpha\le2$ under those
conditions.  At $n=1/2$ the exponent is
$2\alpha/[3(3-\alpha)]$, equal to $0.13,\,0.33,\,1.33$ for $\alpha=0.5,\,1,\,2$: the separation is
weakest for the longest-ranged interactions, but at fixed interaction strength it never closes;
Sec.~\ref{sec:Vdep} gives the condition on $V$ under which the hierarchy holds.  The two exponents coincide only
at $\alpha=0$, i.e.\ a distance-independent (infinite-range) coupling.  There the uniform case is
exactly $H_{\rm int}=(V/2)(N^2-N)$, a constant at fixed particle number, so no pair network exists at all;
for random $V_{ij}$ no such cancellation occurs and the coupling remains $\order(V)$ at all
separations.

The two criteria define parametrically distinct length scales, separating without bound as the
potential is made stronger.  A system whose length lies between them contains local resonances but no long-range pair
network, and that is the regime any tractable simulation occupies.

\subsection{Dependence on interaction strength}
\label{sec:Vdep}

Their $V$ dependences differ.

\emph{The supply of resonant bonds becomes $V$-independent.}  Expanding Eq.~\eqref{eq:p-int-robust} and summing
along the chain with $A_i\simeq2\pi\beta nh\,i^{n-1}$ gives
\begin{equation}
 N_{\rm res}(L,h,V)=N_0
 \Big[1+\kappa_n\frac{\bar\sigma_w^2L^{2-2n}}{h^2}+\cdots\Big],
\label{eq:NresV}
\end{equation}
where $\bar\sigma_w^2=\sum_i\sigma_{{\rm int},i}^2\,i^{3(1-n)}/\sum_i i^{3(1-n)}$ is the variance
weighted by the same power that controls the sum,
with $\kappa_n=(2-n)/[2(4-3n)(2\pi\beta n)^2]$, $\kappa_{1/2}=0.0796$.
The single coefficient $\kappa_n$ follows only because the weighting has been separated out in
the weighted variance defined after Eq.~\eqref{eq:NresV}: $\sigma_{{\rm int},i}^2$ is position dependent on an open chain, markedly so for
$\alpha\le1/2$, and Eq.~\eqref{eq:NresV} should be read as exact in the weighted variance and as a
scaling statement otherwise.
Along the $\order(1)$-resonant-bond trajectory $h\propto L^{2-n}$ the relative correction is
\begin{equation}
 \frac{\delta N_{\rm res}}{N_{\rm res}}\sim
 \begin{cases}
  V^2/L^{2}, & \alpha>\tfrac12,\\[2pt]
  V^2\ln L/L^{2}, & \alpha=\tfrac12,\\[2pt]
  V^2L^{-1-2\alpha}, & \alpha<\tfrac12,
 \end{cases}
\label{eq:NresV-scaling}
\end{equation}
using Eq.~\eqref{eq:sigmaint}.  Evaluated at $\alpha=1$, $V=t_0$ and
$h=h_{\rm FB}$ with $N_c=2.3$, the correction is $2.2\%$ at $L=16$ and $0.15\%$ at $L=64$; it is
$\alpha$ dependent through Eq.~\eqref{eq:sigmaint} and these numbers should not be read as
universal.  The
supply of resonant bonds, and with it the size exponent, is therefore asymptotically blind to interaction
strength as well as to interaction range: $h_{\rm FB}(L,V,\alpha)\propto L^{2-n}$ to leading
power.
The crossover condition is
$N_{\rm res}=N_c(V,\alpha)$, and the table in Sec.~\ref{sec:crossover} shows that $N_c$ retains an
$\order(1)$ dependence on $V/t_0$, falling from $2.57$ at $V=0$ to $2.17$ at $V=2t_0$.  The
\emph{amplitude} of $h_{\rm FB}$ therefore remains $V$-dependent even though its exponent does not.
The first scale carries the $L^{2-n}$ resonant-bond exponent with an interaction-dependent
threshold amplitude; the long-range pair-network scale is explicitly interaction controlled.

\emph{The pair-network scale depends on $V$ as a power.}  By contrast Eq.~\eqref{eq:xres-ratio} gives
$x_{\rm pair}\sim(h^2/Vt_0)^{1/(4-2n-\alpha)}$, so that
\begin{equation}
\begin{aligned}
 \frac{x_{\rm pair}}{x_{\rm bond}}&\sim
 \Big(\frac{h}{t_0}\Big)^{\frac{\alpha}{(2-n)(4-2n-\alpha)}}
 \Big(\frac{V}{t_0}\Big)^{-\frac{1}{4-2n-\alpha}},\\
 x_{\rm bond}\ll x_{\rm pair}
 &\;\Longrightarrow\;
 \frac{V}{t_0}\ll\Big(\frac{h}{t_0}\Big)^{\alpha/(2-n)},
\end{aligned}
\label{eq:Vstar}
\end{equation}
or $V/t_0\ll(h/t_0)^{2\alpha/3}$ at $n=1/2$, up to logarithms and $\order(1)$ constants.  Equation~\eqref{eq:Vstar} is a parametric condition for scale separation rather than a phase
boundary: where
the two scales meet, the pair-network construction is being extrapolated to a region containing only
$\order(1)$ resonant bonds, exactly where its many-bond coarse graining ceases to be controlled.  The resonant bond
expansion itself independently requires $\sigma_{\rm int}\ll A$, i.e.\ $V\ll h/\sqrt L$ at $n=1/2$
for $\alpha>1/2$.  The interpolated thresholds satisfy these conditions moderately well and improve rapidly with $L$;
the lowest-field scan points do not, and are used only to trace $B(N_{\rm res})$.

The two scales thus respond to $V$ differently: the finite-size exponent is insensitive to $V$
while its threshold amplitude carries an $\order(1)$ $V$ dependence, and the pair-network scale carries
an explicit power-law dependence on $V$.  A scan in $V$ at fixed $h$
therefore separates the two mechanisms more cleanly than a scan in $\alpha$, because
$N_{\rm res}$ is $V$-blind at leading order while $x_{\rm pair}\propto V^{-1/(4-2n-\alpha)}$.

\subsection{Direct fixed-shell test}
\label{sec:lamtest}

The assembled law Eq.~\eqref{eq:xres-ratio} is tested directly, with the exponent fixed in
advance: on a fixed macroscopic shell $R=0.45\,x$, the measured pair count
$\Lambda=n_RR$, rescaled as $Y=\Lambda h^2x^{-(4-2n-\alpha)}/(Vt_0)$ at fixed dilute matching
window, must be $h$-independent and grow linearly in $\ln x$ at fixed $(\alpha,V)$.  Over
$x=4000$--$24\,000$ and $h/t_0=100$--$340$ at $V=t_0$: within the resonance-starved window
$w_0/\ell_{\rm comb}\le0.15$ the $h^{-2}$ collapse holds to $7$--$14\%$ at $\alpha=1.5$ and $1$
(and to $33\%$ at the small-$w_0$ end of the $\alpha=0.5$ control), the fitted logarithmic slope
is positive at every $\alpha$, and free-exponent fits give $1.65/2.15/2.74$ against the
prescribed $1.5/2/2.5$ --- consistent in sign and scale with the logarithmic correction plus
finite-window effects, though the accessible window does not separate $B\ln x+C$ from a more
general slowly varying correction.  The fragmented side collapses at least as well as the
bond-side control, with no $w_0$-correlated deterioration --- the failure mode that a
tail-induced breakdown of the island construction would produce.  Outside the starved window
the collapse fails in both directions ($+60\%$ at $\alpha=1.5$, $-3\times$ at $\alpha=0.5$),
and the failure sits entirely in the active-object density: the matching fraction at fixed
window is $h$-independent to $10$--$15\%$ throughout, while
$\rho_{\rm act}\,h/(t_0\sqrt{x})$ first rises toward a plateau with increasing $w_0$
(small-$w_0$ object convergence) and then falls once $w_0/\ell_{\rm comb}\gtrsim0.2$ (loss of
starvation).  The observed breakdown is thus a loss of the object-density asymptotics at the
domain boundaries, not of the long-range matching mechanism; protocol, figures, and the
$h=100$ convergence control are in the Supplemental Material~\cite{SM}.

\section{The marginal case $\alpha=2$}

\label{sec:alpha2}

At $\alpha=2$ the shell sum becomes marginal, so a second logarithm joins the matching logarithm
of Eq.~\eqref{eq:Pmix}.  Since $\alpha=2>2n$ for every $0<n<1$, the objects at $\alpha=2$ are the
reconstructed mixture of Sec.~\ref{sec:islands} whenever Eq.~\eqref{eq:fragcond} holds: islands
carrying the linear law, plus wing bonds carrying the logarithmic channel at weight $f_1^2$.  The
marginal result is therefore a \emph{double} logarithm, whose coefficient --- not whose form ---
is set by the reconstruction.

\paragraph*{The marginal shell sum.}  The shell integral
$\int_0^1(\dd\kappa/\kappa)G_{n,\alpha}=\Gamma(2-\alpha)\Gamma(2-n)/\Gamma(4-\alpha-n)$
(Supplemental Material~\cite{SM}) is finite for $\alpha<2$ and logarithmically divergent as
$\alpha\to2^-$.  At $\alpha=2$ the
integrand reduces to $G_{n,2}(\kappa)=(1-\kappa)^{1-n}$ and the divergence is cut not at a lattice
constant but at the coarse-graining boundary set by the comb, Eq.~\eqref{eq:comb}: the smallest
resolvable fractional separation is $\kappa_0\sim\ell_{\rm comb}(x)/x\sim x^{-n}$.  Hence, directly,
\begin{equation}
 \int_{\kappa_0}^{1}\frac{\dd\kappa}{\kappa}(1-\kappa)^{1-n}
 =\ln\frac{1}{\kappa_0}+\order(1)=n\ln x+\order(1).
\label{eq:alpha2-log}
\end{equation}
Note that the cutoff $R_0\sim\ell_{\rm comb}(x)\sim x^{1-n}$ itself diverges with $x$; the logarithm
is generated at the small-$\kappa$ boundary, not at a fixed microscopic scale.

\paragraph*{The physical marginal result.}  With $\mathcal L$ given by Eq.~\eqref{eq:Pmix}, the
matching bracket contributes $b_{\rm all}[2\ln x+2\ln\kappa+\text{const}]+C_{\rm all}$ inside the
shell sum.  The island (linear) part multiplies Eq.~\eqref{eq:alpha2-log} by a constant, giving
$C_{\rm all}\,n\,x^{2(1-n)}\ln x$; the surviving bond channel supplies a second logarithm through
the closed form of Eq.~\eqref{eq:alpha2-coeff}.  Hence
\begin{equation}
 \ \Lambda_{\alpha=2}(x)\propto \frac{Vt_0}{h^2}\,x^{2(1-n)}
 \big[b_{\rm all}\,n(2-n)(\ln x)^2+C_{\rm all}\,n\ln x\big]\ ,
\label{eq:alpha2-lambda}
\end{equation}
so that asymptotically
$x_{\rm pair}^{2(1-n)}(\ln x_{\rm pair})^2\propto h^2/(Vt_0)$, i.e.\
$x_{\rm pair}(\ln x_{\rm pair})^2\propto h^2/(Vt_0)$ at $n=1/2$.  The two marginalities ---
shell and matching --- meet and compound at $\alpha=2$; the reconstruction of the object ensemble
suppresses the $(\ln x)^2$ coefficient by the factor $b_{\rm all}$ but does not remove the term.
The construction is self-consistent: along the resulting boundary
$h^2\propto x^{2(1-n)}(\ln x)^2$, the starvation ratio obeys
$(x/i_*)^{2(1-n)}\propto1/(\ln x)^2\to0$, so the resonance-starved approximation becomes better
at equality, and faster than in the single-logarithm case.  A fixed macroscopic shell, by
contrast, drops the geometric logarithm and its boundary sits at
$(x/i_*)^{2(1-n)}=\order(1/\ln x)$; at $\alpha=2$ it is the shell-integrated construction that is
controlled.

\paragraph*{The double-logarithm coefficient.}  The bond channel's contribution follows from the
closed form
\begin{align}
 \int_{\kappa_0}^{1}\frac{\dd\kappa}{\kappa}(1-\kappa)^{1-n}
 \big[2\ln x+2\ln\kappa+\text{const}\big] \nonumber\\
 =\;\underbrace{2n(\ln x)^2}_{\text{shell}\,\times\,\ln x}
 \;\underbrace{-\,n^2(\ln x)^2}_{\langle\ln\kappa\rangle}
 \;+\;\order(\ln x)
 \;\propto\;n(2-n)\,(\ln x)^2,
\label{eq:alpha2-coeff}
\end{align}
using $\int_{\kappa_0}(\dd\kappa/\kappa)\ln\kappa=-\tfrac12(\ln\kappa_0)^2$ with
$\ln(1/\kappa_0)=n\ln x+\order(1)$: the $\ln\kappa$ inside the matching law removes part of the
naive product, leaving the coefficient $n(2-n)$ quoted in Eq.~\eqref{eq:alpha2-lambda}.  Had the
ensemble been purely islands ($f_1\to0$, $b_{\rm all}\to0$) only the single geometric logarithm
would remain; the measured $f_1=\order(1)$ of Eq.~\eqref{eq:f1} is what keeps the
$(\ln x)^2$ term alive, at strongly reduced amplitude.

\paragraph*{Fixed-shell versus shell-integrated limits.}  Maximizing $G_{n,\alpha}$ gives
$\kappa_*=(2-\alpha)/(3-n-\alpha)$, so $\kappa_*(\alpha=2)=0$: at equality there is no nonzero
preferred macroscopic shell fraction, which is why choosing a shell and summing over shells give
different answers, and why the shell-integrated result remains sensitive to the small-$\kappa$
boundary rather than to the long-distance saddle that organizes $\alpha<2$.

\section{Three distinct long-range thresholds and finite-size implications}

\label{sec:liom}

The present calculation does not construct a globally convergent LIOM expansion to all orders.  Its role is instead to identify where a finite-size LIOM description should become vulnerable to an algebraic resonant network.

At $t_0=0$ the physical occupations $n_i$ are conserved exactly, long-range interactions included.
Turning on the hopping, and staying away from resonances, a Schrieffer--Wolff or Liouvillian
construction dresses each local occupation by higher-body sectors, and the random long-range terms
it generates --- schematically $d_p^\dagger d_c(n_b-\nu)$, with $p,c$ near the LIOM core and $b$
far away --- inherit algebraic $R^{-\alpha}$ numerators, so a core-restricted squared operator
weight scales as $R^{-2\alpha}$ and becomes non-square-summable below $\alpha=1/2$.  The same sum $\sum_R R^{-2\alpha}$ controls the random-interaction Fock-energy variance derived in
the Supplemental Material~\cite{SM}.  This higher-body tail does not, by itself, establish delocalization; resonant denominators and pair proliferation are the decisive additional ingredients.

Three distinct thresholds in $\alpha$ should therefore not be conflated, and they mean three
different things: $\alpha=1/2$ is a square-summability threshold --- of the induced operator
weights above, and of the random-interaction Fock-energy variance~\cite{SM}; $\alpha=2n$ marks
the \emph{reconstruction of the local resonant-object ensemble}, from asymptotically isolated
bonds to a mixture of fragmented islands and surviving wing bonds, Eq.~\eqref{eq:regimes}, the
equality itself lying marginally on the bond side; and
$\alpha=2$ is the marginality threshold of the long-range \emph{shell sum},
Sec.~\ref{sec:alpha2}, where the shell integral acquires a second logarithm.  None
of the three is, by itself, a localization transition, and keeping them separate is what prevents
the generic $\alpha<2d$ resonance counting from being applied blindly to this deterministic
model.  The pair-network boundary $x_{\rm pair}$, Eq.~\eqref{eq:xres-ratio}, is not a threshold
in $\alpha$ at all but a scale, finite for every $\alpha$ in the controlled range.  The standard symmetry classification of MBL phases is formulated for LIOM couplings
decaying exponentially with operator support~\cite{Serbyn2013,HuseNO2014,Ros2015,Imbrie2016}; the algebraic dressing already
places the model outside a strictly exponentially quasi-local LIOM ansatz for every $\alpha$, with
$\alpha=1/2$ the point where the induced weights cease to be square summable.

The exact continuity equation, Eq.~\eqref{eq:continuity}, also motivates a separation of dynamical sectors.  Long-range interactions can delocalize the resonant pseudospin/energy sector without introducing a bare long-range charge current.  Consequently, strong finite-time charge memory can coexist with much faster loss of pseudospin memory.  Whether charge memory remains nonzero asymptotically is a separate question and is not decided by the present static theory.

\subsection{Finite-size spectral phenomenology}
\label{sec:reinterp}

The resonance-supply scale is the scale on which any tractable simulation actually sits.  For
$n=1/2$ the resonance-starvation length is $i_*=(\pi\beta h/t_0)^2$, so $i_*\simeq377$ at
$h/t_0=10$ and $i_*\simeq1825$ at $h/t_0=22$; chains reachable by exact diagonalization are one to
two orders of magnitude shorter and contain of order \emph{one} resonant bond, far too few for the
real-space pair network of Sec.~\ref{sec:cascade}.  Since $N_{\rm res}\propto L^{2-n}/h$, any
crossover defined by a fixed resonance count obeys
\begin{equation}
 \ h_{\rm FB}(L;V,\alpha)=\frac{C\,t_0}{N_c(V,\alpha)}\,L^{2-n}\ ,\qquad
 C=\frac{1}{\pi^2\beta n(2-n)},
\label{eq:hx-res}
\end{equation}
up to the subleading correction of Sec.~\ref{sec:Vdep}: an $\alpha$- and $V$-blind exponent with an
interaction-dependent amplitude, because the potential supplies the resonances and the interaction
sets only how many are needed.  A finite-size drift of this kind was reported in
Ref.~\cite{yp_nee}, and Eq.~\eqref{eq:hx-res} is \emph{compatible} with it.  We state the
limitation sharply: the exact-graph calculation of Sec.~\ref{sec:fockgraph} shows that $B=1$ is not
a percolation threshold of the resonant Fock graph, so the present work does not establish that the
observed spectral crossover is a fixed-resonance-count phenomenon; only the exponent statement ---
whatever physics is tied to a fixed number of local resonant objects must drift as $L^{2-n}$ ---
is robust.

Two numerical tests follow.  \emph{(i)} Random and uniform long-range interactions should differ
qualitatively at fixed $V$, with the reversal of range dependence described in
Sec.~\ref{sec:uniform}, at identical potential, filling and protocol.  \emph{(ii)} The visibility
law Eq.~\eqref{eq:xres-ratio} refers to a scale on which many resonant objects coexist and should
not be sought in a level-statistics crossing at all; its natural observables are spatially
resolved, the expected signature being faster loss of pseudospin memory than of charge memory, as
permitted by the exact continuity equation~\eqref{eq:continuity}.

\section{Conclusion}

Reference~\cite{letter} states the physical outcome of this theory: the elementary
resonant-object ensemble reconstructs across $\alpha=2n$, the multi-site objects it produces lose
the two-level matching singularity while surviving wing bonds keep a reduced ensemble logarithm,
the scaling of the pair-resonance scale is common to both sides, and only independently
fluctuating long-range couplings drive the network at $VR^{-\alpha}$.  What this paper establishes is
the microscopic structure underlying those statements, and its domain of validity.

On the statistics side: the exact phase-averaged supply $N_0\propto L^{2-n}/h$ with its discrete
corrections and nonuniform $n\to1^-$ limit (Fig.~\ref{fig:supply}); the closed-form Hartree
variance in every $(V,\alpha,L)$ regime, with the proof that the supply law survives the
interaction; and the common-phase comb structure that makes two-bond statistics irreducibly
correlated.  On the local-object side: the run width $w_0$ and its trajectory law, the marginal
logarithm at $\alpha=2n$, the exact support bound $S(\alpha)V$ and threshold $V_*(\alpha)$, the
large-deviation distinction between $V>V_*$ and operative fragmentation
$w_0\gg\xi_{\rm frag}$, Eq.~\eqref{eq:fragcond} --- and the diagonalization evidence
that fragmented clusters match linearly while the wings keep $f_1=\order(1)$ isolated bonds, so
that the ensemble law is $b_{\rm all}q\ln(1/q)+C_{\rm all}q$ with $b_{\rm all}=w_{11}b_1\simeq f_1^2b_1$ quantitatively verified (Fig.~\ref{fig:lresolved}).  On the
network side: the hierarchy separating the strictly linear fixed-pair law, the shell-averaged
$q\ln(1/q)$ bond law with exact reference constants, and the linear island law that combines with
it in the reconstructed mixture; the mesoscopic
shell theory built on the comb; the marginal $\alpha=2$ construction, in which physical islands
compound the shell marginality with the surviving wing-bond matching logarithm; and the pair-resonance and visibility scales with all prefactors and
domain conditions, typical rather than rare-event dominated in representative tests in the
bond, marginal, and fragmented regimes.  On the Fock-space side, an exact finite-size negative: forward branching at
$B=1$ does not produce a giant component, so $h_{\rm FB}\propto L^{2-n}$, with its
$\alpha$-independent exponent, is a one-step branching scale, not a percolation threshold.  And
three long-range thresholds --- $\alpha=1/2$, $\alpha=2n$, $\alpha=2$ --- emerged with strictly
different meanings, none of them by itself a localization transition.

Two steps remain heuristic: the connection of any of these scales to the spectral crossover, and
the promotion of the order-one pair density to an all-generation instability.  Both are stated as
conjectures, and both are testable --- most directly through the random-versus-uniform contrast at
fixed $V$ and through spatially resolved pseudospin-versus-charge memory.

\begin{acknowledgments}
This work is supported by the National Research Foundation of Korea
(NRF), funded by the Ministry of Science and ICT (MSIT), Korea, under
Grant Nos.~NRF-2022R1A2C1011646, RS-2024-00416036, RS-2025-03392969,
and RS-2026-25490114; by the Creation of the Quantum Information
Science R\&D Ecosystem program through the NRF funded by the Korean
government (MSIT) (Grant No.~RS-2023-NR068116); by the Quantum
Simulator Development Project for Materials Innovation through the NRF
funded by MSIT, Korea (Grant No.~RS-2023-NR119931); by the Brain Pool
Program funded by MSIT through the NRF (Grant No.~RS-2025-25446099);
and by the Institute of Information \& Communications Technology
Planning \& Evaluation (IITP) through the Information Technology
Research Center (ITRC) program, funded by MSIT, under the project
Quantum--AI Convergence for Innovation and Talent Development
(Grant No.~RS-2026-25519864).
\end{acknowledgments}

\emph{Data availability.}---The numerical data supporting this article, together with the
analysis and simulation code used to generate the reported protocols and figures, are available
from the contact author upon reasonable request.

\appendix

\section{Numerical resonance counting: protocol and errors}
\label{app:numerics}

All quantities in Sec.~\ref{sec:crossover} come from direct Monte Carlo evaluation of the resonance
functional; no diagonalization is involved and no fit is made to spectral data.

\paragraph*{Ensemble.}  One sample consists of three independent draws.  (i) A single global phase
$\phi$ uniform on $[-\pi,\pi)$, which fixes the entire potential $h_i=h\cos(2\pi\beta i^n+\phi)$,
$i=1,\dots,L$, with $\beta=(\sqrt5-1)/2$ and $n=1/2$.  (ii) A symmetric matrix of interaction
coefficients $V_{ij}=V_{ji}$, $i<j$, drawn independently and uniformly on $[-V,V]$, giving
$U_{ij}=V_{ij}/|i-j|^\alpha$; for the uniform control $V_{ij}\equiv V$.  (iii) One Fock configuration
$\{n_i\}$ drawn uniformly from the $\binom{L}{L/2}$ half-filled states, i.e.\ a uniform random subset
of $L/2$ occupied sites.  The three draws are independent, and successive samples are independent, so
the points of Fig.~\ref{fig:collapse} are statistically independent.  Open boundaries throughout;
$t_0=1$ and $V=t_0$ unless stated.

\paragraph*{Detuning and resonant bonds.}  For each bond $a=1,\dots,L-1$ the quantity evaluated is the exact
diagonal transition energy.  For a flippable bond, writing $s=n_a-n_{a+1}=\pm1$,
\begin{align}
 \Delta_a\equiv s\big[E(S_a)-E(S)\big]
 =\;&(h_{a+1}-h_a)\nonumber\\
 &+\!\!\sum_{j\neq a,a+1}\!\!(U_{a+1,j}-U_{a,j})n_j,
\label{eq:Delta-numeric}
\end{align}
identical to Eq.~\eqref{eq:full-detuning}.  The right-hand side is defined for \emph{every} bond,
flippable or not, and it is that expression --- not the transition energy, which vanishes
identically when $n_a=n_{a+1}$ --- that is evaluated when counting $N_{\rm res}$ below.  The bond's own interaction $U_{a,a+1}n_an_{a+1}$ makes
\emph{no} contribution: the product $n_an_{a+1}$ is invariant under the exchange
$n_a\leftrightarrow n_{a+1}$ (and vanishes identically on a flippable bond), so the shared-site term
cancels exactly and must be absent from $\Delta_a$.

A natural implementation does not produce this directly.
Evaluating the Hartree field as a matrix product $(U\bm n)_i=\sum_jU_{ij}n_j$ gives
\begin{align}
 (U\bm n)_{a+1}-(U\bm n)_a
 =&\!\!\sum_{j\neq a,a+1}\!\!(U_{a+1,j}-U_{a,j})n_j\nonumber\\
 &+U_{a,a+1}(n_a-n_{a+1}),
\label{eq:matvec}
\end{align}
since the unrestricted sum retains $j=a$ and $j=a+1$.  The implementation therefore adds the
counter-term $U_{a,a+1}(n_{a+1}-n_a)$, which removes that contamination and returns
Eq.~\eqref{eq:Delta-numeric} exactly.  Both forms were checked against direct evaluation of
$E(S_a)-E(S)$ from the full diagonal Hamiltonian and agree to machine precision.  A bond is \emph{flippable}
if $n_a\neq n_{a+1}$, and is an \emph{active edge} $e_a$ if it is flippable and $|\Delta_a|<t_0$.
The quantity plotted as $N_{\rm res}$ in Fig.~\ref{fig:collapse} is the measured
interacting count $\langle\#\{a:|\Delta_a|<t_0\}\rangle$ --- the detuning criterion \emph{without} the
flippability mask --- so that the flippability factor $q_L$ appears explicitly in
Eq.~\eqref{eq:Nc0} rather than being absorbed silently.  It is not the $V=0$ asymptotic form
Eq.~\eqref{eq:Nres}; the discrete sum exceeds the asymptotic form by $10\%$ at the smallest $h$ used at $L=12$, falling
to $1.4\%$ at the largest and to below $2\%$ throughout at $L=32$, which contributes to the finite-$L$ deviation from the asymptotic $L^{3/2}$ behavior quoted in the
main text.

\paragraph*{Branching.}  For every active edge $a$ in a sample we form the configuration $S_a$
obtained by exchanging $n_a\leftrightarrow n_{a+1}$, recompute all $\Delta_b(S_a)$ exactly from
Eq.~\eqref{eq:Delta-numeric}, and count the active edges $b\neq a$ of $S_a$.  Equation
\eqref{eq:Bdef} is then the ratio of the accumulated forward count to the accumulated number of
active edges; the backtracking edge $b=a$ is excluded by construction.

\paragraph*{Parameters.}  Sample counts are $N=4\times10^5,\,3\times10^5,\,2.2\times10^5,
\,1.6\times10^5,\,9\times10^4$ for $L=12,16,20,24,32$.  For each $L$ the field is scanned as
$h=f\,h_0(L)$ with $h_0(L)=0.2186\,L^{3/2}/2.3$ and $f\in\{0.55,0.7,0.85,1.0,1.2,1.5\}$, at
$\alpha\in\{0.5,1,2\}$: $5\times3\times6=90$ points.  Each point uses an independent random-number seed.  The lowest-field points
($f=0.55$ at $L=12$--$16$) lie outside the asymptotic resonance-starved regime --- there
$L/i_*\simeq0.67$ and $\sigma_{\rm int}/\min_iA_i\simeq0.9$ --- and are retained only to trace the
numerical $B(N_{\rm res})$ curve; they are \emph{not} used to infer any scaling law.  At the interpolated $B=1$ thresholds themselves the corresponding worst-case values are
$L/i_*\le0.20$ and $\sigma_{\rm int}/\min_iA_i\le0.44$, both attained at $L=12$; for $L\ge16$ they
tighten to $0.12$ and $0.36$.  The thresholds are therefore inside the controlled regime, least
comfortably at the smallest size.

\paragraph*{Errors.}  Uncertainties are obtained by batch means: each run is split into $K=20$
independent batches, $B$ is formed separately in each, and the standard error is the batch standard
deviation divided by $\sqrt K$.  Scaled to the production sample counts this gives a relative error
on $B$ of $0.12\%$ ($L=12$), $0.20\%$ ($L=20$) and $0.44\%$ ($L=32$), and $\simeq0.2\%$ on
$N_{\rm res}$.  These are smaller than the plot markers and more than an order of magnitude below
the systematic $(L,\alpha)$ structure visible in Fig.~\ref{fig:collapse}, which is therefore
genuine and not sampling noise.

\section{The coarse-grained matching law}
\label{app:pmatch}

Section~\ref{sec:matching} establishes that for a \emph{single} bond pair the bare common-phase
matching probability is linear in $J$, Eq.~\eqref{eq:coarea}.  The pair-scale construction of
Sec.~\ref{sec:cascade}, however, uses the matching law only after coarse graining over the pair
ensemble --- in the isolated-bond regime, where the logarithmic law applies --- and that object
behaves differently.  We measure it directly.

Writing the matching condition of Eq.~\eqref{eq:exact-pairmatch} in dimensionless form,
$X_{AB}=|\Omega_A-\Omega_B|/(s_As_Bt_0)$, the matching probability at coupling $q$ is
$P^{\rm match}(q)={\rm Prob}(X_{AB}<q)$ over the ensemble of resonant-bond pairs.  Sampling an $L=100$, $h/t_0=25$ chain at $\alpha=1$ over $\simeq2\times10^6$ resonant-bond pairs
\emph{pooled over all separations} $R\ge2$ gives Fig.~\ref{fig:pmatch}.

First, at fixed $(A,B)$ and $V=0$ the ratio $P/q$ is flat over four decades in $q$ --- four representative pairs have plateaus $P/q=3.8,\,4.7,\,4.6,\,6.7$, each flat to within a few
percent across the sampled window; the pooled value $\simeq4.8$ is their participation-weighted
average.  This confirms Eq.~\eqref{eq:coarea}: a single pair traces a
one-dimensional curve in the $(\Delta_A,\Delta_B)$ plane as $\phi$ varies, the crossing is
transverse, and no logarithm is generated.

Second, the \emph{ensemble} of pairs does generate one, and it does so \emph{whether or not the
interaction is present}.  Fitting $P/q=a+b\ln(1/q)$ over the well-sampled window (all fits here are unweighted least
squares over the plotted points) gives $b=2.00\pm0.05$, $a=1.64$ at $V=0$
and $b=1.95\pm0.04$, $a=1.87$ at $V=t_0$,
against the independent-bond benchmark of Appendix~\ref{app:matching}, $b=2$ and
$a=2c_{\rm mix}=1.7706\ldots$.  The slopes agree within $2.5\%$ and the constants within about $8\%$; the leading slopes of the
bare and interacting ensembles are statistically consistent with each other.

Pooling over all separations is not by itself a test of the mesoscopic-shell hypothesis
Eq.~\eqref{eq:mesoscopic}.  We therefore repeat the measurement shell by shell.  The condition the
argument actually needs is slightly stronger than $\ell_{\rm comb}\ll\Delta R\ll x$: because the pair
density involves envelopes at both $x$ and $x-R$, the shell must satisfy
\begin{equation}
 \ell_{\rm comb}(x)\ll\Delta R\ll\min\{R,\,x-R,\,x\}.
\label{eq:shellcond}
\end{equation}
On a chain long enough to open such a window ($L=4000$, $h/t_0=150$, so $i_*\simeq8.5\times10^4\gg L$),
with the anchor bond confined to $x_0=2000\pm150$ and shells of width
$\Delta R=4\,\ell_{\rm comb}(x_0)\simeq290$, fitting $P/q=a+b\ln(1/q)$ within each shell separately
at $V=0$ gives
\begin{center}
\footnotesize
\setlength{\tabcolsep}{3.5pt}
\begin{tabular}{l|cccc}
\hline\hline
 & \multicolumn{3}{c}{$L=4000$} & $L=8000$\\
$\bar R$ & $400$ & $800$ & $1600$ & $1800$\\
$\Delta R/\min(R,x{-}R)$ & $0.72$ & $0.36$ & $0.72$ & $0.23$\\
pairs $/10^6$ & $2.3$ & $2.4$ & $2.8$ & $2.6$\\
$b$ & $1.930(13)$ & $1.983(14)$ & $2.016(7)$ & $1.944(24)$\\
$a$ & $2.00$ & $1.76$ & $1.61$ & $1.92$\\
\hline\hline
\end{tabular}
\end{center}
against the predicted $b=2$ and $a=2c_{\rm mix}=1.771$; the last column uses $L=8000$,
$x_0=4000\pm60$ and $\bar R=1800$, for which both $R$ and $x-R$ are large compared with $\Delta R$.

Four finite mesoscopic shells, spanning different fractional separations and different degrees of
conditioning, independently reproduce the predicted leading logarithmic coefficient to within $3.5\%$
of $2$.  The finite-shell variation shows up in the $\order(1)$ constant, which scatters by about
$10\%$ about $2c_{\rm mix}$ and is not monotone in
$\Delta R/\min\{R,x-R\}$ --- so the residual corrections are not governed by that ratio alone.  The
two quantities should be read differently: $b\to2$ is the asymptotic prediction that drives the
pair-scale construction, and it is what Eq.~\eqref{eq:joint-shell} requires, whereas $a\to2c_{\rm mix}$ is a
stronger test of the finite constant and carries larger finite-$A$, finite-shell and discrepancy
corrections.  The analytic result is in any case conditional on equidistribution rather than derived
from this measurement.

The mechanism is the quadratic band edge.  Near $\Omega=2t_0$ one has
$\Omega-2t_0\simeq\Delta^2/4t_0$, so the matching condition is the hyperbolic strip
$|\Delta_A^2-\Delta_B^2|\lesssim4t_0J$, whose area in any \emph{smooth two-dimensional} measure is
$\order(J\ln(t_0/J))$.  A single pair does not supply such a measure; the pair ensemble does,
because different $(A,B)$ sample different parts of the comb.  Interaction broadening is therefore
one route to the two-dimensional measure, not the only one, and at the physical $V=t_0$ it changes
the fitted leading logarithmic coefficient by about $2.5\%$.  The logarithm in the bond-regime case of Eq.~\eqref{eq:hcapp} does not rest on being in a ``resolved'' regime.

\section{Independent-bond matching benchmark}

\label{app:matching}

Set $t_0=1$ temporarily and draw two resonant-bond detunings independently from the resonance-starved single-bond distribution.  Parameterize them by $u,v\in[0,1]$, with $\Omega(u)=\sqrt{4+u^2}$ and $s(u)=2/\Omega(u)$.  The independent-draw resonance domain is
\begin{equation}
 |\Omega(u)-\Omega(v)|<q\,s(u)s(v).
\end{equation}
Expanding its exact two-dimensional area for $q\to0$ gives
\begin{equation}
 P_{\rm mix}^{\rm(ind)}(q)=2q\left[\ln(1/q)+c_{\rm mix}\right]+o(q),
\end{equation}
with
\begin{equation}
 c_{\rm mix}=1+\ln\left[\frac{16(5\sqrt5-8)}{28+13\sqrt5}\right].
\end{equation}
If the mixing factors are set to unity first,
\begin{equation}
 c_\Omega=1+(\sqrt5-2)+\ln\left[\frac{16(\sqrt5-2)}{\sqrt5+2}\right]
 =1.1214\ldots,
\end{equation}
so $C_\Omega=e^{c_\Omega}=3.0691\ldots$.  These constants are exact for the independent-bond benchmark.  The actual SV pair problem is the one-dimensional common-phase integral Eq.~\eqref{eq:exact-pairmatch}, so this appendix is the exact fixed-$q$ form of the coarse-grained matching law of the
isolated-bond regime, Eq.~\eqref{eq:shell-to-ind}.

\section{Random coupling moments}

\label{app:Jmoments}

At $R\gg1$, the four distance denominators in Eq.~\eqref{eq:Jzz} are equal at leading order and the signed sum $\varsigma$ has the same distribution as the sum of four independent $U[-1,1]$ variables.  Direct integration of the order-four Irwin--Hall density gives
\begin{equation}
 \mathbb E|\varsigma|=\frac{14}{15},
\end{equation}
while
\begin{equation}
 \mathbb E[|\varsigma|\ln|\varsigma|]=\ln4-\frac{539}{450}.
\end{equation}
These moments fix the random long-distance coupling normalization exactly.  Combined with the
coarse-grained matching law of Sec.~\ref{sec:matching} they give Eq.~\eqref{eq:Pmatch-random}, with
\begin{equation}
\begin{aligned}
 c_\star&=c_{\rm mix}-\frac{\mathbb E[|\varsigma|\ln|\varsigma|]}{\mathbb E|\varsigma|}\\
 &=c_{\rm mix}-\frac{15}{14}\Big(\ln4-\frac{539}{450}\Big)=0.68334\ldots
\end{aligned}
\label{eq:cstar}
\end{equation}
the random amplitude preserving, in the isolated-bond regime, both the $R^{-\alpha}$ power and the accompanying logarithm.  These moments apply to the four-term bond amplitude only; in the
fragmented regime the coupling carries the island form-factor sum, and the corresponding average
is absorbed into the nonuniversal constant of Eq.~\eqref{eq:Pmatch-isl}.

\section{Flippability combinatorics on Fock space}

\label{app:flip}

For $M$ particles on $L$ sites, $P(n_i\!=\!1,n_{i+1}\!=\!0)=M(L-M)/[L(L-1)]$, and doubling for the
opposite ordering gives Eq.~\eqref{eq:qL}.  Conditioned on $F_i=1$, exactly one of the sites $i,i+1$
is occupied, leaving $M-1$ particles on $L-2$ sites.  For the adjacent bond, $F_{i+1}=1$ requires
$n_{i+2}\neq n_{i+1}$, and averaging the two equally likely orderings of $(n_i,n_{i+1})$,
\begin{equation}
 P(F_{i+1}=1\mid F_i=1)=\tfrac12\left[\frac{M-1}{L-2}+\frac{L-M-1}{L-2}\right]=\tfrac12,
\end{equation}
independent of $L$ and of filling.  For a disjoint bond $B$,
$P(F_B=1\mid F_A=1)=2(M-1)(L-M-1)/[(L-2)(L-3)]$, which at half filling is $(L-2)/[2(L-3)]$.

Relative to the unconditional $q_L$ these are corrections of $-5.6$ to $-10\%$ (adjacent) and $+0.7$
to $+2.9\%$ (disjoint), evaluated over $10\le L\le18$; the identities themselves are exact at any
$L$, so the trend extends to the sizes of the main analysis, $12\le L\le32$.  Because a bulk bond has two adjacent neighbors and
$L-4$ disjoint ones, the weighted total is
$2\cdot\tfrac12+(L-4)(L-2)/[2(L-3)]$, which differs from $q_L(L-2)$ by less than $0.4\%$ at $L=10$ and
less than $0.1\%$ by $L=18$.  The two corrections very nearly cancel, justifying
Eq.~\eqref{eq:Bgeom}.

\bibliography{TwoScales}

\begin{thebibliography}{52}%
\makeatletter
\providecommand \@ifxundefined [1]{%
 \@ifx{#1\undefined}
}%
\providecommand \@ifnum [1]{%
 \ifnum #1\expandafter \@firstoftwo
 \else \expandafter \@secondoftwo
 \fi
}%
\providecommand \@ifx [1]{%
 \ifx #1\expandafter \@firstoftwo
 \else \expandafter \@secondoftwo
 \fi
}%
\providecommand \natexlab [1]{#1}%
\providecommand \enquote  [1]{``#1''}%
\providecommand \bibnamefont  [1]{#1}%
\providecommand \bibfnamefont [1]{#1}%
\providecommand \citenamefont [1]{#1}%
\providecommand \href@noop [0]{\@secondoftwo}%
\providecommand \href [0]{\begingroup \@sanitize@url \@href}%
\providecommand \@href[1]{\@@startlink{#1}\@@href}%
\providecommand \@@href[1]{\endgroup#1\@@endlink}%
\providecommand \@sanitize@url [0]{\catcode `\\12\catcode `\$12\catcode
  `\&12\catcode `\#12\catcode `\^12\catcode `\_12\catcode `\%12\relax}%
\providecommand \@@startlink[1]{}%
\providecommand \@@endlink[0]{}%
\providecommand \url  [0]{\begingroup\@sanitize@url \@url }%
\providecommand \@url [1]{\endgroup\@href {#1}{\urlprefix }}%
\providecommand \urlprefix  [0]{URL }%
\providecommand \Eprint [0]{\href }%
\providecommand \doibase [0]{https://doi.org/}%
\providecommand \selectlanguage [0]{\@gobble}%
\providecommand \bibinfo  [0]{\@secondoftwo}%
\providecommand \bibfield  [0]{\@secondoftwo}%
\providecommand \translation [1]{[#1]}%
\providecommand \BibitemOpen [0]{}%
\providecommand \bibitemStop [0]{}%
\providecommand \bibitemNoStop [0]{.\EOS\space}%
\providecommand \EOS [0]{\spacefactor3000\relax}%
\providecommand \BibitemShut  [1]{\csname bibitem#1\endcsname}%
\let\auto@bib@innerbib\@empty
\bibitem [{\citenamefont {Prasad}(2026)}]{letter}%
  \BibitemOpen
  \bibfield  {author} {\bibinfo {author} {\bibfnamefont {Y.}~\bibnamefont
  {Prasad}},\ }\href@noop {} {\bibinfo {title} {{H}artree fragmentation and
  long-range pair networks in aperiodic chains}} (\bibinfo {year}
  {2026})\BibitemShut {NoStop}%
\bibitem [{\citenamefont {Basko}\ \emph {et~al.}(2006)\citenamefont {Basko},
  \citenamefont {Aleiner},\ and\ \citenamefont {Altshuler}}]{Basko}%
  \BibitemOpen
  \bibfield  {author} {\bibinfo {author} {\bibfnamefont {D.}~\bibnamefont
  {Basko}}, \bibinfo {author} {\bibfnamefont {I.}~\bibnamefont {Aleiner}},\
  and\ \bibinfo {author} {\bibfnamefont {B.}~\bibnamefont {Altshuler}},\
  }\bibfield  {title} {\bibinfo {title} {Metal–insulator transition in a
  weakly interacting many-electron system with localized single-particle
  states},\ }\href {https://doi.org/10.1016/j.aop.2005.11.014} {\bibfield
  {journal} {\bibinfo  {journal} {Ann. Phys.}\ }\textbf {\bibinfo {volume}
  {321}},\ \bibinfo {pages} {1126 } (\bibinfo {year} {2006})}\BibitemShut
  {NoStop}%
\bibitem [{\citenamefont {Oganesyan}\ and\ \citenamefont
  {Huse}(2007)}]{Huse_2007}%
  \BibitemOpen
  \bibfield  {author} {\bibinfo {author} {\bibfnamefont {V.}~\bibnamefont
  {Oganesyan}}\ and\ \bibinfo {author} {\bibfnamefont {D.~A.}\ \bibnamefont
  {Huse}},\ }\bibfield  {title} {\bibinfo {title} {Localization of interacting
  fermions at high temperature},\ }\href
  {https://doi.org/10.1103/PhysRevB.75.155111} {\bibfield  {journal} {\bibinfo
  {journal} {Phys. Rev. B}\ }\textbf {\bibinfo {volume} {75}},\ \bibinfo
  {pages} {155111} (\bibinfo {year} {2007})}\BibitemShut {NoStop}%
\bibitem [{\citenamefont {Serbyn}\ \emph {et~al.}(2013)\citenamefont {Serbyn},
  \citenamefont {Papi{\'c}},\ and\ \citenamefont {Abanin}}]{Serbyn2013}%
  \BibitemOpen
  \bibfield  {author} {\bibinfo {author} {\bibfnamefont {M.}~\bibnamefont
  {Serbyn}}, \bibinfo {author} {\bibfnamefont {Z.}~\bibnamefont {Papi{\'c}}},\
  and\ \bibinfo {author} {\bibfnamefont {D.~A.}\ \bibnamefont {Abanin}},\
  }\bibfield  {title} {\bibinfo {title} {Local conservation laws and the
  structure of the many-body localized states},\ }\href
  {https://doi.org/10.1103/PhysRevLett.111.127201} {\bibfield  {journal}
  {\bibinfo  {journal} {Phys. Rev. Lett.}\ }\textbf {\bibinfo {volume} {111}},\
  \bibinfo {pages} {127201} (\bibinfo {year} {2013})}\BibitemShut {NoStop}%
\bibitem [{\citenamefont {Huse}\ \emph {et~al.}(2014)\citenamefont {Huse},
  \citenamefont {Nandkishore},\ and\ \citenamefont {Oganesyan}}]{HuseNO2014}%
  \BibitemOpen
  \bibfield  {author} {\bibinfo {author} {\bibfnamefont {D.~A.}\ \bibnamefont
  {Huse}}, \bibinfo {author} {\bibfnamefont {R.}~\bibnamefont {Nandkishore}},\
  and\ \bibinfo {author} {\bibfnamefont {V.}~\bibnamefont {Oganesyan}},\
  }\bibfield  {title} {\bibinfo {title} {Phenomenology of fully
  many-body-localized systems},\ }\href
  {https://doi.org/10.1103/PhysRevB.90.174202} {\bibfield  {journal} {\bibinfo
  {journal} {Phys. Rev. B}\ }\textbf {\bibinfo {volume} {90}},\ \bibinfo
  {pages} {174202} (\bibinfo {year} {2014})}\BibitemShut {NoStop}%
\bibitem [{\citenamefont {Ros}\ \emph {et~al.}(2015)\citenamefont {Ros},
  \citenamefont {M{\"u}ller},\ and\ \citenamefont {Scardicchio}}]{Ros2015}%
  \BibitemOpen
  \bibfield  {author} {\bibinfo {author} {\bibfnamefont {V.}~\bibnamefont
  {Ros}}, \bibinfo {author} {\bibfnamefont {M.}~\bibnamefont {M{\"u}ller}},\
  and\ \bibinfo {author} {\bibfnamefont {A.}~\bibnamefont {Scardicchio}},\
  }\bibfield  {title} {\bibinfo {title} {Integrals of motion in the many-body
  localized phase},\ }\href {https://doi.org/10.1016/j.nuclphysb.2014.12.014}
  {\bibfield  {journal} {\bibinfo  {journal} {Nucl. Phys. B}\ }\textbf
  {\bibinfo {volume} {891}},\ \bibinfo {pages} {420} (\bibinfo {year}
  {2015})}\BibitemShut {NoStop}%
\bibitem [{\citenamefont {Imbrie}(2016)}]{Imbrie2016}%
  \BibitemOpen
  \bibfield  {author} {\bibinfo {author} {\bibfnamefont {J.~Z.}\ \bibnamefont
  {Imbrie}},\ }\bibfield  {title} {\bibinfo {title} {On many-body localization
  for quantum spin chains},\ }\href {https://doi.org/10.1007/s10955-016-1508-x}
  {\bibfield  {journal} {\bibinfo  {journal} {J. Stat. Phys.}\ }\textbf
  {\bibinfo {volume} {163}},\ \bibinfo {pages} {998} (\bibinfo {year}
  {2016})}\BibitemShut {NoStop}%
\bibitem [{\citenamefont {Chandran}\ \emph {et~al.}(2015)\citenamefont
  {Chandran}, \citenamefont {Kim}, \citenamefont {Vidal},\ and\ \citenamefont
  {Abanin}}]{Chandran2015}%
  \BibitemOpen
  \bibfield  {author} {\bibinfo {author} {\bibfnamefont {A.}~\bibnamefont
  {Chandran}}, \bibinfo {author} {\bibfnamefont {I.~H.}\ \bibnamefont {Kim}},
  \bibinfo {author} {\bibfnamefont {G.}~\bibnamefont {Vidal}},\ and\ \bibinfo
  {author} {\bibfnamefont {D.~A.}\ \bibnamefont {Abanin}},\ }\bibfield  {title}
  {\bibinfo {title} {Constructing local integrals of motion in the many-body
  localized phase},\ }\href {https://doi.org/10.1103/PhysRevB.91.085425}
  {\bibfield  {journal} {\bibinfo  {journal} {Phys. Rev. B}\ }\textbf {\bibinfo
  {volume} {91}},\ \bibinfo {pages} {085425} (\bibinfo {year}
  {2015})}\BibitemShut {NoStop}%
\bibitem [{\citenamefont {Mierzejewski}\ \emph {et~al.}(2018)\citenamefont
  {Mierzejewski}, \citenamefont {Kozarzewski},\ and\ \citenamefont {Prelov{\v
  s}ek}}]{Mierzejewski2018}%
  \BibitemOpen
  \bibfield  {author} {\bibinfo {author} {\bibfnamefont {M.}~\bibnamefont
  {Mierzejewski}}, \bibinfo {author} {\bibfnamefont {M.}~\bibnamefont
  {Kozarzewski}},\ and\ \bibinfo {author} {\bibfnamefont {P.}~\bibnamefont
  {Prelov{\v s}ek}},\ }\bibfield  {title} {\bibinfo {title} {Counting local
  integrals of motion in disordered spinless-fermion and {Hubbard} chains},\
  }\href {https://doi.org/10.1103/PhysRevB.97.064204} {\bibfield  {journal}
  {\bibinfo  {journal} {Phys. Rev. B}\ }\textbf {\bibinfo {volume} {97}},\
  \bibinfo {pages} {064204} (\bibinfo {year} {2018})}\BibitemShut {NoStop}%
\bibitem [{\citenamefont {Abanin}\ \emph {et~al.}(2019)\citenamefont {Abanin},
  \citenamefont {Altman}, \citenamefont {Bloch},\ and\ \citenamefont
  {Serbyn}}]{Abanin_rev}%
  \BibitemOpen
  \bibfield  {author} {\bibinfo {author} {\bibfnamefont {D.~A.}\ \bibnamefont
  {Abanin}}, \bibinfo {author} {\bibfnamefont {E.}~\bibnamefont {Altman}},
  \bibinfo {author} {\bibfnamefont {I.}~\bibnamefont {Bloch}},\ and\ \bibinfo
  {author} {\bibfnamefont {M.}~\bibnamefont {Serbyn}},\ }\bibfield  {title}
  {\bibinfo {title} {Colloquium: Many-body localization, thermalization, and
  entanglement},\ }\href {https://doi.org/10.1103/RevModPhys.91.021001}
  {\bibfield  {journal} {\bibinfo  {journal} {Rev. Mod. Phys.}\ }\textbf
  {\bibinfo {volume} {91}},\ \bibinfo {pages} {021001} (\bibinfo {year}
  {2019})}\BibitemShut {NoStop}%
\bibitem [{\citenamefont {Sierant}\ \emph {et~al.}(2025)\citenamefont
  {Sierant}, \citenamefont {Lewenstein}, \citenamefont {Scardicchio},
  \citenamefont {Vidmar},\ and\ \citenamefont {Zakrzewski}}]{Sierant_rev}%
  \BibitemOpen
  \bibfield  {author} {\bibinfo {author} {\bibfnamefont {P.}~\bibnamefont
  {Sierant}}, \bibinfo {author} {\bibfnamefont {M.}~\bibnamefont {Lewenstein}},
  \bibinfo {author} {\bibfnamefont {A.}~\bibnamefont {Scardicchio}}, \bibinfo
  {author} {\bibfnamefont {L.}~\bibnamefont {Vidmar}},\ and\ \bibinfo {author}
  {\bibfnamefont {J.}~\bibnamefont {Zakrzewski}},\ }\bibfield  {title}
  {\bibinfo {title} {Many-body localization in the age of classical
  computing},\ }\href {https://doi.org/10.1088/1361-6633/ad9756} {\bibfield
  {journal} {\bibinfo  {journal} {Rep. Prog. Phys.}\ }\textbf {\bibinfo
  {volume} {88}},\ \bibinfo {pages} {026502} (\bibinfo {year}
  {2025})}\BibitemShut {NoStop}%
\bibitem [{\citenamefont {Burin}(2006)}]{Burin2006}%
  \BibitemOpen
  \bibfield  {author} {\bibinfo {author} {\bibfnamefont {A.~L.}\ \bibnamefont
  {Burin}},\ }\bibfield  {title} {\bibinfo {title} {Energy delocalization in
  strongly disordered systems induced by the long-range many-body
  interaction},\ }\href {https://arxiv.org/abs/cond-mat/0611387} {\  (\bibinfo
  {year} {2006})},\ \Eprint {https://arxiv.org/abs/cond-mat/0611387}
  {arXiv:cond-mat/0611387} \BibitemShut {NoStop}%
\bibitem [{\citenamefont {Burin}(2015)}]{Burin2015}%
  \BibitemOpen
  \bibfield  {author} {\bibinfo {author} {\bibfnamefont {A.~L.}\ \bibnamefont
  {Burin}},\ }\bibfield  {title} {\bibinfo {title} {Many-body delocalization in
  a strongly disordered system with long-range interactions: Finite-size
  scaling},\ }\href {https://doi.org/10.1103/PhysRevB.91.094202} {\bibfield
  {journal} {\bibinfo  {journal} {Phys. Rev. B}\ }\textbf {\bibinfo {volume}
  {91}},\ \bibinfo {pages} {094202} (\bibinfo {year} {2015})}\BibitemShut
  {NoStop}%
\bibitem [{\citenamefont {Yao}\ \emph {et~al.}(2014)\citenamefont {Yao},
  \citenamefont {Laumann}, \citenamefont {Gopalakrishnan}, \citenamefont
  {Knap}, \citenamefont {M{\"u}ller}, \citenamefont {Demler},\ and\
  \citenamefont {Lukin}}]{Yao2014}%
  \BibitemOpen
  \bibfield  {author} {\bibinfo {author} {\bibfnamefont {N.~Y.}\ \bibnamefont
  {Yao}}, \bibinfo {author} {\bibfnamefont {C.~R.}\ \bibnamefont {Laumann}},
  \bibinfo {author} {\bibfnamefont {S.}~\bibnamefont {Gopalakrishnan}},
  \bibinfo {author} {\bibfnamefont {M.}~\bibnamefont {Knap}}, \bibinfo {author}
  {\bibfnamefont {M.}~\bibnamefont {M{\"u}ller}}, \bibinfo {author}
  {\bibfnamefont {E.~A.}\ \bibnamefont {Demler}},\ and\ \bibinfo {author}
  {\bibfnamefont {M.~D.}\ \bibnamefont {Lukin}},\ }\bibfield  {title} {\bibinfo
  {title} {Many-body localization in dipolar systems},\ }\href
  {https://doi.org/10.1103/PhysRevLett.113.243002} {\bibfield  {journal}
  {\bibinfo  {journal} {Phys. Rev. Lett.}\ }\textbf {\bibinfo {volume} {113}},\
  \bibinfo {pages} {243002} (\bibinfo {year} {2014})}\BibitemShut {NoStop}%
\bibitem [{\citenamefont {Tikhonov}\ and\ \citenamefont
  {Mirlin}(2018)}]{Mirlin}%
  \BibitemOpen
  \bibfield  {author} {\bibinfo {author} {\bibfnamefont {K.~S.}\ \bibnamefont
  {Tikhonov}}\ and\ \bibinfo {author} {\bibfnamefont {A.~D.}\ \bibnamefont
  {Mirlin}},\ }\bibfield  {title} {\bibinfo {title} {Many-body localization
  transition with power-law interactions: Statistics of eigenstates},\ }\href
  {https://doi.org/10.1103/PhysRevB.97.214205} {\bibfield  {journal} {\bibinfo
  {journal} {Phys. Rev. B}\ }\textbf {\bibinfo {volume} {97}},\ \bibinfo
  {pages} {214205} (\bibinfo {year} {2018})}\BibitemShut {NoStop}%
\bibitem [{\citenamefont {Hauke}\ and\ \citenamefont {Heyl}(2015)}]{Hauke2015}%
  \BibitemOpen
  \bibfield  {author} {\bibinfo {author} {\bibfnamefont {P.}~\bibnamefont
  {Hauke}}\ and\ \bibinfo {author} {\bibfnamefont {M.}~\bibnamefont {Heyl}},\
  }\bibfield  {title} {\bibinfo {title} {Many-body localization and quantum
  ergodicity in disordered long-range {Ising} models},\ }\href
  {https://doi.org/10.1103/PhysRevB.92.134204} {\bibfield  {journal} {\bibinfo
  {journal} {Phys. Rev. B}\ }\textbf {\bibinfo {volume} {92}},\ \bibinfo
  {pages} {134204} (\bibinfo {year} {2015})}\BibitemShut {NoStop}%
\bibitem [{\citenamefont {Nandkishore}\ and\ \citenamefont
  {Sondhi}(2017)}]{NandkishoreSondhi2017}%
  \BibitemOpen
  \bibfield  {author} {\bibinfo {author} {\bibfnamefont {R.~M.}\ \bibnamefont
  {Nandkishore}}\ and\ \bibinfo {author} {\bibfnamefont {S.~L.}\ \bibnamefont
  {Sondhi}},\ }\bibfield  {title} {\bibinfo {title} {Many-body localization
  with long-range interactions},\ }\href
  {https://doi.org/10.1103/PhysRevX.7.041021} {\bibfield  {journal} {\bibinfo
  {journal} {Phys. Rev. X}\ }\textbf {\bibinfo {volume} {7}},\ \bibinfo {pages}
  {041021} (\bibinfo {year} {2017})}\BibitemShut {NoStop}%
\bibitem [{\citenamefont {De~Tomasi}(2019)}]{DeTomasi2019}%
  \BibitemOpen
  \bibfield  {author} {\bibinfo {author} {\bibfnamefont {G.}~\bibnamefont
  {De~Tomasi}},\ }\bibfield  {title} {\bibinfo {title} {Algebraic many-body
  localization and its implications on information propagation},\ }\href
  {https://doi.org/10.1103/PhysRevB.99.054204} {\bibfield  {journal} {\bibinfo
  {journal} {Phys. Rev. B}\ }\textbf {\bibinfo {volume} {99}},\ \bibinfo
  {pages} {054204} (\bibinfo {year} {2019})}\BibitemShut {NoStop}%
\bibitem [{\citenamefont {Thomson}\ and\ \citenamefont
  {Schir{\'o}}(2020)}]{Thomson2020}%
  \BibitemOpen
  \bibfield  {author} {\bibinfo {author} {\bibfnamefont {S.~J.}\ \bibnamefont
  {Thomson}}\ and\ \bibinfo {author} {\bibfnamefont {M.}~\bibnamefont
  {Schir{\'o}}},\ }\bibfield  {title} {\bibinfo {title} {Quasi-many-body
  localization of interacting fermions with long-range couplings},\ }\href
  {https://doi.org/10.1103/PhysRevResearch.2.043368} {\bibfield  {journal}
  {\bibinfo  {journal} {Phys. Rev. Research}\ }\textbf {\bibinfo {volume}
  {2}},\ \bibinfo {pages} {043368} (\bibinfo {year} {2020})}\BibitemShut
  {NoStop}%
\bibitem [{\citenamefont {Maksymov}\ and\ \citenamefont
  {Burin}(2020)}]{Maksymov2020}%
  \BibitemOpen
  \bibfield  {author} {\bibinfo {author} {\bibfnamefont {A.~O.}\ \bibnamefont
  {Maksymov}}\ and\ \bibinfo {author} {\bibfnamefont {A.~L.}\ \bibnamefont
  {Burin}},\ }\bibfield  {title} {\bibinfo {title} {Many-body localization in
  spin chains with long-range transverse interactions: Scaling of critical
  disorder with system size},\ }\href
  {https://doi.org/10.1103/PhysRevB.101.024201} {\bibfield  {journal} {\bibinfo
   {journal} {Phys. Rev. B}\ }\textbf {\bibinfo {volume} {101}},\ \bibinfo
  {pages} {024201} (\bibinfo {year} {2020})}\BibitemShut {NoStop}%
\bibitem [{\citenamefont {Botzung}\ \emph {et~al.}(2021)\citenamefont
  {Botzung}, \citenamefont {Hagenm{\"u}ller}, \citenamefont {Masella},
  \citenamefont {Dubail}, \citenamefont {Defenu}, \citenamefont {Trombettoni},\
  and\ \citenamefont {Pupillo}}]{Botzung2021}%
  \BibitemOpen
  \bibfield  {author} {\bibinfo {author} {\bibfnamefont {T.}~\bibnamefont
  {Botzung}}, \bibinfo {author} {\bibfnamefont {D.}~\bibnamefont
  {Hagenm{\"u}ller}}, \bibinfo {author} {\bibfnamefont {G.}~\bibnamefont
  {Masella}}, \bibinfo {author} {\bibfnamefont {J.}~\bibnamefont {Dubail}},
  \bibinfo {author} {\bibfnamefont {N.}~\bibnamefont {Defenu}}, \bibinfo
  {author} {\bibfnamefont {A.}~\bibnamefont {Trombettoni}},\ and\ \bibinfo
  {author} {\bibfnamefont {G.}~\bibnamefont {Pupillo}},\ }\bibfield  {title}
  {\bibinfo {title} {Effects of energy extensivity on the quantum phases of
  long-range interacting systems},\ }\href
  {https://doi.org/10.1103/PhysRevB.103.155139} {\bibfield  {journal} {\bibinfo
   {journal} {Phys. Rev. B}\ }\textbf {\bibinfo {volume} {103}},\ \bibinfo
  {pages} {155139} (\bibinfo {year} {2021})}\BibitemShut {NoStop}%
\bibitem [{\citenamefont {Sierant}\ \emph {et~al.}(2019)\citenamefont
  {Sierant}, \citenamefont {Biedro{\'n}}, \citenamefont {Morigi},\ and\
  \citenamefont {Zakrzewski}}]{Sierant2019}%
  \BibitemOpen
  \bibfield  {author} {\bibinfo {author} {\bibfnamefont {P.}~\bibnamefont
  {Sierant}}, \bibinfo {author} {\bibfnamefont {K.}~\bibnamefont
  {Biedro{\'n}}}, \bibinfo {author} {\bibfnamefont {G.}~\bibnamefont
  {Morigi}},\ and\ \bibinfo {author} {\bibfnamefont {J.}~\bibnamefont
  {Zakrzewski}},\ }\bibfield  {title} {\bibinfo {title} {Many-body localization
  in presence of cavity mediated long-range interactions},\ }\href
  {https://doi.org/10.21468/SciPostPhys.7.1.008} {\bibfield  {journal}
  {\bibinfo  {journal} {SciPost Phys.}\ }\textbf {\bibinfo {volume} {7}},\
  \bibinfo {pages} {008} (\bibinfo {year} {2019})}\BibitemShut {NoStop}%
\bibitem [{\citenamefont {Vu}\ \emph {et~al.}(2022)\citenamefont {Vu},
  \citenamefont {Huang}, \citenamefont {Li},\ and\ \citenamefont
  {Das~Sarma}}]{Vu2022}%
  \BibitemOpen
  \bibfield  {author} {\bibinfo {author} {\bibfnamefont {D.}~\bibnamefont
  {Vu}}, \bibinfo {author} {\bibfnamefont {K.}~\bibnamefont {Huang}}, \bibinfo
  {author} {\bibfnamefont {X.}~\bibnamefont {Li}},\ and\ \bibinfo {author}
  {\bibfnamefont {S.}~\bibnamefont {Das~Sarma}},\ }\bibfield  {title} {\bibinfo
  {title} {Fermionic many-body localization for random and quasiperiodic
  systems in the presence of short- and long-range interactions},\ }\href
  {https://doi.org/10.1103/PhysRevLett.128.146601} {\bibfield  {journal}
  {\bibinfo  {journal} {Phys. Rev. Lett.}\ }\textbf {\bibinfo {volume} {128}},\
  \bibinfo {pages} {146601} (\bibinfo {year} {2022})}\BibitemShut {NoStop}%
\bibitem [{\citenamefont {Defenu}\ \emph {et~al.}(2023)\citenamefont {Defenu},
  \citenamefont {Donner}, \citenamefont {Macr{\`i}}, \citenamefont {Pagano},
  \citenamefont {Ruffo},\ and\ \citenamefont {Trombettoni}}]{Defenu2023}%
  \BibitemOpen
  \bibfield  {author} {\bibinfo {author} {\bibfnamefont {N.}~\bibnamefont
  {Defenu}}, \bibinfo {author} {\bibfnamefont {T.}~\bibnamefont {Donner}},
  \bibinfo {author} {\bibfnamefont {T.}~\bibnamefont {Macr{\`i}}}, \bibinfo
  {author} {\bibfnamefont {G.}~\bibnamefont {Pagano}}, \bibinfo {author}
  {\bibfnamefont {S.}~\bibnamefont {Ruffo}},\ and\ \bibinfo {author}
  {\bibfnamefont {A.}~\bibnamefont {Trombettoni}},\ }\bibfield  {title}
  {\bibinfo {title} {Long-range interacting quantum systems},\ }\href
  {https://doi.org/10.1103/RevModPhys.95.035002} {\bibfield  {journal}
  {\bibinfo  {journal} {Rev. Mod. Phys.}\ }\textbf {\bibinfo {volume} {95}},\
  \bibinfo {pages} {035002} (\bibinfo {year} {2023})}\BibitemShut {NoStop}%
\bibitem [{\citenamefont {Padhan}\ \emph {et~al.}(2026)\citenamefont {Padhan},
  \citenamefont {Colbois}, \citenamefont {Alet},\ and\ \citenamefont
  {Laflorencie}}]{Padhan2026}%
  \BibitemOpen
  \bibfield  {author} {\bibinfo {author} {\bibfnamefont {A.}~\bibnamefont
  {Padhan}}, \bibinfo {author} {\bibfnamefont {J.}~\bibnamefont {Colbois}},
  \bibinfo {author} {\bibfnamefont {F.}~\bibnamefont {Alet}},\ and\ \bibinfo
  {author} {\bibfnamefont {N.}~\bibnamefont {Laflorencie}},\ }\bibfield
  {title} {\bibinfo {title} {Long-range resonances in quasiperiodic many-body
  localization},\ }\href {https://doi.org/10.1103/PhysRevLett.136.197101}
  {\bibfield  {journal} {\bibinfo  {journal} {Phys. Rev. Lett.}\ }\textbf
  {\bibinfo {volume} {136}},\ \bibinfo {pages} {197101} (\bibinfo {year}
  {2026})}\BibitemShut {NoStop}%
\bibitem [{\citenamefont {Griniasty}\ and\ \citenamefont
  {Fishman}(1988)}]{Fishman}%
  \BibitemOpen
  \bibfield  {author} {\bibinfo {author} {\bibfnamefont {M.}~\bibnamefont
  {Griniasty}}\ and\ \bibinfo {author} {\bibfnamefont {S.}~\bibnamefont
  {Fishman}},\ }\bibfield  {title} {\bibinfo {title} {Localization by
  pseudorandom potentials in one dimension},\ }\href
  {https://doi.org/10.1103/PhysRevLett.60.1334} {\bibfield  {journal} {\bibinfo
   {journal} {Phys. Rev. Lett.}\ }\textbf {\bibinfo {volume} {60}},\ \bibinfo
  {pages} {1334} (\bibinfo {year} {1988})}\BibitemShut {NoStop}%
\bibitem [{\citenamefont {Das~Sarma}\ \emph {et~al.}(1990)\citenamefont
  {Das~Sarma}, \citenamefont {He},\ and\ \citenamefont {Xie}}]{Sarma1990}%
  \BibitemOpen
  \bibfield  {author} {\bibinfo {author} {\bibfnamefont {S.}~\bibnamefont
  {Das~Sarma}}, \bibinfo {author} {\bibfnamefont {S.}~\bibnamefont {He}},\ and\
  \bibinfo {author} {\bibfnamefont {X.~C.}\ \bibnamefont {Xie}},\ }\bibfield
  {title} {\bibinfo {title} {Localization, mobility edges, and metal-insulator
  transition in a class of one-dimensional slowly varying deterministic
  potentials},\ }\href {https://doi.org/10.1103/PhysRevB.41.5544} {\bibfield
  {journal} {\bibinfo  {journal} {Phys. Rev. B}\ }\textbf {\bibinfo {volume}
  {41}},\ \bibinfo {pages} {5544} (\bibinfo {year} {1990})}\BibitemShut
  {NoStop}%
\bibitem [{\citenamefont {Li}\ \emph {et~al.}(2025)\citenamefont {Li},
  \citenamefont {Tu},\ and\ \citenamefont {Das~Sarma}}]{LiTuDasSarma2025}%
  \BibitemOpen
  \bibfield  {author} {\bibinfo {author} {\bibfnamefont {Z.-J.}\ \bibnamefont
  {Li}}, \bibinfo {author} {\bibfnamefont {Y.-T.}\ \bibnamefont {Tu}},\ and\
  \bibinfo {author} {\bibfnamefont {S.}~\bibnamefont {Das~Sarma}},\ }\bibfield
  {title} {\bibinfo {title} {Many-body localization in a slowly varying
  potential},\ }\href {https://doi.org/10.1103/PhysRevB.112.014203} {\bibfield
  {journal} {\bibinfo  {journal} {Phys. Rev. B}\ }\textbf {\bibinfo {volume}
  {112}},\ \bibinfo {pages} {014203} (\bibinfo {year} {2025})}\BibitemShut
  {NoStop}%
\bibitem [{\citenamefont {Aubry}\ and\ \citenamefont {André}(1980)}]{AA}%
  \BibitemOpen
  \bibfield  {author} {\bibinfo {author} {\bibfnamefont {S.}~\bibnamefont
  {Aubry}}\ and\ \bibinfo {author} {\bibfnamefont {G.}~\bibnamefont {André}},\
  }\bibfield  {title} {\bibinfo {title} {Analyticity breaking and {Anderson}
  localization in incommensurate lattices},\ }\href@noop {} {\bibfield
  {journal} {\bibinfo  {journal} {Ann. Isr. Phys. Soc.}\ }\textbf {\bibinfo
  {volume} {3}},\ \bibinfo {pages} {133} (\bibinfo {year} {1980})}\BibitemShut
  {NoStop}%
\bibitem [{\citenamefont {Iyer}\ \emph {et~al.}(2013)\citenamefont {Iyer},
  \citenamefont {Oganesyan}, \citenamefont {Refael},\ and\ \citenamefont
  {Huse}}]{Huse}%
  \BibitemOpen
  \bibfield  {author} {\bibinfo {author} {\bibfnamefont {S.}~\bibnamefont
  {Iyer}}, \bibinfo {author} {\bibfnamefont {V.}~\bibnamefont {Oganesyan}},
  \bibinfo {author} {\bibfnamefont {G.}~\bibnamefont {Refael}},\ and\ \bibinfo
  {author} {\bibfnamefont {D.~A.}\ \bibnamefont {Huse}},\ }\bibfield  {title}
  {\bibinfo {title} {Many-body localization in a quasiperiodic system},\ }\href
  {https://doi.org/10.1103/PhysRevB.87.134202} {\bibfield  {journal} {\bibinfo
  {journal} {Phys. Rev. B}\ }\textbf {\bibinfo {volume} {87}},\ \bibinfo
  {pages} {134202} (\bibinfo {year} {2013})}\BibitemShut {NoStop}%
\bibitem [{\citenamefont {Mac{\'e}}\ \emph {et~al.}(2019)\citenamefont
  {Mac{\'e}}, \citenamefont {Laflorencie},\ and\ \citenamefont
  {Alet}}]{Mace2019}%
  \BibitemOpen
  \bibfield  {author} {\bibinfo {author} {\bibfnamefont {N.}~\bibnamefont
  {Mac{\'e}}}, \bibinfo {author} {\bibfnamefont {N.}~\bibnamefont
  {Laflorencie}},\ and\ \bibinfo {author} {\bibfnamefont {F.}~\bibnamefont
  {Alet}},\ }\bibfield  {title} {\bibinfo {title} {Many-body localization in a
  quasiperiodic {Fibonacci} chain},\ }\href
  {https://doi.org/10.21468/SciPostPhys.6.4.050} {\bibfield  {journal}
  {\bibinfo  {journal} {SciPost Phys.}\ }\textbf {\bibinfo {volume} {6}},\
  \bibinfo {pages} {050} (\bibinfo {year} {2019})}\BibitemShut {NoStop}%
\bibitem [{\citenamefont {Khemani}\ \emph {et~al.}(2017)\citenamefont
  {Khemani}, \citenamefont {Sheng},\ and\ \citenamefont {Huse}}]{Khemani2017}%
  \BibitemOpen
  \bibfield  {author} {\bibinfo {author} {\bibfnamefont {V.}~\bibnamefont
  {Khemani}}, \bibinfo {author} {\bibfnamefont {D.~N.}\ \bibnamefont {Sheng}},\
  and\ \bibinfo {author} {\bibfnamefont {D.~A.}\ \bibnamefont {Huse}},\
  }\bibfield  {title} {\bibinfo {title} {Two universality classes for the
  many-body localization transition},\ }\href
  {https://doi.org/10.1103/PhysRevLett.119.075702} {\bibfield  {journal}
  {\bibinfo  {journal} {Phys. Rev. Lett.}\ }\textbf {\bibinfo {volume} {119}},\
  \bibinfo {pages} {075702} (\bibinfo {year} {2017})}\BibitemShut {NoStop}%
\bibitem [{\citenamefont {Varma}\ and\ \citenamefont
  {{\v{Z}}nidari{\v{c}}}(2019)}]{Varma2019}%
  \BibitemOpen
  \bibfield  {author} {\bibinfo {author} {\bibfnamefont {V.~K.}\ \bibnamefont
  {Varma}}\ and\ \bibinfo {author} {\bibfnamefont {M.}~\bibnamefont
  {{\v{Z}}nidari{\v{c}}}},\ }\bibfield  {title} {\bibinfo {title} {Diffusive
  transport in a quasiperiodic {Fibonacci} chain: Absence of many-body
  localization at weak interactions},\ }\href
  {https://doi.org/10.1103/PhysRevB.100.085105} {\bibfield  {journal} {\bibinfo
   {journal} {Phys. Rev. B}\ }\textbf {\bibinfo {volume} {100}},\ \bibinfo
  {pages} {085105} (\bibinfo {year} {2019})}\BibitemShut {NoStop}%
\bibitem [{\citenamefont {Falc{\~a}o}\ \emph {et~al.}(2024)\citenamefont
  {Falc{\~a}o}, \citenamefont {Aramthottil}, \citenamefont {Sierant},\ and\
  \citenamefont {Zakrzewski}}]{Falcao2024}%
  \BibitemOpen
  \bibfield  {author} {\bibinfo {author} {\bibfnamefont {P.~R.~N.}\
  \bibnamefont {Falc{\~a}o}}, \bibinfo {author} {\bibfnamefont {A.~S.}\
  \bibnamefont {Aramthottil}}, \bibinfo {author} {\bibfnamefont
  {P.}~\bibnamefont {Sierant}},\ and\ \bibinfo {author} {\bibfnamefont
  {J.}~\bibnamefont {Zakrzewski}},\ }\bibfield  {title} {\bibinfo {title}
  {Many-body localization crossover is sharper in a quasiperiodic potential},\
  }\href {https://doi.org/10.1103/PhysRevB.110.184209} {\bibfield  {journal}
  {\bibinfo  {journal} {Phys. Rev. B}\ }\textbf {\bibinfo {volume} {110}},\
  \bibinfo {pages} {184209} (\bibinfo {year} {2024})}\BibitemShut {NoStop}%
\bibitem [{\citenamefont {Singh}\ \emph {et~al.}(2021)\citenamefont {Singh},
  \citenamefont {Ware}, \citenamefont {Vasseur},\ and\ \citenamefont
  {Gopalakrishnan}}]{Singh2021}%
  \BibitemOpen
  \bibfield  {author} {\bibinfo {author} {\bibfnamefont {H.}~\bibnamefont
  {Singh}}, \bibinfo {author} {\bibfnamefont {B.}~\bibnamefont {Ware}},
  \bibinfo {author} {\bibfnamefont {R.}~\bibnamefont {Vasseur}},\ and\ \bibinfo
  {author} {\bibfnamefont {S.}~\bibnamefont {Gopalakrishnan}},\ }\bibfield
  {title} {\bibinfo {title} {Local integrals of motion and the quasiperiodic
  many-body localization transition},\ }\href
  {https://doi.org/10.1103/PhysRevB.103.L220201} {\bibfield  {journal}
  {\bibinfo  {journal} {Phys. Rev. B}\ }\textbf {\bibinfo {volume} {103}},\
  \bibinfo {pages} {L220201} (\bibinfo {year} {2021})}\BibitemShut {NoStop}%
\bibitem [{\citenamefont {Thomson}\ and\ \citenamefont
  {Schir{\'o}}(2023)}]{Thomson2023}%
  \BibitemOpen
  \bibfield  {author} {\bibinfo {author} {\bibfnamefont {S.~J.}\ \bibnamefont
  {Thomson}}\ and\ \bibinfo {author} {\bibfnamefont {M.}~\bibnamefont
  {Schir{\'o}}},\ }\bibfield  {title} {\bibinfo {title} {Local integrals of
  motion in quasiperiodic many-body localized systems},\ }\href
  {https://doi.org/10.21468/SciPostPhys.14.5.125} {\bibfield  {journal}
  {\bibinfo  {journal} {SciPost Phys.}\ }\textbf {\bibinfo {volume} {14}},\
  \bibinfo {pages} {125} (\bibinfo {year} {2023})}\BibitemShut {NoStop}%
\bibitem [{\citenamefont {Prasad}\ and\ \citenamefont
  {Garg}(2024)}]{Prasad2024}%
  \BibitemOpen
  \bibfield  {author} {\bibinfo {author} {\bibfnamefont {Y.}~\bibnamefont
  {Prasad}}\ and\ \bibinfo {author} {\bibfnamefont {A.}~\bibnamefont {Garg}},\
  }\bibfield  {title} {\bibinfo {title} {Single-particle excitations across the
  localization and many-body localization transition in quasiperiodic
  systems},\ }\href {https://doi.org/10.1103/PhysRevB.109.094204} {\bibfield
  {journal} {\bibinfo  {journal} {Phys. Rev. B}\ }\textbf {\bibinfo {volume}
  {109}},\ \bibinfo {pages} {094204} (\bibinfo {year} {2024})}\BibitemShut
  {NoStop}%
\bibitem [{\citenamefont {Schreiber}\ \emph {et~al.}(2015)\citenamefont
  {Schreiber}, \citenamefont {Hodgman}, \citenamefont {Bordia}, \citenamefont
  {L{\"u}schen}, \citenamefont {Fischer}, \citenamefont {Vosk}, \citenamefont
  {Altman}, \citenamefont {Schneider},\ and\ \citenamefont
  {Bloch}}]{Schreiber2015}%
  \BibitemOpen
  \bibfield  {author} {\bibinfo {author} {\bibfnamefont {M.}~\bibnamefont
  {Schreiber}}, \bibinfo {author} {\bibfnamefont {S.~S.}\ \bibnamefont
  {Hodgman}}, \bibinfo {author} {\bibfnamefont {P.}~\bibnamefont {Bordia}},
  \bibinfo {author} {\bibfnamefont {H.~P.}\ \bibnamefont {L{\"u}schen}},
  \bibinfo {author} {\bibfnamefont {M.~H.}\ \bibnamefont {Fischer}}, \bibinfo
  {author} {\bibfnamefont {R.}~\bibnamefont {Vosk}}, \bibinfo {author}
  {\bibfnamefont {E.}~\bibnamefont {Altman}}, \bibinfo {author} {\bibfnamefont
  {U.}~\bibnamefont {Schneider}},\ and\ \bibinfo {author} {\bibfnamefont
  {I.}~\bibnamefont {Bloch}},\ }\bibfield  {title} {\bibinfo {title}
  {Observation of many-body localization of interacting fermions in a
  quasirandom optical lattice},\ }\href
  {https://doi.org/10.1126/science.aaa7432} {\bibfield  {journal} {\bibinfo
  {journal} {Science}\ }\textbf {\bibinfo {volume} {349}},\ \bibinfo {pages}
  {842} (\bibinfo {year} {2015})}\BibitemShut {NoStop}%
\bibitem [{\citenamefont {Smith}\ \emph {et~al.}(2016)\citenamefont {Smith},
  \citenamefont {Lee}, \citenamefont {Richerme}, \citenamefont {Neyenhuis},
  \citenamefont {Hess}, \citenamefont {Hauke}, \citenamefont {Heyl},
  \citenamefont {Huse},\ and\ \citenamefont {Monroe}}]{Smith2016}%
  \BibitemOpen
  \bibfield  {author} {\bibinfo {author} {\bibfnamefont {J.}~\bibnamefont
  {Smith}}, \bibinfo {author} {\bibfnamefont {A.}~\bibnamefont {Lee}}, \bibinfo
  {author} {\bibfnamefont {P.}~\bibnamefont {Richerme}}, \bibinfo {author}
  {\bibfnamefont {B.}~\bibnamefont {Neyenhuis}}, \bibinfo {author}
  {\bibfnamefont {P.~W.}\ \bibnamefont {Hess}}, \bibinfo {author}
  {\bibfnamefont {P.}~\bibnamefont {Hauke}}, \bibinfo {author} {\bibfnamefont
  {M.}~\bibnamefont {Heyl}}, \bibinfo {author} {\bibfnamefont {D.~A.}\
  \bibnamefont {Huse}},\ and\ \bibinfo {author} {\bibfnamefont
  {C.}~\bibnamefont {Monroe}},\ }\bibfield  {title} {\bibinfo {title}
  {Many-body localization in a quantum simulator with programmable random
  disorder},\ }\href {https://doi.org/10.1038/nphys3783} {\bibfield  {journal}
  {\bibinfo  {journal} {Nat. Phys.}\ }\textbf {\bibinfo {volume} {12}},\
  \bibinfo {pages} {907} (\bibinfo {year} {2016})}\BibitemShut {NoStop}%
\bibitem [{\citenamefont {Altshuler}\ \emph {et~al.}(1997)\citenamefont
  {Altshuler}, \citenamefont {Gefen}, \citenamefont {Kamenev},\ and\
  \citenamefont {Levitov}}]{Altshuler1997}%
  \BibitemOpen
  \bibfield  {author} {\bibinfo {author} {\bibfnamefont {B.~L.}\ \bibnamefont
  {Altshuler}}, \bibinfo {author} {\bibfnamefont {Y.}~\bibnamefont {Gefen}},
  \bibinfo {author} {\bibfnamefont {A.}~\bibnamefont {Kamenev}},\ and\ \bibinfo
  {author} {\bibfnamefont {L.~S.}\ \bibnamefont {Levitov}},\ }\bibfield
  {title} {\bibinfo {title} {Quasiparticle lifetime in a finite system: A
  nonperturbative approach},\ }\href
  {https://doi.org/10.1103/PhysRevLett.78.2803} {\bibfield  {journal} {\bibinfo
   {journal} {Phys. Rev. Lett.}\ }\textbf {\bibinfo {volume} {78}},\ \bibinfo
  {pages} {2803} (\bibinfo {year} {1997})}\BibitemShut {NoStop}%
\bibitem [{\citenamefont {Mossi}\ and\ \citenamefont
  {Scardicchio}(2017)}]{Mossi2017}%
  \BibitemOpen
  \bibfield  {author} {\bibinfo {author} {\bibfnamefont {G.}~\bibnamefont
  {Mossi}}\ and\ \bibinfo {author} {\bibfnamefont {A.}~\bibnamefont
  {Scardicchio}},\ }\bibfield  {title} {\bibinfo {title} {Ergodic and localized
  regions in quantum spin glasses on the {B}ethe lattice},\ }\href
  {https://doi.org/10.1098/rsta.2016.0424} {\bibfield  {journal} {\bibinfo
  {journal} {Philos. Trans. R. Soc. A}\ }\textbf {\bibinfo {volume} {375}},\
  \bibinfo {pages} {20160424} (\bibinfo {year} {2017})}\BibitemShut {NoStop}%
\bibitem [{\citenamefont {Ghosh}\ \emph {et~al.}(2019)\citenamefont {Ghosh},
  \citenamefont {Acharya}, \citenamefont {Sahu},\ and\ \citenamefont
  {Mukerjee}}]{Subroto_corr}%
  \BibitemOpen
  \bibfield  {author} {\bibinfo {author} {\bibfnamefont {S.}~\bibnamefont
  {Ghosh}}, \bibinfo {author} {\bibfnamefont {A.}~\bibnamefont {Acharya}},
  \bibinfo {author} {\bibfnamefont {S.}~\bibnamefont {Sahu}},\ and\ \bibinfo
  {author} {\bibfnamefont {S.}~\bibnamefont {Mukerjee}},\ }\bibfield  {title}
  {\bibinfo {title} {Many-body localization due to correlated disorder in
  {Fock} space},\ }\href {https://doi.org/10.1103/PhysRevB.99.165131}
  {\bibfield  {journal} {\bibinfo  {journal} {Phys. Rev. B}\ }\textbf {\bibinfo
  {volume} {99}},\ \bibinfo {pages} {165131} (\bibinfo {year}
  {2019})}\BibitemShut {NoStop}%
\bibitem [{\citenamefont {Roy}\ \emph {et~al.}(2019)\citenamefont {Roy},
  \citenamefont {Chalker},\ and\ \citenamefont {Logan}}]{RoyChalkerLogan2019}%
  \BibitemOpen
  \bibfield  {author} {\bibinfo {author} {\bibfnamefont {S.}~\bibnamefont
  {Roy}}, \bibinfo {author} {\bibfnamefont {J.~T.}\ \bibnamefont {Chalker}},\
  and\ \bibinfo {author} {\bibfnamefont {D.~E.}\ \bibnamefont {Logan}},\
  }\bibfield  {title} {\bibinfo {title} {Percolation in {F}ock space as a proxy
  for many-body localization},\ }\href
  {https://doi.org/10.1103/PhysRevB.99.104206} {\bibfield  {journal} {\bibinfo
  {journal} {Phys. Rev. B}\ }\textbf {\bibinfo {volume} {99}},\ \bibinfo
  {pages} {104206} (\bibinfo {year} {2019})}\BibitemShut {NoStop}%
\bibitem [{\citenamefont {Prelov\v{s}ek}\ \emph {et~al.}(2021)\citenamefont
  {Prelov\v{s}ek}, \citenamefont {Mierzejewski}, \citenamefont {Krsnik},\ and\
  \citenamefont {Bari\v{s}i\'{c}}}]{Prelovsek2021}%
  \BibitemOpen
  \bibfield  {author} {\bibinfo {author} {\bibfnamefont {P.}~\bibnamefont
  {Prelov\v{s}ek}}, \bibinfo {author} {\bibfnamefont {M.}~\bibnamefont
  {Mierzejewski}}, \bibinfo {author} {\bibfnamefont {J.}~\bibnamefont
  {Krsnik}},\ and\ \bibinfo {author} {\bibfnamefont {O.~S.}\ \bibnamefont
  {Bari\v{s}i\'{c}}},\ }\bibfield  {title} {\bibinfo {title} {Many-body
  localization as a percolation phenomenon},\ }\href
  {https://doi.org/10.1103/PhysRevB.103.045139} {\bibfield  {journal} {\bibinfo
   {journal} {Phys. Rev. B}\ }\textbf {\bibinfo {volume} {103}},\ \bibinfo
  {pages} {045139} (\bibinfo {year} {2021})}\BibitemShut {NoStop}%
\bibitem [{\citenamefont {Roy}\ and\ \citenamefont
  {Logan}(2020)}]{RoyLogan2020}%
  \BibitemOpen
  \bibfield  {author} {\bibinfo {author} {\bibfnamefont {S.}~\bibnamefont
  {Roy}}\ and\ \bibinfo {author} {\bibfnamefont {D.~E.}\ \bibnamefont
  {Logan}},\ }\bibfield  {title} {\bibinfo {title} {Fock-space correlations and
  the origins of many-body localization},\ }\href
  {https://doi.org/10.1103/PhysRevB.101.134202} {\bibfield  {journal} {\bibinfo
   {journal} {Phys. Rev. B}\ }\textbf {\bibinfo {volume} {101}},\ \bibinfo
  {pages} {134202} (\bibinfo {year} {2020})}\BibitemShut {NoStop}%
\bibitem [{\citenamefont {Prasad}\ and\ \citenamefont {Garg}(2021)}]{yp_nee}%
  \BibitemOpen
  \bibfield  {author} {\bibinfo {author} {\bibfnamefont {Y.}~\bibnamefont
  {Prasad}}\ and\ \bibinfo {author} {\bibfnamefont {A.}~\bibnamefont {Garg}},\
  }\bibfield  {title} {\bibinfo {title} {Many-body localization and enhanced
  nonergodic subdiffusive regime in the presence of random long-range
  interactions},\ }\href {https://doi.org/10.1103/PhysRevB.103.064203}
  {\bibfield  {journal} {\bibinfo  {journal} {Phys. Rev. B}\ }\textbf {\bibinfo
  {volume} {103}},\ \bibinfo {pages} {064203} (\bibinfo {year}
  {2021})}\BibitemShut {NoStop}%
\bibitem [{\citenamefont {Nag}\ and\ \citenamefont {Garg}(2019)}]{Nag2019}%
  \BibitemOpen
  \bibfield  {author} {\bibinfo {author} {\bibfnamefont {S.}~\bibnamefont
  {Nag}}\ and\ \bibinfo {author} {\bibfnamefont {A.}~\bibnamefont {Garg}},\
  }\bibfield  {title} {\bibinfo {title} {Many-body localization in the presence
  of long-range interactions and long-range hopping},\ }\href
  {https://doi.org/10.1103/PhysRevB.99.224203} {\bibfield  {journal} {\bibinfo
  {journal} {Phys. Rev. B}\ }\textbf {\bibinfo {volume} {99}},\ \bibinfo
  {pages} {224203} (\bibinfo {year} {2019})}\BibitemShut {NoStop}%
\bibitem [{\citenamefont {De~Roeck}\ and\ \citenamefont
  {Huveneers}(2017)}]{DeRoeck2017}%
  \BibitemOpen
  \bibfield  {author} {\bibinfo {author} {\bibfnamefont {W.}~\bibnamefont
  {De~Roeck}}\ and\ \bibinfo {author} {\bibfnamefont {F.}~\bibnamefont
  {Huveneers}},\ }\bibfield  {title} {\bibinfo {title} {Stability and
  instability towards delocalization in many-body localization systems},\
  }\href {https://doi.org/10.1103/PhysRevB.95.155129} {\bibfield  {journal}
  {\bibinfo  {journal} {Phys. Rev. B}\ }\textbf {\bibinfo {volume} {95}},\
  \bibinfo {pages} {155129} (\bibinfo {year} {2017})}\BibitemShut {NoStop}%
\bibitem [{\citenamefont {Thiery}\ \emph {et~al.}(2018)\citenamefont {Thiery},
  \citenamefont {Huveneers}, \citenamefont {M\"uller},\ and\ \citenamefont
  {De~Roeck}}]{avalanche2018}%
  \BibitemOpen
  \bibfield  {author} {\bibinfo {author} {\bibfnamefont {T.}~\bibnamefont
  {Thiery}}, \bibinfo {author} {\bibfnamefont {F.}~\bibnamefont {Huveneers}},
  \bibinfo {author} {\bibfnamefont {M.}~\bibnamefont {M\"uller}},\ and\
  \bibinfo {author} {\bibfnamefont {W.}~\bibnamefont {De~Roeck}},\ }\bibfield
  {title} {\bibinfo {title} {Many-body delocalization as a quantum avalanche},\
  }\href {https://doi.org/10.1103/PhysRevLett.121.140601} {\bibfield  {journal}
  {\bibinfo  {journal} {Phys. Rev. Lett.}\ }\textbf {\bibinfo {volume} {121}},\
  \bibinfo {pages} {140601} (\bibinfo {year} {2018})}\BibitemShut {NoStop}%
\bibitem [{\citenamefont {{\v S}untajs}\ \emph {et~al.}(2020)\citenamefont {{\v
  S}untajs}, \citenamefont {Bon{\v c}a}, \citenamefont {Prosen},\ and\
  \citenamefont {Vidmar}}]{Suntajs2020}%
  \BibitemOpen
  \bibfield  {author} {\bibinfo {author} {\bibfnamefont {J.}~\bibnamefont {{\v
  S}untajs}}, \bibinfo {author} {\bibfnamefont {J.}~\bibnamefont {Bon{\v c}a}},
  \bibinfo {author} {\bibfnamefont {T.}~\bibnamefont {Prosen}},\ and\ \bibinfo
  {author} {\bibfnamefont {L.}~\bibnamefont {Vidmar}},\ }\bibfield  {title}
  {\bibinfo {title} {Quantum chaos challenges many-body localization},\ }\href
  {https://doi.org/10.1103/PhysRevE.102.062144} {\bibfield  {journal} {\bibinfo
   {journal} {Phys. Rev. E}\ }\textbf {\bibinfo {volume} {102}},\ \bibinfo
  {pages} {062144} (\bibinfo {year} {2020})}\BibitemShut {NoStop}%
\bibitem [{\citenamefont {Tu}\ \emph {et~al.}(2023)\citenamefont {Tu},
  \citenamefont {Vu},\ and\ \citenamefont {Das~Sarma}}]{Tu2023}%
  \BibitemOpen
  \bibfield  {author} {\bibinfo {author} {\bibfnamefont {W.-H.}\ \bibnamefont
  {Tu}}, \bibinfo {author} {\bibfnamefont {D.}~\bibnamefont {Vu}},\ and\
  \bibinfo {author} {\bibfnamefont {S.}~\bibnamefont {Das~Sarma}},\ }\bibfield
  {title} {\bibinfo {title} {Avalanche stability transition in interacting
  quasiperiodic systems},\ }\href {https://doi.org/10.1103/PhysRevB.107.014203}
  {\bibfield  {journal} {\bibinfo  {journal} {Phys. Rev. B}\ }\textbf {\bibinfo
  {volume} {107}},\ \bibinfo {pages} {014203} (\bibinfo {year}
  {2023})}\BibitemShut {NoStop}%
\bibitem [{SM()}]{SM}%
  \BibitemOpen
  \href@noop {} {}\bibinfo {note} {See Supplemental Material at [URL will be
  inserted by publisher] for the range dependence of the forward-branching
  scale, the Hartree-fragmentation data (breaking probability, support
  threshold, island statistics), the island-diagonalization protocol and
  matching-law diagnostic, the exact resonant Fock graph at $L\le18$,
  island-level shell pair-count distributions, shell-averaging diagnostics,
  conditional resonance distributions, coarse-grained shell geometry, and exact
  Fock-space covariance identities, and the direct fixed-shell test of the
  pair-count scaling with its $h=100$ convergence control.}\BibitemShut {Stop}%
\end{thebibliography}%

\clearpage
\onecolumngrid
\begin{center}\textbf{\large Supplemental Material for ``Resonance statistics, Fock-space branching, and long-range pair networks in slowly varying interacting chains''}\end{center}
\vspace{2mm}\twocolumngrid
\setcounter{equation}{0}\setcounter{figure}{0}\setcounter{table}{0}\setcounter{section}{0}
\providecommand{\e}{\mathrm{e}}\providecommand{\ii}{\mathrm{i}}\providecommand{\dd}{\mathrm{d}}
\renewcommand{\thesection}{S\arabic{section}}\renewcommand{\theequation}{S\arabic{equation}}\renewcommand{\thefigure}{S\arabic{figure}}\renewcommand{\thetable}{S\Roman{table}}

\noindent This Supplemental Material collects the numerical protocols and supporting
calculations behind the main text: the range dependence of the forward-branching scale
(Sec.~\ref{sm:nc}); the Hartree-fragmentation data --- breaking probability, support threshold,
and island-size statistics (Sec.~\ref{sm:frag}); the island-diagonalization protocol and the
matching-law diagnostic (Sec.~\ref{sm:islmatch}); the exact resonant Fock graph at $L\le18$
(Sec.~\ref{sm:fock}); the island-level shell pair-count distributions (Sec.~\ref{sm:typ}); the
common-phase shell diagnostic (Sec.~\ref{sm:Cxr}); conditional resonance distributions and the
coarse-grained shell geometry (Secs.~\ref{sm:resdist} and~\ref{sm:ratio}); and exact Fock-space
covariance identities (Secs.~\ref{sm:cov} and~\ref{sm:covderiv}); and the direct fixed-shell
test of the pair-count scaling (Sec.~\ref{sm:lamtest}).  Equation
and figure numbers without an ``S'' refer to the main text.

\section{Range dependence of the forward-branching scale}
\label{sm:nc}

\begin{figure}[t]\centering
\includegraphics{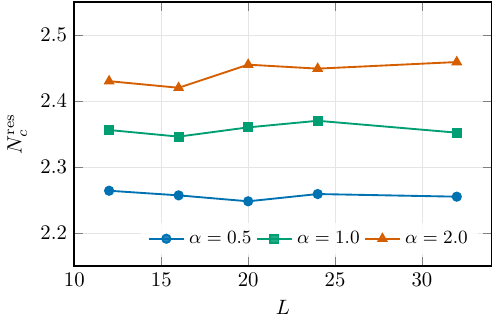}
\caption{Resonant-bond number at the branching criterion $B=1$, obtained from the same
evaluation of Eq.~(15) of the main text as Fig.~2 of the main text and therefore not independent of it.
The three interaction ranges give closely spaced curves that are essentially flat in $L$; the spread across
$\alpha$ is small and, if anything, increases slightly: $1.073$ at $L=12$ against $1.090$ at $L=32$.
The uniform-marginal baseline, Eq.~(17) of the main text, is
$\alpha$-independent.}
\label{sm:fignc}\end{figure}

Solving $B(L,\alpha,h)=1$ for the forward branching of Eq.~(15) of the main text gives the forward-branching resonant-bond
number $N_c^{\rm res}(L,\alpha)$ shown in Fig.~\ref{sm:fignc}.  Because it is obtained from the same evaluation
as Fig.~2 of the main text, it is a presentation of that calculation rather than
independent evidence for it.  Over $12\le L\le32$ and $0.5\le\alpha\le2$ the scale stays in the
range $N_c^{\rm res}\simeq2.25$--$2.46$ and is essentially independent of $L$; the ratio
$N_c^{\rm res}(\alpha{=}2)/N_c^{\rm res}(\alpha{=}0.5)$ is $1.073$ at $L=12$ and $1.090$ at $L=32$.  This weak trend is equally consistent with a small persistent range dependence; the data
constrain the spread to be $\order(10\%)$ without determining its limit.

\section{Hartree fragmentation: breaking probability, support threshold, and island statistics}
\label{sm:frag}

\subsection{Breaking probability and support threshold}

The breaking probability of the main text, $p_{\rm br}(V,\alpha)=\Pr(|\eta_i|>t_0)$, is evaluated
by direct sampling of the Hartree detuning
$\eta_i=\sum_{j\neq i,i+1}(U_{i+1,j}-U_{i,j})n_j$ over $2\times10^6$ independent draws of the
couplings and half-filled occupations, with the spatial sums cut at range $2000$ (the omitted tail
is negligible for the $\alpha$ values shown).  For $\alpha>1$ the support of $\eta_i$ is strictly
bounded, $\operatorname{ess\,sup}|\eta_i|=S(\alpha)V$ with $S(\alpha)=2[2\zeta(\alpha)-1]$, giving
the support threshold $V_*(\alpha)=t_0/S(\alpha)$ of the main text:

\begin{table}[h]
\caption{Breaking probability $p_{\rm br}$ and support threshold.  Entries $<5\times10^{-7}$ are
unresolved at the sampling depth; for $\alpha=1.25$, $V=0.25\,t_0$ the ratio $V/V_*\simeq4.1$, so
$p_{\rm br}>0$ strictly, yet $\xi_{\rm frag}=p_{\rm br}^{-1}\gtrsim10^6$ --- fragmentation is
mathematically allowed but operationally absent, which is why the controlled-domain condition of
the main text is $w_0\gg\xi_{\rm frag}$ and not merely $V>V_*$.}
\label{smtab:pbr}
\begin{ruledtabular}
\begin{tabular}{ccccccc}
$\alpha$ & $S(\alpha)$ & $V_*/t_0$ & \multicolumn{4}{c}{$p_{\rm br}$ at $V/t_0=$}\\
 & & & $0.25$ & $0.5$ & $1$ & $2$\\
\colrule
$1.25$ & $16.38$ & $0.061$ & $<5{\times}10^{-7}$ & $0.0101$ & $0.180$ & $0.486$\\
$1.50$ & $8.45$  & $0.118$ & $<5{\times}10^{-7}$ & $0.0051$ & $0.146$ & $0.442$\\
$2.00$ & $4.58$  & $0.218$ & $<5{\times}10^{-7}$ & $0.0021$ & $0.113$ & $0.402$\\
\end{tabular}
\end{ruledtabular}
\end{table}

An independent measurement of the same quantity inside the turning-point windows of the island
ensemble below agrees to $\sim10\%$ (e.g.\ $0.0058$, $0.147$, and $0.437$ at $V/t_0=0.5$, $1$,
and $2$, respectively, at $\alpha=1.5$), so
$\xi_{\rm frag}=p_{\rm br}^{-1}$ therefore provides a quantitative characteristic
fragmentation scale, tracking the measured snapshot island scale at the $10$--$20\%$ level.

\subsection{Island-size statistics}

Islands are maximal runs of consecutive bonds with $|\Delta_i|<t_0$, $\Delta_i$ the full detuning.
The ensemble uses chains of $W=4\times10^4$ bonds centered at $x_0=10^6$, the exact potential at
$n=1/2$, $\beta=(\sqrt5-1)/2$, random couplings with range cutoff $300$, random half-filled
occupations, and a fresh global phase per realization ($100$ realizations per parameter set).  The
field $h$ is chosen to set the bare run width $w_0=2t_0/(h\psi'^2)$; increasing $w_0$ at fixed $V$
is equivalent to moving outward along the pair-network trajectory of the main text.
Table~\ref{smtab:isl} gives the bond-weighted mean island size
$\langle\ell\rangle_B=\langle\ell^2\rangle/\langle\ell\rangle$; these data are plotted in
Fig.~4 of the main text.  In every parameter set the resonant-bond density agrees with the bare
density to within $0.3\%$, and the count-weighted mean $\langle\ell\rangle_{\rm isl}$ is much
smaller than $\langle\ell\rangle_B$ (e.g.\ $2.8$ against $7.6$ at $\alpha=2$, $V=t_0$,
$w_0=100$): the island-size distribution is broad, and the tail carries the bond weight.

\begin{table}[h]
\caption{Bond-weighted mean island size $\langle\ell\rangle_B$ against bare run width $w_0$.
Saturation (boldface rows) sets in when $w_0\gg\xi_{\rm frag}$, at a level consistent with
$\xi_{\rm frag}$: $6.8$ and $2.3$ at $\alpha=1.5$, and $8.8$ and $2.5$ at $\alpha=2$, for
$V/t_0=1$ and $2$, respectively.}
\label{smtab:isl}
\begin{ruledtabular}
\begin{tabular}{cccccc}
$\alpha$ & $V/t_0$ & $w_0{=}10$ & $30$ & $100$ & $300$\\
\colrule
$1.5$ & $0.25$ & $9.40$ & $24.3$ & $64.7$ & $155.6$\\
$1.5$ & $0.5$  & $7.54$ & $15.3$ & $28.8$ & $44.8$\\
$1.5$ & $1$    & $4.49$ & $5.83$ & $6.27$ & $6.25$\\
$1.5$ & $2$    & $2.35$ & $2.43$ & $2.45$ & $2.33$\\
$2$   & $0.25$ & $9.59$ & $25.1$ & $68.9$ & $167.9$\\
$2$   & $0.5$  & $7.97$ & $17.0$ & $33.9$ & $59.1$\\
$2$   & $1$    & $5.00$ & $6.66$ & $7.60$ & $7.65$\\
$2$   & $2$    & $2.59$ & $2.69$ & $2.75$ & $2.63$\\
\end{tabular}
\end{ruledtabular}
\end{table}

\subsection{Boundary stability of the island partition}
\label{sm:heal}

Because internal hopping changes occupations, and occupations feed the Hartree detunings of the
boundary bonds, the snapshot partition must be checked against the islands' own dynamics.  For
every cut bond between two adjacent islands (both $\ell\le6$; larger neighbors are skipped and
counted, $2$--$27\%$ of cuts depending on $w_0$), all internal Fock configurations of
\emph{both} adjacent islands (fixed particle number each) are enumerated; for each configuration
pair the cut bond's spectator-only Hartree detuning
$\Delta_{\rm cut}=\Delta_{\rm cut}^{\rm ref}+\sum_{j\in I\cup J}g_j\,(n_j-n_j^{\rm ref})$, with
$g_j=U_{c+1,j}-U_{c,j}$ and $j\neq c,c+1$, is recomputed, and the cut is \emph{healable} if any
configuration makes it active ($|\Delta_{\rm cut}|<t_0$ and flippable).  Two disorder
realizations $\times$ six phases of $4\times10^4$-bond chains per parameter set.

Results ($w_0=10/30/100/300$): the per-cut healing probability is $w_0$-independent ---
$P_{\rm heal}=0.65/0.61/0.58/0.57$ at $(\alpha,V)=(2,t_0)$; $0.58/0.53/0.53/0.53$ at
$(1.5,t_0)$; $0.41/0.45/0.45/0.44$ at $(2,2t_0)$; $0.36/0.41/0.40/0.39$ at $(1.5,2t_0)$ --- the
per-configuration healed fraction is small and flat ($f_{\rm heal}=0.09$--$0.18$), and merging
islands across every healable cut (the most permissive rule: one configuration suffices)
saturates the bond-weighted merged-cluster size: $7.1/9.2/9.8/10.2$ bonds at $(2,t_0)$ against
the unhealed $\langle\ell\rangle_B\simeq7.6$, $6.8/8.2/9.3/9.1$ at $(1.5,t_0)$, and
$4$--$5$ at $V=2t_0$.  Healing re-draws boundaries locally and renormalizes the island scale by
an $\order(1)$ factor.

\emph{Recursive closure.}  A merged cluster has a larger internal configuration space than its
constituents, so it can in principle reopen further cuts.  We therefore iterate: after merging,
the remaining cuts are re-evaluated against the \emph{merged} clusters and merged again, to a
fixed point.  To avoid any enumeration gap, each cluster is represented at a cut by the
$w_{\rm cut}\le8$ sites nearest that cut.  We stress that this truncation is \emph{not} justified
by a decay bound: because the $V_{ij}$ are independent, $g_j=V_{c+1,j}|j-c-1|^{-\alpha}
-V_{c,j}|j-c|^{-\alpha}$ has no derivative cancellation, and realization by realization
$|g_j|=\order(|j-c|^{-\alpha})$, not $\order(|j-c|^{-\alpha-1})$ --- the smooth-coupling estimate
would contradict the very multipole mechanism of Sec.~V of the main text.  The worst-case remote
influence is correspondingly \emph{not} small ($0.22\,t_0$ at $(\alpha,V)=(2,t_0)$ and
$2.65\,t_0$ at $(1.5,2t_0)$, exceeding the window).  What is small is the ensemble scale of the
omitted tail, rms $0.009\,t_0$ and $0.064\,t_0$ respectively, and the truncation is validated
\emph{empirically} by window-size convergence: with no analytic tail added at all,
$w_{\rm cut}=4,8,10$ give the same observables to better than $1.5\%$
(Table~\ref{smtab:win}).

\begin{table}[h]
\caption{Window-size convergence of the recursive-healing fixed point (no analytic tail added).
The truncation at $w_{\rm cut}$ nearest sites is validated by convergence, not by a bound.}
\label{smtab:win}
\begin{ruledtabular}
\begin{tabular}{lcccc}
$(\alpha,V/t_0)$, $w_0$ & quantity & $w_{\rm cut}{=}4$ & $8$ & $10$\\
\colrule
$(2,1)$, $w_0{=}30$   & $A_{\rm isl}$ & $0.1476$ & $0.1455$ & $0.1454$\\
$(2,1)$, $w_0{=}30$   & $\langle\ell\rangle_{\rm count}$ & $4.81$ & $4.90$ & $4.92$\\
$(2,1)$, $w_0{=}300$  & $A_{\rm isl}$ & $0.1518$ & $0.1509$ & $0.1507$\\
$(1.5,2)$, $w_0{=}30$ & $A_{\rm isl}$ & $0.3002$ & $0.2978$ & $0.2980$\\
$(1.5,2)$, $w_0{=}300$& $A_{\rm isl}$ & $0.3224$ & $0.3211$ & $0.3209$\\
\end{tabular}
\end{ruledtabular}
\end{table}  A cluster that fits inside
the window is enumerated at its own fixed particle number; a window that is part of a larger
cluster may exchange one particle with the remainder.  A cut reopens if \emph{any} admissible
configuration pair returns it to the window, with no energetic accessibility imposed --- the
rule therefore over-counts reopening.  Validation: the first-pass reopening probabilities
reproduce the exact fixed-particle-number values above ($0.66/0.61/0.59$ versus
$0.65/0.61/0.58$ at $(2,t_0)$).

The recursion converges in $4$--$7$ passes.  The \emph{bond-weighted} cluster mean then drifts
upward with $w_0$ wherever fragmentation is weak, but this is a tail effect: the
\emph{count}-weighted mean and the active-object density $A_{\rm isl}=\rho_{\rm act}/p$ (measured
as the number of active clusters per resonant bond) saturate (Table~\ref{smtab:heal}), with
cluster medians of $1$--$2$ bonds throughout.  Three statements of different strength must be
kept apart here.  The \emph{object-density} scaling is controlled: $\rho_{\rm act}$, not
$\langle\ell\rangle_B$, enters the pair density, and it is $w_0$-independent.  The
\emph{integrated} transition weight per unit length is bounded by the sum rule of the main text.
The \emph{low-frequency} matching density of the asymptotic rare tail is not bounded by either
of these; along the accessible cluster sizes it shows no growth, and its uniform boundedness
along the tail is the one assumption of the construction (main text, Sec.~V\,D).  Within that
accessible range the tail is weakly coupled: the doublet transition norm falls with cluster size: at $(\alpha,V)=(2,t_0)$,
$\langle\lVert m\rVert\rangle=0.672$, $0.538$, $0.459$, $0.420$ for $\ell=1$--$4$, falling to
$0.319$, $0.283$ at $\ell=9$, $12$ (at $(1.5,2t_0)$: $0.679$, $0.528$, $0.462$, $0.421$ for
$\ell=1$--$4$, and $0.296$ at $\ell=9$), so the rare large clusters are also the most weakly
coupled.

\emph{The surviving single-bond fraction.}  The reconstruction leaves a finite population of
isolated wing bonds at every $w_0$.  Measuring the fraction $f_1$ of \emph{active} objects that
are single bonds, for the raw partition and at the recursive fixed point:
$f_1^{\rm isl}=0.352,0.365,0.376,0.403$ and $f_1^{\rm fix}=0.320,0.427,0.461,0.502$ at
$(\alpha,V)=(2,t_0)$ for $w_0=10,30,100,300$; $f_1^{\rm isl}=0.576,0.545,0.567,0.599$ and
$f_1^{\rm fix}=0.536,0.544,0.554,0.592$ at $(1.5,2t_0)$.  There is no downward trend: $f_1$ is
flat or rising, consistent with the wing argument of Sec.~V of the main text, in which the wings
and the fragmented interior both scale with $w_0$.  This is what keeps $b_{\rm all}=f_1^2b_1>0$
in the ensemble matching law of the main text.

\emph{Sector weights on the physical shell.}  Because the comb correlates objects, the wing--wing
pair fraction on a mesoscopic shell need not equal $f_1^2$.  Measuring it directly on the shells
used for the typicality scans (active objects only, center window at $x_0=4000$, shell at
$\bar R=1800$, $V=t_0$): $f_{11}^{\rm shell}=0.1779$ against $f_1(x_0)f_1(x_0-\bar R)=0.1682$ at
$\alpha=1.5$ (ratio $1.058$; $5.1\times10^4$ pairs), and $0.1468$ against $0.1430$ at $\alpha=2$
(ratio $1.027$; $3.9\times10^4$ pairs).  The factorized sector weights are therefore accurate to
a few percent, and any residual is absorbed into $b_{\rm all}$.

\begin{table*}[t]
\caption{Recursive boundary healing: fixed point of the permissive merge rule against bare run
width.  $\langle\ell\rangle_B$ is tail dominated and drifts at $V=t_0$; the count-weighted mean
and the active-object density $A_{\rm isl}$, which is what enters $\rho_{\rm act}=A_{\rm isl}p$,
are $w_0$-independent.  Unhealed-island values are given for reference.}
\label{smtab:heal}
\begin{ruledtabular}
\begin{tabular}{lccccc}
$(\alpha,V/t_0)$ & quantity & $w_0{=}10$ & $30$ & $100$ & $300$\\
\colrule
$(2,1)$ & islands: $\langle\ell\rangle_{\rm count}$ / $A_{\rm isl}$
 & $2.8$/$0.254$ & $2.8$/$0.246$ & $2.8$/$0.239$ & $2.7$/$0.247$\\
$(2,1)$ & fixed pt: $\langle\ell\rangle_B$ & $8.8$ & $18.3$ & $33.1$ & $47.5$\\
$(2,1)$ & fixed pt: $\langle\ell\rangle_{\rm count}$ / $A_{\rm isl}$
 & $4.6$/$0.169$ & $4.9$/$0.145$ & $4.8$/$0.142$ & $4.6$/$0.150$\\
\colrule
$(1.5,1)$ & fixed pt: $\langle\ell\rangle_B$ & $8.6$ & $17.1$ & $28.0$ & $36.0$\\
$(1.5,1)$ & fixed pt: $\langle\ell\rangle_{\rm count}$ / $A_{\rm isl}$
 & $4.3$/$0.184$ & $4.5$/$0.161$ & $4.3$/$0.162$ & $4.2$/$0.165$\\
\colrule
$(2,2)$ & fixed pt: $\langle\ell\rangle_B$ & $5.6$ & $7.2$ & $7.9$ & $7.5$\\
$(2,2)$ & fixed pt: $\langle\ell\rangle_{\rm count}$ / $A_{\rm isl}$
 & $2.5$/$0.284$ & $2.4$/$0.279$ & $2.4$/$0.274$ & $2.2$/$0.289$\\
\colrule
$(1.5,2)$ & fixed pt: $\langle\ell\rangle_B$ & $4.9$ & $6.3$ & $6.4$ & $6.1$\\
$(1.5,2)$ & fixed pt: $\langle\ell\rangle_{\rm count}$ / $A_{\rm isl}$
 & $2.1$/$0.301$ & $2.2$/$0.294$ & $2.1$/$0.307$ & $2.0$/$0.317$\\
\end{tabular}
\end{ruledtabular}
\end{table*}

\section{Island diagonalization and the matching-law diagnostic}
\label{sm:islmatch}

\subsection{Protocol}

Each island harvested from the ensemble above (at $w_0=100$) is diagonalized exactly in its own
particle-number sector: an island of $\ell$ bonds is $\ell+1$ sites carrying the sampled particle
number $m$; the local Hamiltonian contains the hopping $-t_0$, the exact onsite energies
$h_i$ plus the frozen exterior Hartree field of the sampled configuration, and the exact internal
couplings $U_{ij}$ of the island sites.  Islands with $m=0$ or $m=\ell+1$ are frozen (no internal
doublet) and are excluded --- they are roughly half of all islands.  Dense diagonalization is used
for $\ell\le10$ and sparse Lanczos (two lowest states) for $\ell=11$--$20$, Hilbert-space
dimensions up to $352\,716$; islands beyond $\ell=20$ carry $5.2\%/7.4\%$ of the resonant-bond
weight at $V=t_0$ ($\alpha=1.5/2$) and $\le0.1\%$ at $V=2t_0$, and are excluded.  Each island
yields its splitting $\Omega=E_1-E_0$ and transition-density norm
$\lVert m\rVert=[\sum_a|\langle E_0|n_a|E_1\rangle|^2]^{1/2}$.

The splitting distribution of genuine islands is regular: for $\ell\ge3$, the median of $\Omega$
is $1.02$--$1.13\,t_0$, a fraction $0.37$--$0.49$ lies below $t_0$, and over $98\%$ lies below
$2t_0$ --- the two-level edge at $2t_0$ is absent.  For $\ell=1$ objects the edge is exact
($\Omega_{\min}=2t_0$), but the conditional detuning of the \emph{surviving} isolated bonds is
reshaped by fragmentation (they live preferentially in the run wings, biased toward
$|\Delta|\to t_0$), which reduces their matching-log slope below the ideal benchmark.

\subsection{Matching-law diagnostic}

The matching probability over island pairs,
$P^{\rm match}(q)=\Pr(|\Omega_A-\Omega_B|<2qt_0|\zeta|)$ with
$\zeta$ the coupling amplitude carrying the island form factors, is evaluated over $2\times10^7$
independently sampled pairs; the diagnostic $D(w)$ of the main text uses the splittings alone.
Fitting $P/q=a+b\ln(1/q)$ over $q\le10^{-2}$:

\begin{table}[h]
\caption{Matching-log slopes $b$ by pair class.  The full-ensemble slope obeys
$b_{\rm all}\simeq f_{\ell=1}^2\,b_{\ell=1}$, identifying the residual logarithm as the
contribution of the surviving single-bond fraction $f_{\ell=1}$.}
\label{smtab:slopes}
\begin{ruledtabular}
\begin{tabular}{ccccccc}
$\alpha$ & $V/t_0$ & $f_{\ell=1}$ & $b_{\ell=1}$ & $b_{\ell\ge3}$ & $b_{\rm all}$ & $f_{\ell=1}^2b_{\ell=1}$\\
\colrule
$1.5$ & $1$ & $0.44$ & $0.60$ & $-0.02$ & $0.11$ & $0.11$\\
$2$   & $1$ & $0.41$ & $0.28$ & $0.00$  & $0.048$ & $0.047$\\
$1.5$ & $2$ & $0.56$ & $0.75$ & $0.01$  & $0.24$ & $0.24$\\
$2$   & $2$ & $0.52$ & $0.74$ & $0.00$  & $0.20$ & $0.20$\\
\end{tabular}
\end{ruledtabular}
\end{table}

\subsection{Matching law on the recursively healed ensemble}

The diagonalization above is performed on the snapshot island partition.  To verify that the
sector structure survives the recursive boundary healing of Sec.~\ref{sm:heal}, the healed
fixed-point clusters themselves were diagonalized for two representative parameter sets at
$w_0=30$ and $100$, with the same protocol (dense diagonalization, here for $\ell\le14$).  The
healed pool differs from the snapshot pool in exactly the expected way --- the bond-weighted tail
is broader, so a larger fraction of resonant-bond weight sits in clusters beyond the
diagonalization cap ($0.62$--$0.66$ at $(2,t_0)$ versus $0.07$ for $\ell>20$ in the snapshot;
$0.12$--$0.13$ at $(1.5,2t_0)$) --- but the matching statistics are unchanged in structure
(Table~\ref{smtab:healmatch}): the $\ell\ge2$ sector is slope-free, $D(w)$ shows no systematic low-frequency growth
over three decades, and the full-ensemble slope obeys $b_{\rm all}=f_1^2b_{\ell=1}$ to within $3\%$ with
$f_1$ the healed single-bond fraction, which at $w_0=100$, $(2,t_0)$ is $0.47$ against the
$0.461$ of Table~\ref{smtab:heal}.  The $b_{\ell=1}$ values are somewhat below the snapshot ones
(the surviving singletons after healing sit still closer to the window edge), so the residual
logarithm is if anything weaker, but it is present at the fixed point with the same sector
origin.  The excluded large clusters are the rare tail whose low-frequency matching density is
the stated assumption of the construction; within $\ell\le14$ it shows no growth.

\begin{table*}[t]
\caption{Matching law recomputed on the recursively healed fixed-point ensemble.  Healed
clusters of $\ell\le14$ bonds are diagonalized ($f_1$ is the single-bond fraction of all active
healed clusters; the fraction of resonant-bond weight in excluded clusters $\ell>14$ is listed);
$D(w)$ is the pair-density diagnostic of the $\ell\ge3$ sector at $w=10^{-4},10^{-3},10^{-2}$ in
units of $t_0^{-1}$; slopes fitted over $q\le10^{-2}$; $48$ chains of $4\times10^4$ bonds per row.}
\label{smtab:healmatch}
\begin{ruledtabular}
\begin{tabular}{ccccccccc}
$(\alpha,V/t_0)$ & $w_0$ & excl. & $f_1$ & $b_{\ell=1}$ & $b_{\ell\ge2}$ & $b_{\rm all}$ & $f_1^2b_{\ell=1}$ & $D(w)$\\
\colrule
$(2,1)$   & $30$  & $0.66$ & $0.44$ & $0.23$ & $-0.007$ & $0.069$ & $0.071$ & $0.67,0.83,0.79$\\
$(2,1)$   & $100$ & $0.62$ & $0.47$ & $0.19$ & $-0.001$ & $0.055$ & $0.055$ & $0.87,0.90,0.88$\\
$(1.5,2)$ & $30$  & $0.12$ & $0.53$ & $0.92$ & $-0.002$ & $0.264$ & $0.266$ & $0.62,0.62,0.62$\\
$(1.5,2)$ & $100$ & $0.13$ & $0.54$ & $0.78$ & $0.002$  & $0.237$ & $0.235$ & $0.66,0.62,0.62$\\
\end{tabular}
\end{ruledtabular}
\end{table*}

The $D(w)$ saturation quoted in the main text was additionally verified for the $\ell=11$--$20$
sector separately (flat at $D\simeq1.7$--$2.1$, larger than the small-island value but bounded)
and, at $(\alpha,V)=(1.5,2t_0)$, by exact all-pairs counting over $8868$ islands:
$D=0.739,0.743,0.737,0.738,0.738,0.739,0.735$ at
$w/t_0=10^{-4},3{\times}10^{-4},10^{-3},3{\times}10^{-3},10^{-2},3{\times}10^{-2},10^{-1}$, with
statistical errors below $1\%$; $D$ is quoted in units of $t_0^{-1}$ throughout.

\subsection{Two-bond sector and the all-transition check}

Two diagnostics close the island ensemble.  \emph{(i) The $\ell=2$ sector.}  Two-bond islands
make up $f_{\ell=2}=0.234$ and $0.245$ of the population at $(\alpha,V)=(2,t_0)$ and
$(1.5,2t_0)$; restricting the matching pairs to $\ell_A=\ell_B=2$ gives slopes
$b_{\ell=2}=-0.008$ and $-0.005$ ($1.09\times10^6$ and $1.20\times10^6$ pairs), zero within
noise, and the combined genuine-island class satisfies $b_{\ell\ge2}=-0.001$ in both cases: the
matching logarithm is carried exclusively by $\ell=1$.  \emph{(ii) All transitions.}  Because
the parent problem is a high-energy one, the doublet-based diagnostic was repeated using the
island's \emph{full} many-body spectrum: dense spectra for every $\ell\le6$ island plus a
$20\%$ subsample of $\ell=7$--$10$ ($10\,852$ and $14\,546$ islands at the two parameter sets),
every eigenstate pair $(r,s)$ with local transition density
$W_{rs}=\sum_a|\langle r|n_a|s\rangle|^2>10^{-10}$ retained ($7.0\times10^6$ and
$1.7\times10^6$ transitions), and infinite-temperature weighting: the initial eigenstate is
uniform, so a channel is sampled with probability proportional to $W_{rs}$.  Sampling
$6\times10^6$ cross-island transition pairs gives, at
$w/t_0=10^{-4},3\times10^{-4},10^{-3},3\times10^{-3},10^{-2},3\times10^{-2},10^{-1}$,
$D=0.274$, $0.270$, $0.272$, $0.272$, $0.273$, $0.272$, $0.270$ at $(\alpha,V)=(2,t_0)$ and
$D=0.286$, $0.292$, $0.289$, $0.283$, $0.286$, $0.285$, $0.284$ at $(1.5,2t_0)$, with
small-$w$ slopes $+0.0001$ and $+0.0006$ and weighted-median transition frequencies
$1.67\,t_0$ and $1.73\,t_0$.  The all-transition pair density is flat over three decades: no
singularity appears in the weighted infinite-temperature all-transition ensemble, and the
lowest doublet used in the main analysis is representative, not special.

\section{Exact resonant Fock graph at $L\le18$}
\label{sm:fock}

For each realization $(\phi,\{V_{ij}\})$ of the ensemble of the main-text Appendix on numerical
protocol, all $\binom{L}{L/2}$ half-filled configurations are enumerated, the full detuning
$\Delta_a(S)$ is evaluated for every bond of every configuration, and every active edge (flippable
and $|\Delta_a|<t_0$) is inserted; connected components are found exactly, and the forward
branching $B$ is evaluated from the identical samples.  Realization counts are
$300/200/100/50$ for $L=12/14/16/18$; $V=t_0$; the field grid is that of the main text.

The $B=1$ crossings are at $N_{\rm res}=2.41/2.41/2.42/2.42$ ($\alpha=0.5$, $L=12/14/16/18$),
$2.51/2.54/2.52/2.55$ ($\alpha=1$), and $2.60/2.61/2.62/2.65$ ($\alpha=2$).  At these crossings
the largest component holds $13$--$53$ vertices and the mean finite-cluster size is
$\chi_{\rm cl}=3.3$--$5.0$, essentially independent of $L$.  Table~\ref{smtab:fock} shows the
percolation diagnostic at $\alpha=0.5$ and $2$; the $\alpha=1$ data are in Fig.~3 of the main
text.  At every fixed $N_{\rm res}$ studied the diagnostic falls with $L$.  The finite-size
graphs are far from any giant-component regime and are incompatible with ordinary percolation at
fixed $\order(1)$ resonance count over the accessible sizes; extrapolating to strict
thermodynamic subcriticality would exceed what $L\le18$ can establish.  The same trend is
visible without reference to a percolation scaling form: $S_{\rm max}/N_F$ itself falls from
$3.5\%$ ($1.2\%$) at $L=12$ to $0.11\%$ ($0.03\%$) at $L=18$ at the $B=1$ crossing for
$\alpha=0.5$ ($\alpha=2$).

\begin{table}[h]
\caption{Percolation diagnostic $S_{\rm max}/N_F^{2/3}$ of the exact resonant Fock graph
($V=t_0$).  Rows are the field grid $f=h/h_0(L)$; $N_{\rm res}$ varies weakly with $L$ at fixed
$f$ (values quoted for $L=18$).}
\label{smtab:fock}
\begin{ruledtabular}
\begin{tabular}{ccccccc}
$\alpha$ & $f$ & $N_{\rm res}$ & $L{=}12$ & $14$ & $16$ & $18$\\
\colrule
$0.5$ & $0.55$ & $4.75$ & $2.63$ & $2.34$ & $2.20$ & $1.52$\\
$0.5$ & $0.70$ & $3.64$ & $1.15$ & $0.77$ & $0.56$ & $0.27$\\
$0.5$ & $0.85$ & $2.87$ & $0.58$ & $0.33$ & $0.20$ & $0.09$\\
$0.5$ & $1.00$ & $2.36$ & $0.34$ & $0.17$ & $0.09$ & $0.04$\\
$2$   & $0.55$ & $4.69$ & $0.92$ & $0.43$ & $0.17$ & $0.07$\\
$2$   & $0.70$ & $3.48$ & $0.36$ & $0.17$ & $0.06$ & $0.03$\\
$2$   & $0.85$ & $2.81$ & $0.19$ & $0.09$ & $0.03$ & $0.01$\\
$2$   & $1.00$ & $2.39$ & $0.12$ & $0.06$ & $0.02$ & $0.01$\\
\end{tabular}
\end{ruledtabular}
\end{table}

\section{Island-level shell pair-count distributions}
\label{sm:typ}

The typicality statistics of the main text are evaluated at the level of the physical objects.  On
chains of $L=8000$ sites at $h/t_0=150$ (so $i_*\simeq8.5\times10^4\gg L$), with the full detuning
at $V=t_0$, $\alpha=1$, every maximal resonant run in the region of interest is identified,
frozen runs are excluded, and each remaining island is diagonalized exactly (dense to $\ell=10$,
sparse to $\ell=14$; longer runs are $\ll1\%$ and skipped).  Pairs are counted between islands
centered in $x_0=4000\pm60$ and islands in the one-sided shell $R=1800\pm2\ell_{\rm comb}$, with
the matching condition $|\Omega_A-\Omega_B|<2\lambda(V/R^\alpha)|g|\,\lVert m_A\rVert\lVert
m_B\rVert$, $g$ a unit-variance-scale Gaussian amplitude.  The scale factor $\lambda$ is an
auxiliary rescaling parameter used to compare the distributions at matched mean --- it absorbs
the $\order(1)$ matching constants \emph{and} compensates the fixed-shell $R^{-\alpha}$ factor
between different $\alpha$, and is not interpreted as a physical coupling prefactor --- tuned so
that $\langle N_{\rm pair}\rangle$ sweeps through the takeoff value; $600$ realizations.  The full
scan, including the small-$\lambda$ rows, is:

\begin{table}[h]
\caption{Island-level shell pair-count distribution ($\alpha=1$, $V=t_0$).  $P^{\rm Poi}$ and
$r^{\rm Poi}$ denote the Poisson reference values $1-\e^{-\langle N\rangle}$ and
$1/\langle N\rangle$.}
\label{smtab:typ}
\begin{ruledtabular}
\begin{tabular}{cccccc}
$\langle N\rangle$ & median & $P(N{\ge}1)$ & $P^{\rm Poi}$ & ${\rm Var}/\langle N\rangle^2$ & $r^{\rm Poi}$\\
\colrule
$0.23$ & $0$ & $0.19$ & $0.21$ & $5.7$ & $4.3$\\
$0.42$ & $0$ & $0.30$ & $0.34$ & $3.4$ & $2.4$\\
$0.78$ & $0$ & $0.46$ & $0.54$ & $2.0$ & $1.3$\\
$1.41$ & $1$ & $0.64$ & $0.76$ & $1.24$ & $0.71$\\
$2.52$ & $2$ & $0.80$ & $0.92$ & $0.83$ & $0.40$\\
$3.56$ & $3$ & $0.87$ & $0.97$ & $0.66$ & $0.28$\\
\end{tabular}
\end{ruledtabular}
\end{table}

Down to $\langle N\rangle\simeq0.2$ the probability of at least one matched pair tracks the
Poisson reference to within $\simeq15\%$, and the variance exceeds it by less than a factor of
two: the mean is not carried by rare comb coincidences.

The same protocol, repeated in the clean isolated-bond regime ($\alpha=0.5$) and in the clean
fragmented regime ($\alpha=1.5$) at $V=t_0$ --- with the coupling scale $\lambda$ rescaled to
compensate the shell factor $R^{-\alpha}$ and matched so that $\langle N\rangle\simeq1.4$ ---
gives closely comparable distributions
(Table~\ref{smtab:typ2}): takeoff typicality is not a property of the marginal point
$\alpha=2n$; the same typical-not-rare behavior holds on both sides of the reconstruction.

The $\alpha$ labels above are the asymptotic classification; at these parameters the bare runs
have $w_0(x_0)\simeq14$.  Repeating the $\alpha=0.5$ scan in the one--two-bond regime ---
$h=1100\,t_0$, so $w_0(x_0)\simeq1.9$ and the harvested objects are one--two-bond
($2400$ realizations, center window $\pm150$) --- gives the same structure:
$\langle N\rangle=0.18/0.33/0.58/1.00/1.67$ at $\lambda=0.3/0.6/1.2/2.4/4.8$ (medians
$0/0/0/1/1$), and at matched mean $\langle N\rangle=1.41$: median $1$, $P(N\ge1)=0.61$
$[$Poisson $0.76]$, ${\rm Var}/\langle N\rangle^2=1.44$ $[0.71]$.  Typicality is not an
artifact of the multi-bond parameters of the main scan.

The control above uses a Gaussian amplitude $g$.  Repeating the identical scan with the
matching window built from the \emph{physical} conditional island coupling,
$|\Omega_A-\Omega_B|<2\lambda(V/R^\alpha)|\Gamma_{AB}|$ with
$\Gamma_{AB}=\sum_{ab}X_{ab}\,m^A_am^B_b$ and $X_{ab}$ independent uniform on $[-1,1]$ (leading
order in $R$), gives closely comparable distributions at matched mean
(Table~\ref{smtab:typG}): the typicality conclusion does not depend on the surrogate amplitude.

\begin{table}[h]
\caption{Matched-mean shell pair-count distributions with the physical island form-factor
coupling $\Gamma_{AB}$ replacing the Gaussian amplitude ($V=t_0$, $600$ realizations per row).
Reference values as in Table~\ref{smtab:typ}.}
\label{smtab:typG}
\begin{ruledtabular}
\begin{tabular}{cccccc}
$\alpha$ & $\langle N\rangle$ & median & $P(N{\ge}1)$ & $P^{\rm Poi}$ & ${\rm Var}/\langle N\rangle^2$\\
\colrule
$0.5$ & $1.43$ & $1$ & $0.66$ & $0.76$ & $1.11$\\
$1.0$ & $1.39$ & $1$ & $0.64$ & $0.75$ & $1.41$\\
$1.5$ & $1.43$ & $1$ & $0.65$ & $0.76$ & $1.16$\\
\end{tabular}
\end{ruledtabular}
\end{table}

\begin{table}[h]
\caption{Island-level shell pair-count distribution at matched mean across the reconstruction
($V=t_0$, $600$ realizations each; $n=1/2$, so $\alpha=0.5$ is the isolated-bond
regime, $\alpha=1$ the marginal point, and $\alpha=1.5$ the fragmented regime).  Reference
values as in Table~\ref{smtab:typ}.}
\label{smtab:typ2}
\begin{ruledtabular}
\begin{tabular}{ccccccc}
$\alpha$ & $\langle N\rangle$ & median & $P(N{\ge}1)$ & $P^{\rm Poi}$ & ${\rm Var}/\langle N\rangle^2$ & $r^{\rm Poi}$\\
\colrule
$0.5$ & $1.44$ & $1$ & $0.69$ & $0.76$ & $1.13$ & $0.70$\\
$1.0$ & $1.41$ & $1$ & $0.64$ & $0.76$ & $1.24$ & $0.71$\\
$1.5$ & $1.38$ & $1$ & $0.63$ & $0.75$ & $1.23$ & $0.73$\\
\end{tabular}
\end{ruledtabular}
\end{table}

\section{Shell-averaged common-phase correlation factor}
\label{sm:Cxr}

The joint average Eq.~(46) of the main text is \emph{not} factorized in Sec.~VIII of the main text, so
the correlation factor $\mathcal C(x,R)$ --- the ratio of the joint resonance probability at bonds
$x-R$ and $x$ to the product of its marginals --- is not an ingredient of the pair density.  It is an
\emph{independent} diagnostic of the same equidistribution hypothesis, and that is what we evaluate
here.  The pointwise bound is exact; under uniform-offset equidistribution the shell average is
exactly unity.

Writing $\delta_i^{(0)}=-A_i^{\rm d}\sin\psi_i$ with
$\psi_i=2\pi\beta i^n+\tfrac12\Delta\theta_i+\phi$, the resonance condition $|\delta_i^{(0)}|<t_0$ is
\begin{equation}
 \psi_i \bmod \pi\in(-\epsilon_i,\epsilon_i),\qquad
 \epsilon_i=\arcsin\!\Big[\min\Big(1,\frac{t_0}{A_i^{\rm d}}\Big)\Big],
\end{equation}
so $p_i^{(0)}=2\epsilon_i/\pi$, consistent with Eq.~(5) of the main text.  Two bonds therefore
define two arcs on a circle of circumference $\pi$ whose relative offset
$\Delta_{ij}=(\psi_j-\psi_i)\bmod\pi$ is independent of $\phi$.  The joint probability is the exact
arc overlap, and
\begin{equation}
 \mathcal C(x,R)=\frac{P^{\rm res}_{x-R,\,x}}{p_{x-R}^{(0)}p_{x}^{(0)}}
 \le\frac{1}{\max\big(p_{x-R}^{(0)},p_{x}^{(0)}\big)} ,
\label{eq:Cbound}
\end{equation}
the bound being saturated when the arcs coincide.  Since $p^{(0)}\to0$ in the resonance-starved regime,
$\mathcal C$ is \emph{not} bounded pointwise: this is the comb coincidence of Sec.~III\,C of the main text.  We
verified Eq.~\eqref{eq:Cbound} on $1194$ pairs at $x=200,10^3,5\times10^3$ with no violation, the
observed maximum reaching $10.4$ against a bound of $10.7$.

\begin{figure}[t]\centering
\includegraphics{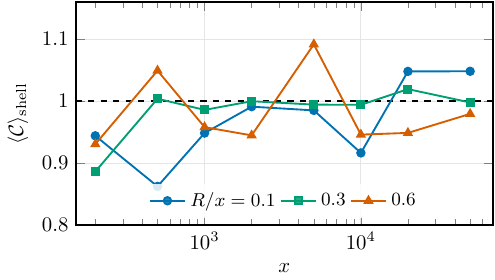}
\caption{Shell-averaged common-phase correlation factor, from the exact arc-overlap construction of
Eq.~\eqref{eq:Cbound}, at shell width $8\,\ell_{\rm comb}(x)$.  The dashed line is the
equidistribution value Eq.~\eqref{eq:Cequi}.  Over two and a half decades in $x$ and three fractional
separations the mean is $0.978$ with a spread of $0.052$ and only a weak upward drift, $0.029(13)$
per decade, whereas the
pointwise factor reaches $10.4$ at comb coincidences.}
\label{sm:figCxr}\end{figure}

Under uniform-offset equidistribution the shell average takes the exact value derived below.  \emph{If} the offsets $\Delta_{x-R,\,x}$ equidistribute
mod $\pi$ as $R$ ranges over a shell, the expected overlap of arcs of lengths $2\epsilon_{x-R}$ and
$2\epsilon_{x}$ at uniform relative offset is $(2\epsilon_{x-R})(2\epsilon_{x})/\pi$, whence
\begin{equation}
 \big\langle\mathcal C(x,R)\big\rangle_{\rm shell}=1
\label{eq:Cequi}
\end{equation}
\emph{exactly}, with corrections governed by the discrepancy of the sequence
$R\mapsto\Delta_{x-R,\,x}$.  A shell wider than several local comb periods,
$\ell_{\rm comb}\ll\Delta R(x)$ of Eq.~(9) of the main text, is expected to sample the offset broadly;
we do not prove equidistribution for the sequence $R\mapsto\Delta_{x-R,\,x}$, and treat it as a
coarse-graining hypothesis validated numerically below.  Evaluating the exact overlap numerically at $n=1/2$, $\beta=(\sqrt5-1)/2$ and $h/t_0=50\sqrt{x/200}$
(chosen to hold the chain deep in the resonance-starved regime as $x$ grows, so that
$p^{(0)}\simeq0.093$ throughout), with shell width $8\,\ell_{\rm comb}(x)$ and fractional
separations $R/x\in\{0.1,0.3,0.6\}$, gives
$\langle\mathcal C\rangle=0.978\pm0.052$ over $200\le x\le5\times10^4$, with only a weak upward
drift of $0.029(13)$ per decade across two and a half decades (Fig.~\ref{sm:figCxr}).  Widening the
shell brings the average to within a few percent of unity but not monotonically: $1.12$ at one comb
period, then $0.96$--$1.04$ for widths $4$--$32$.  Over the tested regime, shell averaging approaches the equidistribution value, independently of the matching-law test
in Appendix~B of the main text; the two are separate consequences of the same hypothesis.  Bond by
bond, by contrast, $\mathcal C$ may not be treated as $\order(1)$ at all.

\section{Conditional detuning and splitting distributions}

\label{sm:resdist}

For the exact $V=0$ bond mismatch, Eq.~(4) of the main text has the form $\delta=A_i^{\rm d}\sin\vartheta$ with uniform $\vartheta$.  The two preimages of $\delta$ in one period give
\begin{equation}
 \rho_i(\delta)=\frac{1}{\pi\sqrt{(A_i^{\rm d})^2-\delta^2}}.
\end{equation}
The probability $|\delta|<t_0$ is Eq.~(5) of the main text, and conditioning on it gives the
truncated arcsine density
\begin{equation}
 \rho_i(\delta\mid{\rm res})=
 \frac{1}{2\arcsin(t_0/A_i^{\rm d})\sqrt{(A_i^{\rm d})^2-\delta^2}},
 \qquad |\delta|<t_0 ,
\label{eq:rho-res-sm}
\end{equation}
which flattens to $1/2t_0$ in the resonance-starved limit.

For two adjacent bonds ($R=1$) the interaction-induced Hartree shifts are intrinsically
anticorrelated: in the bulk thermodynamic limit, for $\alpha>1/2$,
\begin{equation}
 \rho^H_{R=1}(\alpha)=-\,\frac{\zeta(2\alpha)-1+2^{-2\alpha-2}}{2\zeta(2\alpha)-1},
\label{eq:rhoR1-sm}
\end{equation}
approaching $-1/2$ as $\alpha\to1/2^{+}$.  The coefficient is evaluated for the half-filled
configuration ensemble of the model, in which different-site occupation products average to
$\langle n_jn_k\rangle=1/4$; for a fully occupied chain the last term in the numerator doubles to
$2^{-2\alpha-1}$.  Monte Carlo sampling of the model ensemble reproduces the half-filled
coefficient ($-0.31$ against $-0.309$ at $\alpha=1$).  This is a property of the static shifts and does not
enter the conditional detuning update of the main text.  Under $\Omega=\sqrt{4t_0^2+\delta^2}$, the two signs of $\delta$ yield the finite-$A_i^{\rm d}$ splitting density; taking $A_i^{\rm d}/t_0\to\infty$ gives Eq.~(36) of the main text.

The one-bond law is exact but does not imply a product law for two bonds.  Their exact joint measure is generated by the single phase $\phi$ as in Eq.~(8) of the main text and the shell mapping of its Sec.~VII.

\section{Coarse-grained shell baseline and the $\alpha=2$ boundary}

\label{sm:ratio}

After the joint shell average of Eq.~(46) of the main text, the fixed-fraction shell envelope in Eq.~(54) of the main text is
\begin{equation}
 G_{n,\alpha}(\kappa)=(1-\kappa)^{1-n}\kappa^{2-\alpha}.
\end{equation}
This function is useful only as a coarse-grained envelope; it is not a microscopic partner distribution because the exact resonant-bond process has the comb resolution Eq.~(9) of the main text.

A logarithmic-shell integration uses the measure $\dd R/R=\dd\kappa/\kappa$ and gives
\begin{equation}
 \int_0^1\frac{\dd\kappa}{\kappa}G_{n,\alpha}(\kappa)
 =\frac{\Gamma(2-\alpha)\Gamma(2-n)}{\Gamma(4-\alpha-n)}.
\end{equation}
This baseline is finite for $\alpha<2$ and singular at the lower endpoint as $\alpha\to2^-$.  In the actual SV chain the lower limit is set by the position-dependent comb spacing, $\kappa_0\sim x^{-n}/(2\beta n)$, which makes the equality case explicitly cutoff sensitive.

\
\section{Exact Fock-space covariance identities}

\label{sm:cov}

The real-space resonance construction and the Fock-space energy statistics are logically independent.  Consider half-filled configurations $S$ with $M=L/2$ particles and diagonal energy
\begin{equation}
 E(S)=\sum_i h_i n_i^S+\sum_{i<j}\frac{V_{ij}}{|i-j|^\alpha}n_i^S n_j^S.
\end{equation}
Let two configurations have Hamming distance $x$ and therefore $k=(L-x)/2$ common occupied sites.  Two distinct covariance conventions are useful and should not be conflated.

\subsection{Raw phase-averaged one-body second moment}

Averaging over the global phase gives
\begin{equation}
 \overline{h_i h_j}^{\,\phi}=\frac{h^2}{2}\cos(\theta_i-\theta_j).
\end{equation}
After the arithmetic average over all configuration pairs at fixed $k$, define
\begin{equation}
 Z_L=\sum_{j=1}^L e^{\ii2\pi\beta j^n}.
\end{equation}
The raw one-body second moment is exactly
\begin{equation}
 M_h^{\rm raw}(k)=\frac{h^2}{2}\left[
 k+\frac{M^2-k}{L(L-1)}\left(|Z_L|^2-L\right)
 \right].
\label{eq:Chraw}
\end{equation}
It is affine in $k$.  A raw-moment sign change exists over the physical Hamming range if and only if
\begin{equation}
 |Z_L|^2<L.
\label{eq:raw-sign-condition}
\end{equation}
For $0<n<1$, the endpoint asymptotic
\begin{equation}
 |Z_L|^2\sim\frac{L^{2(1-n)}}{(2\pi\beta n)^2}
\label{eq:Zasymp}
\end{equation}
implies the thermodynamic criterion
\begin{equation}
 \begin{cases}
 n>1/2: & \ \text{sign change occurs},\\
 n=1/2: & \ \text{occurs iff }\beta>1/\pi,\\
 n<1/2: & \ \text{no sign change}.
 \end{cases}
\label{eq:raw-sign-asymp}
\end{equation}
Equation~\eqref{eq:Zasymp} is strongly oscillatory at ED sizes and should not be used as a quantitative finite-$L$ approximation; the exact discrete $Z_L$ must be used whenever the raw convention is required.

\subsection{Configuration-connected one-body covariance: an exact sum rule}

For a \emph{fixed} phase realization, let
\begin{equation}
 E_h(S)=\sum_i h_i n_i^S,
 \qquad
 \overline E_h=\frac{M}{L}\sum_i h_i
\end{equation}
be the one-body Fock energy and its uniform half-filled configuration average.  The arithmetic fixed-$k$ connected covariance is
\begin{equation}
 C_{h,\phi}^{\rm conn}(k)
 =\left\langle E_h(S)E_h(S')\right\rangle_k-\overline E_h^{\,2}.
\end{equation}
Elementary counting gives the exact identity
\begin{equation}
 C_{h,\phi}^{\rm conn}(k)=
 \frac{k-M^2/L}{L-1}
 \left[
 \sum_i h_i^2-\frac1L\left(\sum_i h_i\right)^2
 \right].
\label{eq:Chconn}
\end{equation}
No property of the slowly varying potential was used in deriving Eq.~\eqref{eq:Chconn}.  It holds for every finite $L$ and for any fixed set of onsite numbers $\{h_i\}$, deterministic or random.  The potential enters only through the amplitude $\sum_i(h_i-\bar h)^2$, with $\bar h=L^{-1}\sum_i h_i$.

The pair-weight distribution of the overlap $k$ for two independently drawn $M$-particle configurations is hypergeometric,
\begin{equation}
 w_k=\frac{\binom{M}{k}\binom{L-M}{M-k}}{\binom{L}{M}},
 \qquad
 \sum_k w_k k=\frac{M^2}{L}.
\label{eq:hypergeom}
\end{equation}
Permutation symmetry of the fixed-$k$ configuration-pair ensemble gives $\langle n_i n_i'\rangle_k=k/L$ and $\langle n_i n_j'\rangle_k=(M^2-k)/[L(L-1)]$ for $i\neq j$.  Together with $\sum_i(h_i-\bar h)=0$, this is the complete origin of Eq.~\eqref{eq:Chconn}.  A connected correlator has zero $w_k$-weighted mean, and Eq.~\eqref{eq:Chconn} is affine; its zero therefore occurs exactly at the mean overlap.  At half filling,
\begin{equation}
 k=L/4
 \quad\Longleftrightarrow\quad
 x/L=1/2.
\label{eq:half-crossing}
\end{equation}
For $L\equiv2\ (\mathrm{mod}\ 4)$, $L/4$ is not an allowed integer overlap; the affine continuation
then crosses zero between the two neighboring physical sectors.
The previously quoted nonuniversal raw-moment zero is therefore not a connected-covariance invariant.  At half filling, the $x/L=1/2$ zero carries no information about $n$, $\beta$, $\phi$, or even the functional form of $h_i$; all model-specific one-body information resides in the covariance amplitude, while the random interaction contribution $C_V(k)$ supplies the additional quadratic overlap structure.

\subsection{Random interaction contribution}

For the disorder covariance of the zero-mean random interaction, define
\begin{equation}
 S_\alpha(L)=\sum_{r=1}^{L-1}(L-r)r^{-2\alpha}.
\end{equation}
The arithmetic fixed-$k$ average is exactly
\begin{equation}
 C_V(k)=\frac{V^2}{3}\frac{k(k-1)}{L(L-1)}S_\alpha(L).
\label{eq:CV}
\end{equation}
This contribution is quadratic in $k$.  Its asymptotics are
\begin{equation}
 S_\alpha(L)\sim
 \begin{cases}
 L\zeta(2\alpha), & \alpha>1/2,\\[2mm]
 L\ln L, & \alpha=1/2,\\[2mm]
 \displaystyle\frac{L^{2-2\alpha}}{(1-2\alpha)(2-2\alpha)}, & 0<\alpha<1/2.
 \end{cases}
\label{eq:Sasymp}
\end{equation}
The point $\alpha=1/2$ is a variance/square-summability threshold.  It is the same mathematical sum that appears in the core-restricted higher-body LIOM weight discussed in the main text, and it is distinct from the $\alpha=2$ resonance-network boundary.

A comparison to any fitted Fock-space ``covariance'' must therefore specify the convention.  Under the raw phase/disorder convention the relevant quantity is $M_h^{\rm raw}(k)+C_V(k)$, not the one-body piece alone.  Under a configuration-connected convention Eq.~\eqref{eq:Chconn} enforces the exact one-body sum rule at $x/L=1/2$, while the interaction term changes the total connected polynomial.  Without the original pair-aggregation and subtraction convention, a fitted positive form $a(1-x/L)^b$ should not be promoted to a fundamental correlation exponent.

\section{Fock covariance derivation}

\label{sm:covderiv}

For two $M$-particle configurations $S,S'$ with common occupied set of size $k$, the arithmetic fixed-$k$ average can be performed by exact counting.  Among the $M^2$ ordered pairs $(i\in S,j\in S')$, exactly $k$ are diagonal.  By permutation invariance of the fixed-$k$ configuration-pair ensemble, every ordered distinct physical site pair occurs with the same multiplicity.  Therefore
\begin{equation}
 \left\langle E_h(S)E_h(S')\right\rangle_k
 =\frac{k}{L}\sum_i h_i^2
 +\frac{M^2-k}{L(L-1)}\sum_{i\ne j}h_i h_j.
\label{eq:fock-raw-general}
\end{equation}
Phase averaging Eq.~\eqref{eq:fock-raw-general} gives Eq.~\eqref{eq:Chraw}.

Subtracting the square of the uniform configuration mean, $(M\sum_i h_i/L)^2$, and collecting terms gives
\begin{equation}
 C_{h,\phi}^{\rm conn}(k)
 =\frac{k-M^2/L}{L-1}
 \left[\sum_i h_i^2-\frac1L\left(\sum_i h_i\right)^2\right],
\end{equation}
which proves Eq.~\eqref{eq:Chconn}.  This derivation uses only permutation symmetry of the fixed-overlap configuration ensemble and $\sum_i(h_i-\bar h)=0$; it is therefore valid for arbitrary fixed onsite values $h_i$.  The hypergeometric mean $\langle k\rangle=M^2/L$ then makes the zero-crossing sum rule immediate.

For the random interaction contribution, permutation invariance of the same fixed-$k$ ensemble makes the counting factor exact: each unordered physical site pair appears in the common occupied set in the fraction
\begin{equation}
 \frac{\binom{k}{2}}{\binom{L}{2}}=\frac{k(k-1)}{L(L-1)}
\end{equation}
of configuration pairs.  Multiplying by the exact variance $V^2/(3r^{2\alpha})$ and summing over physical separations gives Eq.~\eqref{eq:CV}.  No statistical approximation is involved in this combinatorial factor.

Finally, the endpoint estimate
\begin{equation}
 \int_0^L\dd x\,e^{\ii2\pi\beta x^n}
 \sim\frac{L^{1-n}}{\ii2\pi\beta n}e^{\ii2\pi\beta L^n}
\end{equation}
yields Eq.~\eqref{eq:Zasymp} asymptotically.  Because subleading oscillatory terms are large at ED sizes, this formula should be used only for the thermodynamic sign criterion, not for finite-$L$ numerical normalization.

\section{Direct fixed-shell test of the pair-count scaling}
\label{sm:lamtest}

The pair-count scaling law
$\Lambda(x)\sim(Vt_0/h^2)\,x^{4-2n-\alpha}[B\ln x+C]$, Eq.~(55) of the main text, is
tested directly, with the
algebraic exponent fixed in advance.  On chains with the full detuning, active islands are
harvested in a centre window of $121$ sites at $x_0$ and in a one-sided shell at
$R=\kappa x_0$, $\kappa=0.45$, of width $4\ell_{\rm comb}$; each active island enters with its
exact doublet $(\Omega,\lVert m\rVert)$, and pairs are matched with the physical form-factor
coupling, $|\Omega_A-\Omega_B|<2\lambda(V/R^\alpha)|\Gamma_{AB}|$.  The matching scale $\lambda$
is fixed \emph{per $\alpha$} (not per $x$ or $h$) such that the window is dilute,
$q\le0.02$ at the smallest shell; the measured $\Lambda=n_RR$ then gives
$Y\equiv\Lambda h^2x_0^{-(4-2n-\alpha)}/(Vt_0)$, which the theory predicts to be
$h$-independent and linear in $\ln x$ at fixed $(\alpha,V)$, with nonuniversal
$B_{\alpha,V}>0$ and $C_{\alpha,V}$.  Parameters: $x_0=4000$--$24\,000$,
$h/t_0=100$--$340$ ($3000$ realizations per point), $V=t_0$, $\alpha=0.5$, $1$, $1.5$ ---
the bond side, the marginal point, and the fragmented side.

\begin{figure}[t]\centering
\includegraphics{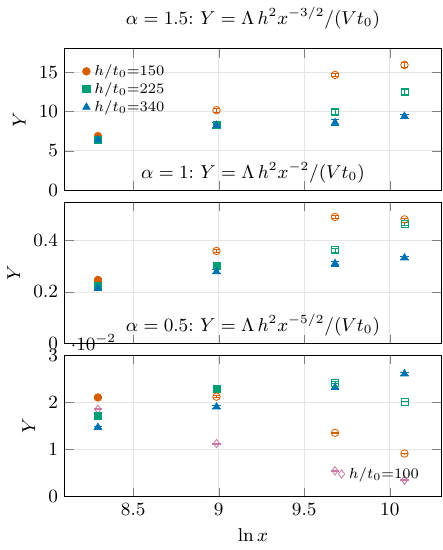}
\caption{Direct fixed-shell test of the pair-count scaling, the takeoff construction of the main text:
$Y\equiv\Lambda\,h^2x^{-(4-2n-\alpha)}/(Vt_0)$ against $\ln x$ at fixed shell fraction
$\kappa=R/x=0.45$ and fixed matching scale per $\alpha$ (dilute window, $q\le0.02$ at the
smallest shell), with the algebraic exponent \emph{fixed in advance}, not fitted.  Filled
symbols: points inside the resonance-starved window $w_0/\ell_{\rm comb}\le0.15$; open symbols:
outside it.  Batch-mean error bars (10 batches; often smaller than the symbols).}
\label{smfig:lamtest}\end{figure}

Two domain conditions bound the test and were fixed before the analysis: diluteness of the
matching window, and the resonance-starved condition $w_0\ll\ell_{\rm comb}$, implemented as
$w_0/\ell_{\rm comb}\le0.15$.  Figure~\ref{smfig:lamtest} shows the result.  Within the starved
window the $h^{-2}$ rescaling collapses the series to $7\%$/$1\%$ at $\alpha=1.5$,
$14\%$/$8\%$ at $\alpha=1$, and $33\%$/$18\%$ at $\alpha=0.5$; $Y$ grows with $\ln x$ with
positive fitted slope at every $\alpha$ ($B=1.53(12)$, $0.059(4)$, $0.0037(1)$); and the free
secondary power-law fit gives slopes $1.65$, $2.15$, $2.74$ against the prescribed $1.5$, $2$,
$2.5$ --- an excess consistent in sign and scale with the logarithmic correction plus
finite-window effects, though the accessible $\ln x$ interval ($\simeq1.8$ inside the window)
does not distinguish $B\ln x+C$ from a more general slowly varying correction.  Most
importantly, the fragmented side collapses at least as well as the bond-side control: there is
no deterioration that grows with $w_0$ at $\alpha=1.5$, which is the failure mode a breakdown
of the island construction by the large-cluster tail would produce.

\begin{figure}[t]\centering
\includegraphics{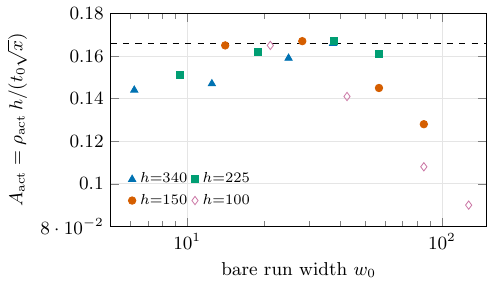}
\caption{Normalized active-object density $A_{\rm act}$ at $\alpha=0.5$, $V=t_0$, against $w_0$.
At small $w_0$, $A_{\rm act}$ rises toward a common plateau $\simeq0.166$ (dashed) reached for
$w_0\gtrsim15$; it then falls once the starvation ratio $w_0/\ell_{\rm comb}$ exceeds
$\simeq0.2$ (the falling points are those with $w_0/\ell_{\rm comb}=0.21$--$0.51$),
independently of how that ratio is reached.  The residual spread of
Fig.~\ref{smfig:lamtest} is carried by this factor squared; the matching fraction itself is
$h$-independent to $10$--$15\%$ throughout.}
\label{smfig:Aact}\end{figure}

Outside the starved window the collapse breaks in both directions --- $Y$ rises by up to
$60\%$ at $\alpha=1.5$ and falls by up to a factor $3$ at $\alpha=0.5$.  The origin was
localized by decomposition: the matching fraction at fixed window is $h$-independent to
$10$--$15\%$ at every $x$ (and its $R^{\alpha}$-scaled value grows with $\ln R$), while the
active-object density departs from its assumed $\rho_{\rm act}\propto t_0\sqrt{x}/h$ form.
Figure~\ref{smfig:Aact} isolates this factor at $\alpha=0.5$, including a dedicated $h=100$
series: at fixed $x$, decreasing $h$ while remaining within the resonance-starved window
increases $w_0$ and brings the normalized active-object density toward its asymptotic value
($A_{\rm act}\simeq0.144\to0.166$ over $w_0=6\to40$), identifying the residual spread as a
local-object finite-$w_0$ correction rather than a failure of the pair-matching scaling; past
$w_0/\ell_{\rm comb}\simeq0.2$ the density falls, marking the end of the starved regime.  The
breakdown is therefore a loss of the object-density asymptotics at the window boundaries, not
of the long-range matching mechanism.

\end{document}